\documentclass{ruthesis}
\usepackage{graphicx}
\usepackage[american]{babel}
\usepackage{float}
\usepackage{upgreek}
\usepackage{etoolbox}
\usepackage[style=apa,sortcites=true,sorting=nyt,backend=biber]{biblatex}
\usepackage{amsmath}
\usepackage{amsfonts}
\usepackage{mathtools}
\usepackage{fixltx2e}
\usepackage{algorithm}
\usepackage[noend]{algpseudocode}
\usepackage{pdflscape}
\usepackage{siunitx}

\makeatletter
\def\BState{\State\hskip-\ALG@thistlm}
\makeatother

\DeclareMathOperator*{\argmax}{\arg\!\max}
\DeclarePairedDelimiter{\norm}{\lVert}{\rVert}

\DeclareLanguageMapping{american}{american-apa}
\begin{document}

\def\mathbi#1{\textbf{\em #1}}
\title{STOCHASTIC APPROXIMATION EM FOR EXPLORATORY ITEM FACTOR ANALYSIS}
\author{Eugene Geis}
\program{Rutgers Graduate School of Education}
\director{Gregory Camilli}
\submissionyear{2019}
\submissionmonth{October}
\abstract{The stochastic approximation EM algorithm (SAEM) is described for the estimation of item and person parameters given test data coded as dichotomous or ordinal variables. The method hinges upon the eigenanalysis of missing variables sampled as augmented data; the augmented data approach was introduced by Albert's seminal work applying Gibbs sampling to Item Response Theory in 1992. Similar to maximum likelihood factor analysis, the factor structure in this Bayesian approach depends only on sufficient statistics, which are computed from the missing latent data. A second feature of the SAEM algorithm is the use of the Robbins-Monro procedure for establishing convergence. Contrary to Expectation Maximization methods where costly integrals must be calculated, this method is well-suited for highly multidimensional data, and an annealing method is implemented to prevent convergence to a local maximum likelihood. Multiple calculations of errors applied within this framework of Markov Chain Monte Carlo are presented to delineate the uncertainty of parameter estimates. Given the nature of EFA (exploratory factor analysis), an algorithm is formalized leveraging the Tracy-Widom distribution for the retention of factors extracted from an eigenanalysis of the sufficient statistic of the covariance of the augmented data matrix. Simulation conditions of dichotomous and polytomous data, from one to ten dimensions of factor loadings, are used to assess statistical accuracy and to gauge computational time of the EFA approach of this IRT-specific implementation of the SAEM algorithm. Finally, three applications of this methodology are also reported that demonstrate the effectiveness of the method for enabling timely analyses as well as substantive interpretations when this method is applied to real data.}

\beforepreface
\acknowledgements{I'd like to thank Greg Camilli, first and foremost, for helping me so much to finish this work at a pace that made life manageable. I'd also like to thank Suzanne Brahmia for her enthusiasm for psychometrics, and Ken Fujimoto for being excited about this research when the rest of the reviewers kept refusing to acknowledge this methodology. Jimmy de la Torre thankfully brought me into his program and taught me the value of pushing back against stubbornness. Drew Gitomer showed me that concepts and ideas that challenge the status quo can be reasonably discussed, but also politically harmful, and it was wonderful to listen to a clear and reasonable person in his sharing of that experience. Thanks to Charlie Iaconangelo for providing the social glue of the program, Nathan Minchen for his painstaking adherence to detail, Wenchao Ma for making all of us look bad, and Kevin Crouse for keeping my cynicism in check. Also, to the inspiring patience of the statistics professors who worked with us and carried our brains over the conceptual hurdles... especially Bill Strawderman (with all of his friendly quips about physicists), Harry Crane (who boxes in order to deal with our stupidity), and Javier Cabrera (who quietly demonstrates his unbelievable intellectual superiority without even trying). Lastly, to the supporters in the hallway, Gordon Stankavage, Matt Winkler, Colleen McDermott, and the assorted professors who allowed me to corrupt the integrity of their classrooms, Melinda Mangin, Chia-Yi Chiu, and Drew again.

From an egoic point of view, the intention of completing a second PhD was whimsical at the outset. Seeking to escape the social myopia of the hard sciences, I found a better salary as a public high school science teacher, and the freedom to enjoy my nights, weekends, and summers had little equivalency in the subterranean world of experimental particle physics. In my summers I found a wonderful opportunity to assist a gifted and inspirationally kind \textit{tabletop} physicist, Thomas Kornack, who demonstrated that growth can be exponential. He invited me into his 500 square-foot lab, taught me about state-of-the-art precision magnetostatics, and most importantly: he taught me that my skills have value. In three summers, he was in a 5000 square-foot experimentation and engineering facility, and I felt blessed that he'd pay me a better hourly rate than my teacher's salary to do some of the coolest shit I've ever worked on.

Simultaneously, my school district did the opposite. Nothing got better. Campbell's Law became my everyday experience. The Hawthorne Effect was the unrepetent consequence of an overhyped, underfunded, and micromanaged governing structure, exacerbated by a unionized workforce that relinquished any semblance of responsibility for quality control. The school was a distillation of the progeny of haves and have-nots. The haves were competitive with the fortunate sons and daughters of millionaire financiers and political legislators, while the have-nots pushed the policy-makers to invoke the assistance of the educational utility players: the science teachers. Testing shows a lack of critical thinking skills: do more labs. Testing shows a lack of reading comprehension: have them write more lab reports. I tried to coach two activities, and the compensation amounted to an average increase of about \$7 per hour, with an aggregate non-negotiable investment of time that amounted to approximately 700 hours each year. If one were to split this hourly wage by the number of students involved, the rate would drop to a little less than 20 cents per student per hour. Please reflect on that and reach your own conclusions about the economic sustainability of public education and extracurricular programs, in general.

As a quantitative practitioner, my distrust of assessment metrics and the administration's reactivity to such measures became too much to bear. The president of our local board advocated for intelligent policy, but seemed to have no perspective from the boots on the ground. Our superintendent was the classic self-aggrandizing fiscal conservative demanding a 20\% increase in his own taxpayer-funded salary. I became critical of the institution, the union, and the teachers who could afford to annex all of the incremental salaried coaching slots since gym class required near-zero lesson planning, preparation, grading, and adaptability. We would train on fairness and equity, but the hypocrisy of the unspoken rules of employment and fealty to the age-ism of union leadership was beyond reproach.

It is an understatement that I am a passionate individual. I sought a window into the operational mechanisms of this broken system, one that is so resitant to functional improvement. Rutgers was a home where I found my passions 15 years earlier, and it again became a home for the next level of my personal inquiry. Statistics was taught by their practitioners with care and concern. Inferences that affect outcomes evoke ethical consequences! Who would've thought!? While exceptions to every rule can be found, my public education liberal arts administrators rarely provided a bridge between quantitative outcomes and qualitative interpretations beyond their own self-referential narrative. Even the Gates Foundation has proven the inadequacy of the most current research to provoke parctical effective change within the current educational environment. This implies \textit{systemic failures}, whether it's the current story being shared by educational researchers, or the futility in applying perturbations to a stagnant environment.

This work, inspired by my advisor, Greg Camilli, comes from the moral motivation to better scrutinize the inference of large-scale assessments that may have the gravitas to shift the educational policies of nation-states. Shanghai's 9-year-olds may have outperformed the rest of the OECD's sample population on the math module of the PISA, but their local emphasis on algorithmic problem solving of fraction-based arithmetic for children between age 8 and 9 would undoubtedly play a role in univariate test scores on an assessment from which a majority of items (nearly 60\%) are comprised of questions about fractions! Would it be of interest to policymakers if Russian students were the best in the world at solving for an unknown variable? Did the OECD mention that the German students scored the highest on items formulated around the skills of data interpretation? 

A methodology for extracting statistically significant factors has not been integrated into the reporting of these global assessments, and it is highly likely that the validity of these falsely univariate assessments will suffer. The work of this dissertation is an attempt to make some headway into a statistical approach that subverts this pathology of psychometric imprecision and political near-sightedness that result in the impossibility of verifiable \textit{ex ante} economic analysis; we cannot afford to make impulsive changes driven by budgets only to check our work and massage the results at a later date... but it is a near certainty that this is now the accepted mode of operation of our ivory tower economists, many of whom are substantially incentivized by their funding institutions rather than our children's collective benefit. 

As economists are always apt to push the objective fundamentals of marginal costs and economies of scale, the \textit{quality} of the human experience is rarely well-quantified nor has an effort been mobilized to adequately measure outcomes \textit{ex ante} of an economic decision. Certainly \textit{ex ante} evaluations are fraught with difficulties, especially as unpredictable consequences arise from new policies or technologies \parencite{marschak1974economic}, but the impacts on learning and higher order thinking from the digital transformation of the classroom experience of college students cannot be understated. Having personally been enrolled in several online courses with more than 280 aggregate college credits across my transcripts, the experience of online education is very poor and quite inimical to critical thinking. It has been my experience that juvenile criterion measures are employed \textit{ad nauseum} and interactivity with higher order concepts is reduced to the struggle of gleaning wisdom from an AOL chatroom. \textcite{harasim2012learning} confirms that my sentiment of substandard discussion in online learning is not unfounded. 

Unfortunately, as our global policies further adhere to purely economic arguments, it is also woefully clear that econo-\textit{misseds} (especially in the United States) earn great incentives for motivating policies of automation and scale that provide monetary rewards for capitalists while reducing the value provided to the proletariat; throughout popular culture and the national news, there is a clear overarching theme of socioeconomic decision-making based on elitist crony capitalism driven by economic math over the interest of our society.

Finally, acknowledgements to the National Science Foundation as this work was supported in part by grant 1433760 from the NSF, Education and Human Resources.}
\dedication{This work is dedicated to smart creatives that haven't let the obscene incentives of corporate greed lull them into contributions to the vampire squid; at some point it is hopeful that your conscientiousness will conquer the value judgments of economics, inspiring other smart creatives to pivot toward an ontological system that values life, liberty, and the pursuit of happiness instead of the worship of war, polarizing politics, manipulative media, click-bait, marketing, pharma, credit bubbles, and debt.}
\figurespage
\tablespage

\afterpreface

%




\chapter{Introduction}

Given the technical nature of this dissertation, an informal discussion of the landscape is warranted for bridging mathematics and the blurry nature of reality. Statistical methods emerged as a byproduct of careful and sometimes predatory statesmanship and thus derives its etymology. As the governing policy of a nation-state is required to increase in its efficiency, therein emerges a natural motivation for predictability, and in its wake, measurement. The motivations of the earliest trailblazers of statistical methods may have been questionable, but those innovations continue to address features, appetites, and behavior of living things; qualitative distinctions and their measurable artifacts are indeed \textit{fuzzy}. The Origin of Species inspired Francis Galton to apply Darwin's concepts to the measurement of differences in humans, going so far as to pioneer the \textit{questionnaire}. The adjacent but separate investigation of psychology emerged from Herbart's measurement of sensation and became further crystallized under Wundt's philosophical outline of psychology as an empirical science. Inseparable from this historical context, a system for measuring human faculties implied to many of these thinkers that there is a potential for enhancing human behavior and development. 

The ability of a human being to improve or progress, in a general sense, is implicit, and both the philosophical and economic gains associated with the relative growth of individual and collective human knowledge are intrinsic to the motivations behind our efforts to refine our psychological measurement toolkits. As such, the implementation of measurement in psychology straddles a rift between qualitative judgement and quantification. Instruments of assessment must be applied with a predictable resolution of these two criteria lest the validity of the instruments' respective measurement goals become too strongly characterized by frivolous whimsy or authoritarian determinism. 

The design of educational assessments is required to adhere to the standards of a meta-assessment defined within the rubrics of reliability and validity. Reliability is a constraint enforcing consistency; a test's scores should correlate with an expression of a psychological characteristic, e.g. an examinee's ability across a domain of knowledge or skill. If a researcher creates items providing statistical information about a test population's ability within a domain, reshuffling, choosing subsets, or replacing questions designed to measure similar content should not result in scores that do not correlate with the original assessment. Validity, within applications of psychological assessments and their--potentially law-abiding definition of--quantitative and qualitative inferences, seeks to constrain which inferences may be permissible as well as the uncertainties of the inferences regarding test scores that may be concluded from such a test. As measurement is the price for reducing uncertainty, psychometrics has emerged as the quantitative application for control of statistical inference in measurement applied to psychological characteristics. A test is reliable if it correlates with other administrations of itself, subsets of itself, or other ``parallel'' tests, i.e. tests already shown to measure what the new test claims to measure. The validity of a test requires a more rigorous study as one must consider the decisions made as a result of a test score. Beyond the informal discussion in this chapter, the argumentation of validity will largely be ignored.

Administration of assessments are only pursued after thorough investigation of the statistical properties of the test, and only after this quantitative investigation can the test have the potential to withstand the rigors of a validity study. In general, while a statistical estimate's p-value may commonly be quoted, the quantitative interpretation constrains this numeric value's application to measurements of randomness alone. Mathematical models or logic structures that may be invoked to assess the quantitative properties of psychological assessments, especially when applied to latent constructs\footnote{The term `latent construct' is loosely defined here as a logically independent psychological descriptor, for example emotional intelligence, riskiness, or IQ. It is not reflective of its well-rehearsed definition within the academic sphere of `construct validity.'}, must first demonstrate robustness against randomness, i.e. the assessment must be reliable. This management of entropy—by first asserting mathematical order and then pressure-testing its organization of randomness—summarizes the intended goals of statistical modeling and simulation. From a bird's eye view the process of (1) positing a model, (2) simulating outcomes, and (3) verifying the outcomes reflect experimental results, is how phenomenology is tested by science. The domain of psychology is firmly entrenched in this difficult niche, not only on a cultural macro-level but also on a micro-level as a clinical practice, and none-the-least within the halls of our local schools. Within the subdomain of an academic progression through educational milestones, we surmise that a student's thoughts and judgements are to become more thorough, rigorous, logical, or refined in their recollections, reflections, predictions, and decisions. 

In the practice of educational assessment, content experts make decisions about test design, psychometricians study the quantitative properties, and institutions administer the tests. While being more politically and socially conspicuous than ever before, the goals of assessment have also become more consequential, and thus the practice of psychometrics must become more precise. This advancement of assessment in education and psychology has motivated psychometricians to expand their methodologies, not only in developing mathematical models for statistical measurement, but also in computing the results of these estimation algorithms. Further, as a test item bank grows and the tested populations increase, statistical computation becomes more difficult and time-consuming, especially as our mathematical models are only an approximation; stochasticity is inherent to measurement and can never be completely avoided. While this work focuses on applied statistical methodology, the stated motivation for the increase in precision not only relates to the applied statistics, but also to the pertinence of construct validity and the interpretation of test scores. 

Psychometrics hails from a statistical foundation that is firmly entrenched in the quantitative disciplines. The dominant implementation is typically characterized by suppositions of latent attributes as applied to measures of student performance for the evaluation of relative outcomes, subject proficiency, and the attainment of specific domains of knowledge or expertise. Psychology has also adapted this framework for many purposes, a few notable instances include the Big Five personality tests, clinical purposes such as assessments of social anxiety, fatigue, or stress, or in business applications such as management coaching and conflict resolution. Items, the individual questions on a test, must undergo rigorous studies to ensure their applicability to the domains being assessed as well as their consistency from one administration of the test to another; another way of stating this, the items must show content relevance within a construct. Towards this end, classical test theory evolved into more complex methods such as IRT (item response theory) and CDM (cognitive diagnosis modeling). In other words, test design has evolved from a narrow quantitative view of the entire assessment to the piecewise construction of items \parencite{loevinger1957objective} to form a measurement tool used to compute informative profiles of the subjects being assessed. The estimation of characteristics of items (like difficulty) and qualities of persons (like proficiency or level of skill) in the framework of IRT is now a common practice within educational and psychological assessments, notably for calibration and statistical information, as well as test design and validity studies.

Univariate latent scales of psychological proficiencies such as mathematical ability or anxiousness conjure an optimistic view of experts' ability to craft an assessment that measures one thing, and one thing alone. For example, many primary school teachers may have the belief that a math quiz constructed with word problems assesses mathematical ability, but a strong reading ability of the English language may be implicit to success on such a quiz. The consequence of this realization enforces the recognition that such an assessment is multidimensional, or it measures more than just mathematical ability. Further, half of the word problems may focus on the rules known as the `order of operations,' while the other items may focus on fractions; for students to earn successful scores on a similar test structure, it would require them to demonstrate (1) strong reading comprehension, (2) mastery of order of operations, and (3) mastery of fractions. The first requirement is important for the entire test, but the second and third requirements are only important for their respective items. This type of assessment would be very different from a focused `order of operations' assessment with very little required reading, as the former contains items dependent on multiple latent proficiencies of each examinee while the latter is far more appropriate to the application of a univariate latent scale. In the word problem assessment, one dimension is important for the entire test, but a second dimension (order of operations) is theoretically\footnote{A test may be designed to assess a specific number of latent traits but may require domain expertise not immediately apparent to the experts constructing the test; for example, our primary school teachers overlooking reading comprehension as a necessary tool for solving a mathematical word problem. Order of Operations inherently requires mastery of multiplication, division, addition, subtraction, and exponents, thus helping elucidate the difficulties of good test design and the subsequent investigations pertinent to make valid inferences of a test's diagnostic implications.} required for the first half, and a third (fractions) for the second half. This type of assessment is referred to as a `bifactor' design where two dimensions load on each item. If the test were redesigned using mathematical formulae rather than framing each question as a word problem, the same test has a subscale design. In other words, eliminating the necessity of reading comprehension, each item may be considered to load on each dimension independently; fractions and order of operations do not comingle within any single test item. The statistical machinery required to analyze these multidimensional designs needs elaboration. Thus, there has been great interest in estimation procedures for MIRT (Multidimensional IRT). The extension of IRT into a multidimensional latent space invokes a more nuanced approach to applications of IRT, and while this is commonly implemented as a confirmatory analysis, exploratory methods for assessing dimensionality should not require a significant investment of time or dollars. 

When transitioning from unidimensional to multidimensional calculations, implementations of MIRT can be fraught with computational difficulties. Traditional EM (expectation-maximization) methods require integration and moving from unidimensional EM into higher dimensions severely impacts the work required for accurate estimation; using an integral by quadrature, the computation expands exponentially in the number of grid points required. This problem is informally referred to as the `curse of dimensionality.' With regards to the hypothetical test of word problems, this computational difficulty not only pertains to the extracting the statistical properties of the test, but also to the statistical information that can be isolated to the content of each item. This work proposes a set of statistical tools for assessing dimensionality, and while the impact on computation is the primary focus, the results of its application also provides quantitative metrics that can be used to inform the content as a test is constructed and submitted for evidence of reliability.

The dominant approach to MIRT, in practice, is confirmatory in nature. Psychometricians invoke an assumed structure, and estimation methods conforming to this structure. In our word problem example, the mathematical formulae governing the item structure contains a linear combination of latent abilities: all items would contain a single `Language Ability' parameter for the influence of reading comprehension, half would contain a second `Order of Operations Ability' parameter, and the other half would contain a second `Fraction Ability' parameter. There may be correlations between these abilities, or more appropriately, a correlation matrix that describes the relationship between these dimensions of ability. Thus, the psychometrician may parameterize these abilities to be orthogonal (or statistically independent), or to estimate each dimension while accounting for a correlation between them. Whether simulating test data or fitting to real data, each choice of an imposed `structure,' e.g. orthogonal abilities vs. correlated abilities, reflects a confirmatory approach over an exploratory one. In the word problem example, we are ab initio invoking three separate abilities, but in practice real data may reveal that there are two substantive dimensions, or even five. The complexity of multidimensional structure and the expertise of the SMEs (subject matter experts) that design such assessments are usually enough to constrain certainty in making confirmatory assumptions. Modern test reports, further, may show subscores that are estimates of an examinee's ability on subsets of interrelated items. While arguments can be made that there is a mathematical justification for reporting overall scores or subscores, it is incontrovertible that the complexity of inferences increases dramatically as the dimensions of proficiency increase. 

In measuring multidimensional assessments, MIRT is implemented as a mathematical model in which parameters may be estimated using a multitude of methods. As the number of parameters increases, certain methods such as the EM (expectation-maximization) algorithm can become computationally unfeasible. Specific to MIRT, convergence to a solution can also be complicated by the indeterminacy of rotations in the multidimensional latent space of abilities. As the number of items and dimensions increase, the estimation of item parameters becomes unwieldy; for this reason many researchers choose a confirmatory approach over an exploratory one, implementing parameterizations guided by the a priori test design rather than data. 
There is a wide range of applications for MIRT models in the field of education, but for other fields as well. For instance, there is currently much interest in understanding the structure of quality of life assessments based on categorical item responses in the health sciences. Other areas of application include clinical measurement related to anxiety disorders, alcohol problems, and physical functioning. While many assessments are developed according to a theoretical blueprint, it is often the case that the empirical item structure diverges from the expected structure. Thus, exploratory, as opposed to confirmatory, MIRT analysis is an important tool for critiquing assessments as well as for the theoretical development of the constructs targeted by those assessments.

Exploratory MIRT models only exist in the very specific sense in which (1) a number of factors is chosen, (2) an unrestricted factor model is obtained (that is, factor loadings are freely estimated assuming uncorrelated latent variables), and (3) factor loadings are rotated for the purpose of interpretation. The first and second steps above are qualitatively different in a typical EFA (exploratory factor analysis) model for continuous variables (though rotation remains an important issue). For example, in EFA maximum likelihood factor analysis, a series of factors are obtained in order of decreasing eigenvalues (e.g., principal axis factoring) prior to rotation. Thus, information is obtained to judge the number of factors as well as the relative strength of those factors. 

In this research, there are three specific goals: (1) to apply a novel method of computational estimation of an MIRT model on highly multidimensional data, (2) to demonstrate its utility in reproducing simulation parameters with accuracy and speed comparable or better than the commercial software most oft used by psychometricians, and (3) to detail the application of the Marchenko-Pastur law as a statistical test for the multidimensionality of dichotomous and polytomous data. 

Following this introduction, the first section of this dissertation is the presentation of the SAEM (stochastic approximation EM algorithm) for MIRT building on the work of Meng and Schilling (\cite{MengSchilling1996}); they extended the gibbs sampling method of the two-parameter ogive [2PNO] introduced by Albert (\cite{Albert1992}).  One important aspect of the algorithm is its computational advantage; that the generation of augmented data for estimation can be computed using ``embarrassingly parallel'' backend sockets. As a consequence, the computational workload is easily parallelized and computation time is reduced. A second important aspect of this algorithm is the flexibility with which the sufficient statistics can be used for parameter estimation. The innovation comes from the inclusion of the RM (Robbins-Monro) procedure, demonstrated to further reduce computation time for stochastic approximation of multidimensional models applied to large scale assessment. The model used to explore this algorithm has a closed form solution for the likelihood though there is no closed form solution for the model parameter estimates.

In the second section, univariate and multidimensional dichotomous and polytomous data are simulated and then estimated within the SAEM method. Simulation conditions will include small and large numbers of items $J$ = \{30, 100\}, small and large numbers of examinees $N$ = \{5K, 10K, 100K\}, and modifications to the RM gain constant. Inclusion of a guessing parameter will also be implemented for a simulation condition of dichotomous data. Small and large numbers of dimensions $Q$ = \{1, 3, 10\} are simulated in bifactor and subscale test designs. The indeterminacy for rotations within the latent subspace allow for the novel solution of monitoring the convergence of the covariance matrix of augmented data; a sufficient statistic of the gibbs sampler. Computation time will be shown to increase only incrementally as dimensionality increases. The studies will also explore the estimation of standard errors for the special case of the univariate SAEM implementation. There are several options available with which to implement the calculations for standard errors. As this approach uses gibbs sampling, multiple approximations of MCMC sampling errors will be compared with the error approximations obtained from the inversion of the Hessian. As the probit distribution function is nearly identical to a special case of the logistic, 2PL (two-parameter logistic) standard error approximations can also be calculated and compared. Empirical standard errors can also be derived using samples from the converged posterior; this estimate is accomplished in similar fashion to the methodology for sampling plausible values in the reporting of NAEP (National Assessment of Educational Progress) results (\cite{MislevyBeatonKaplanSheehan1992}).  The SAEM-MIRT estimation will also be applied to three real assessments, FCI (Force Concept Inventory), CCI (Chemistric Concept Inventory), and QOL (Quality of Life) data. 

In the third section, the application of random matrix theory to the augmented covariance matrix will be explored. The Marcenko-Pastur law (\cite{MarcenkoPastur1967})  can be used to validate the eigenvalue distribution of this covariance validating the dimensionality of the data. Simulated random numbers and simulated and real data from the previous section will be used to examine the properties of the converged eigenvalue distributions. Controlled adjustments to simulation conditions will also be tested to provide sensitivity analyses, i.e., the relative magnitude of parameters and the structure of the assessment will change the outcome of statistical tests of the eigenvalue distribution of augmented data. The extent to which these perturbations are detectable will constrain the utility of this analysis. 

Finally, reviewing applications of this estimation procedure and related literature, the research to be presented here is by no means closed. As implemented today, the code created for this algorithm approaches the computational speed of FlexMIRT, the industry standard for MIRT; as this code is written in R, there is ample opportunity to enhance its speed. More obvious next steps should include a robust approach for polytomous errors, simultaneous estimation of both dichotomous and polytomous data, and missing data imputation. The algorithm's ability to return the converged covariance matrix also presents another opportunity for a didactic on rotations in the latent subspace.

\chapter{MIRT Estimation Approaches} \label{ch:theory}
\setcounter{figure}{0}
\setcounter{table}{0}

\section{Item Response Theory}
Investigations into the extent to which a test measures what it is supposed to measure motivates our statistical toolkit. From the perspective of the frequentist statistician, Classical Test Theory posits a test score as a measurement that has properties of a random variable; it is comprised of a \textit{true score} $T$ and some additive error $E$. This model implicitly defines an assessment as the instrument of measurement. The statistical assumptions are convenient for providing mathematical properties of parallel tests, reliability, and internal consistency (\cite{allen2001introduction}). As the measurement properties of a test need to be investigated, test construction and thus item analyses inevitably follow. A large number of item formats are accessible to the researcher, but we are primarily interested in the properties of the measures provided by test items. For achievement tests, the properties of these test items typically correspond to our conceptions of difficulty and discrimination; while the `difficulty' of a question is intuitive, the discrimination can be less so. For illustrative purposes, the ICC (item-characteristic-curve) is helpful.

\begin{figure}
\centering
\includegraphics[width=5.5in]{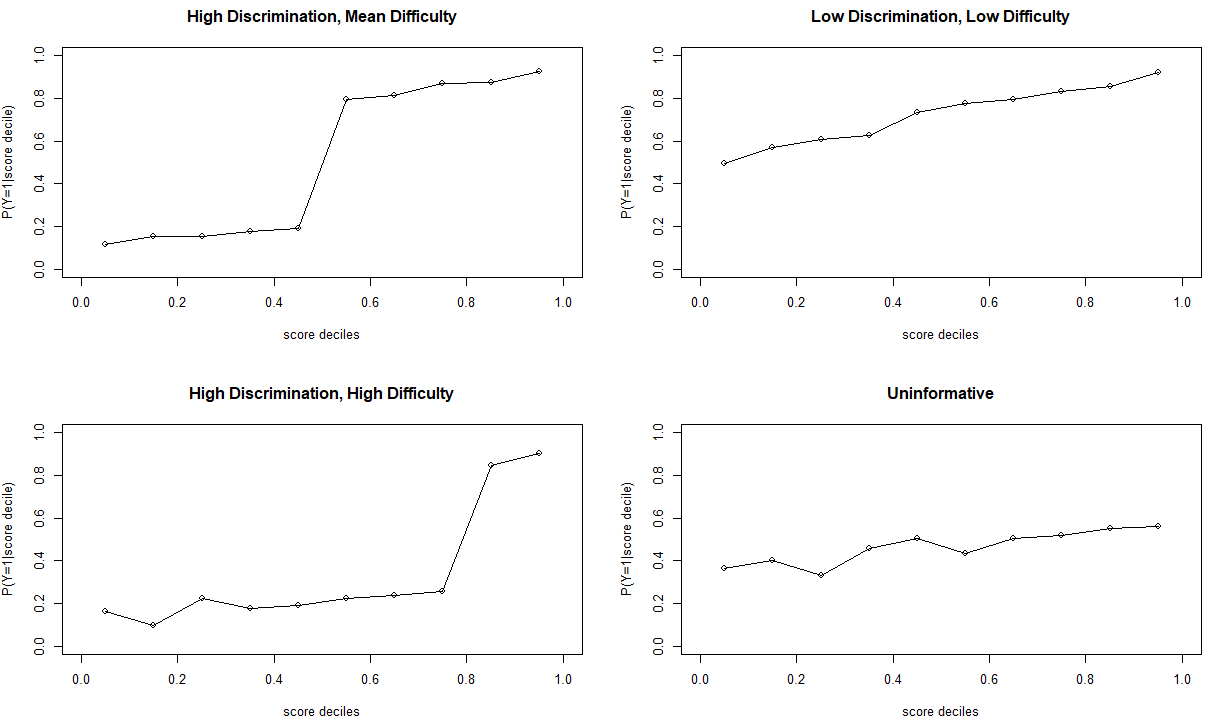}
\caption[Item Characteristic Curves]{
The four plots show item characteristic curves with varying difficulty and high or low discrimination. An ICC with low statistical information is also shown.}\label{fig:iccdecile}
\end{figure}

For an arbitrary multiple-choice test consisting of several items, data can be used to construct ordinal deciles of examinee scores. A bernoulli random variable, $Y$, is chosen to represent a multiple-choice item; 1 indicates a correct answer and 0 is incorrect. An ICC is constructed by plotting the conditional probability of achieving a correct answer to a multiple choice question of the test $P(Y=1|decile)$. In Figure ~\ref{fig:iccdecile} four examples of item characteristic curves are shown. Each point represents an overall probability of a correct answer on a single item for a given decile. There are ten points in each plot corresponding to ordinal deciles of lowest to highest scores. To maintain simplicity of interpretation, achievement tests should imply that examinees with high scores have higher probabilities of answering items correctly than examinees with low scores. In general, the data for the four items in Figure~\ref{fig:iccdecile} show that as an examinee's score increases, there is an increase in the probability for answering each item correctly. At this point, only one assumption is being made about the inferences that can be applied given this score--higher levels of ability correspond to higher probabilities of a correct answers across the test; this score can be used to infer \textit{ability}, or a single dimension of $\theta$.

As intutition might suggest, if an item is considered `difficult' the majority of examinees will have less than a 50\% chance to answer it correctly; this situation is represented by the lower left plot in Figure ~\ref{fig:iccdecile}. In contrast, if the majority  of examinees have a better than 50\% chance to correctly answer an item, it is considered to be of low difficulty and this is shown with the upper right plot in the same figure. In terms of discrimination, both plots on the left side of Figure ~\ref{fig:iccdecile} depict a large increase in the probability of a correct answer; these ICCs denote items considered with a high discrimination. The two plots on the right side of Figure ~\ref{fig:iccdecile} show much lower discrimination, and some might argue that the lower right plot shows no discrimination at all. In the context of the discrimination shown in the top left plot, if a student were to correctly answer this item it is highly likely that this student would fall in the upper half of examinee scores. Similarly, if a student scored in the top two deciles of scores, it is highly likely that this student correctly answered the item which has an ICC like the bottom left. In other words, an item with high discrimination provides a strong statistical relationship with the latent ability that the test is measuring. The ICCs on the right side of the figure do not provide much statistical information in regards to whether an examinee with a correct answer should fall in lower or upper deciles of the scores.

In moving to a psychometric methodology that more appropriately centers on the functioning of an item, ICCs for items on achievement tests typically take the shape of an S-curve as the aforementioned figure shows. Though there are multiple candidates for mathematically approximating this S-curve, two commonly used functions in IRT (item response theory) are the logistic and the normal ogive (the cumulative probability for a normal distribution). If the S-curve is constrained such that it approaches zero at the lowest ability, and one at the greatest ability, each of these cumulative probability functions approach 1. One general form of the logistic is known as the 3PL (three-parameter logistic) and is written as

\begin{equation}
\label{eq:p3pl}
P\left( {{Y} = 1|{\theta}} \right) = \gamma + \frac{1-\gamma}{1+e^{-1.7 \alpha \left( \theta - \beta\right)}}.
\end{equation}

\begin{figure}
\centering
\includegraphics[width=5.5in]{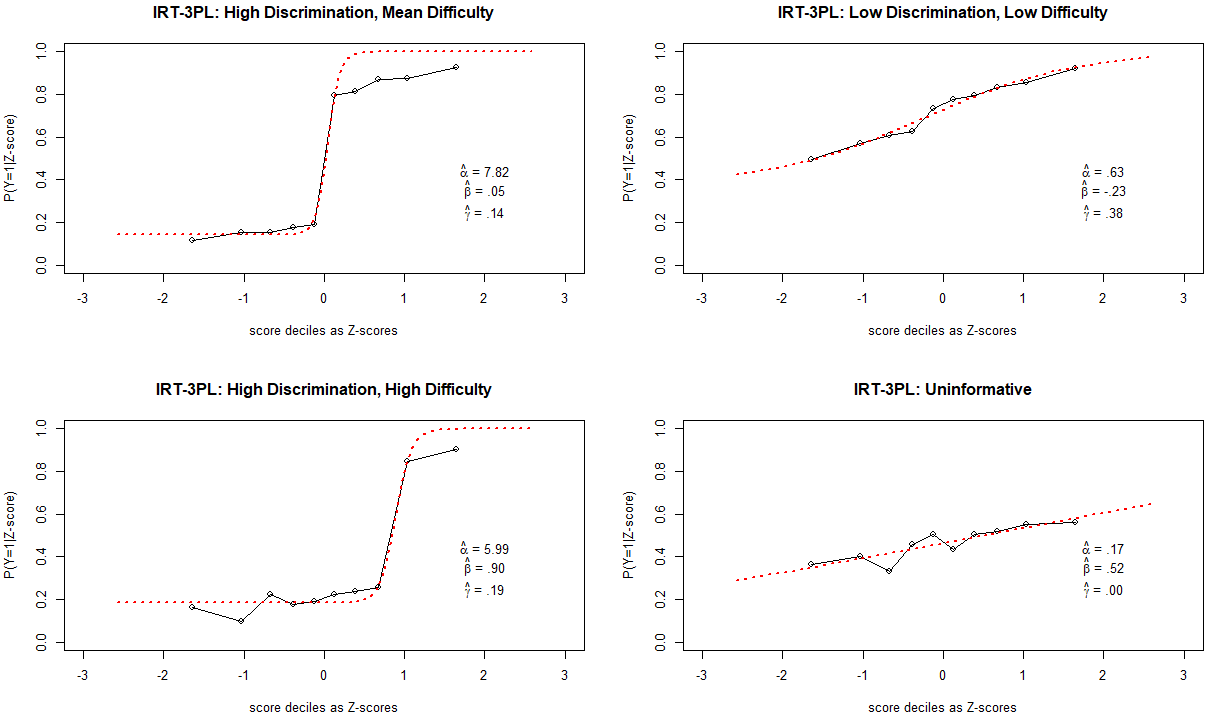}
\caption[Item Characteristic Curves]{
The four plots show the same simulation of data from Figure~\ref{fig:iccdecile} with the deciles converted to quantiles along a normal distribution. Item characteristic curves with varying difficulty and discrimination are shown and fit with the 3PL from Equation~\ref{eq:p3pl}.}\label{fig:icc3pl}
\end{figure}

In this equation, there is a straightforward interpretation; $\alpha$ is the discrimination parameter, $\beta$ is the difficulty parameter, $\gamma$ is a guessing parameter, and the constant of 1.7 is derived by attempting to equate this logistic to the normal cumulative distribution function after setting $\gamma=0$, $\alpha=1$, and $\beta=0$ (\cite{camilli1994teacher}). To better understand the form of this equation, one can attempt to fit this logistic to the data from Figure ~\ref{fig:iccdecile}. In Figure ~\ref{fig:icc3pl}, the deciles have now been converted to ten Z-scores corresponding to p-values of .05, .15, .25, ..., .85, .95.

The results of the fit of the 3PL are illustrative of the values of the parameters of the model; $\alpha$ increases as discrimination increases, $\beta$ defines the point on the horizontal axis where the slope of the curve is the greatest\footnote{In personality assessments and likert scale surveys $\alpha$ may have negative values, in which case the S-curve is reversed. For the purpose of this dissertation and the intended inferences in educational psychometrics, the approach will be constrained to $\alpha>0$.}, and $\gamma$ approximates the chance of randomly selecting the correct answer at the lowest ability. When $\gamma=0$, the logistic would be constrained to approach zero at negative infinity and this special case of Equation~\ref{eq:p3pl} is denoted as the 2PL (two-parameter logitistic). The 2PL is very useful for interpreting the logistic function as $\beta$ approximates the point at which the 2PL intersects with a predicted probability of 50\% and $\alpha$ is the slope of the logistic at that point of intersection. In using the 2PL or 3PL, there are expedient features of this mathematical convention for statistical calculations such as the likelihood and its derivatives, but the integration of the logistic is not as convenient as as a function that has a conjugate distribution. The 2PNO (two-parameter normal ogive) and three-parameter normal ogive are alternatives to the 2PL and 3PL, written as

\begin{equation}
\label{eq:p3pno}
P\left( {{Y} = 1|{\theta}} \right) = g + (1-g) \cdot  {\Phi \left(a \theta - b \right)}
\end{equation}

\noindent where $\Phi$ is the cumulative normal distribution function. In this equation, $a$, $b$, and $c$ are the correlates of $\alpha$, $\beta$, and $\gamma$ in Equation~\ref{eq:p3pl}. As with the logistic model, the 2PNO is a special case of the 3PNO where $g=0$. This model will be expanded to multidimensional latent traits and polytomous items in the next sections.
 
 
\section{Multidimensional Factor Model for Item Responses} \label{sec:MIRT}

\begin{figure}
\centering
\includegraphics[width=5.5in]{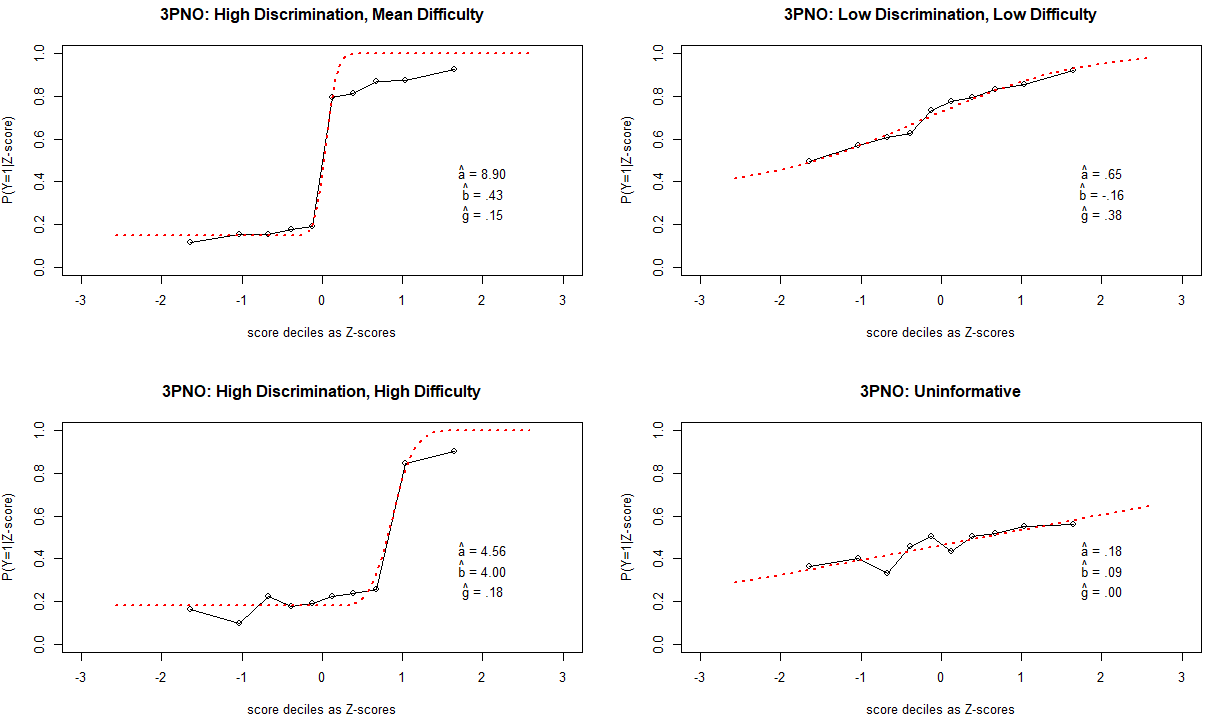}
\caption[Item Characteristic Curves]{
The four plots show the same simulation of data from Figure~\ref{fig:iccdecile} and ~\ref{fig:icc3pl} with the deciles converted to quantiles along a normal distribution. Item characteristic curves with varying difficulty and discrimination are shown and fit with a univariate 3PNO from Equation~\ref{eq:p3pno}.}\label{fig:icc3pno}
\end{figure}

Assume a set of test items $j=1,...,J$ and $i=1,...,N$ examinees. For ease of presentation, item responses of a single assessment are assumed to be dichotomous with correct responses scored as $Y_{ij}=1$ and incorrect responses labeled $Y_{ij}=0$. A more general approach is presented in the next section that is applicable to polytomous\footnote{Polytomous items are defined to allow for partial credit or likert responses, and thus can allow for a more refined assessment of an examinee's knowledge or preferences, respectively. In the case of polytomous scoring, the psychometrician can avoid the troublesome `guessing' parameter in Equation ~\ref{eq:p3pl} and its equivalent form in the 3PNO.} items, for which dichotomous scoring is a special case of binary outcomes. Assume there is a vector of $Q$ latent variables that account for an examinee’s observed item responses. Demonstrated in previous research (\cite{Albert1992}, \cite{beguinglas2001}, and \cite{fox2010bayesian}), the cumulative normal function (also termed the normal ogive) will be used to model item responses rather than the logistic function; both give virtually indistinguishable results when the parameters are properly transformed. In Figure~\ref{fig:icc3pno}, the same simulated data is fit using a 3PNO and it may be seen that the curves only differ in the concavity as they approach the horizontal asymptotes.

Normal ogive models like the 2PNO provide a more flexible tool for stochastic methods than the 2PL due to the complexity of the posterior of the logistic prior; for the 2PNO, a normal prior distribution results in a normal posterior distribution. Using the cumulative normal distribution function $\Phi$, a multidimensional model for a correct response on item $j$ presented to examinee $i$ is given by
\begin{eqnarray}
P\left( {{Y_{ij}} = 1|{\eta _{ij}}} \right) = \Phi \left( {{\eta _{ij}}} \right) \\
{\eta _{ij}} = {\bf{A}}_{j}{\bf{\uptheta}}_{i} - b_j,  \label{eq:p2pno}
\end{eqnarray}
\noindent where $ \bf{\uptheta} $ is a $Q \times 1$ vector of latent factor scores or abilities for examinee $i$, and $\textbf{A}_j$  is the $1 \times Q$ vector of slopes or factor loadings, which signifies that an examinee’s observed response may be affected by combinations of latent skills or abilities. The goal is to discover the structure of the item responses as represented by $\bf{A}$, rather than to assume the structure as in confirmatory factor analysis. For the pedagogical development of this arithmetic and estimations of simulations, there is an assumption that $Q$ is known. In later chapters, it will be shown that the formalism developed here can be used in the estimation of $\hat{Q}$.

In the latent space, the response process is represented as a multivariate regression of missing item responses (also known as propensities) $z$ on $\bf{\uptheta}$:
\begin{eqnarray}
z_{ij} &=& {\bf{A}}_{j} {\bf{\uptheta }}_{i}  - b_{j} + \varepsilon _{ij}
   \\
    &=& \eta_{ij} + \varepsilon _{ij}.
\end{eqnarray}
\noindent This linear form is a factor model. For examinee $i$, $z_{ij}$ is a vector of missing item responses now transformed to a continuous space. Let ${\bf{A}}_{j}$ denote row $j$ of the matrix $\bf{A}$ of slopes (or factor loadings) for an item, and let the $Q \times 1$ vector of (missing) latent abilities (or factor scores) for an examinee be specified with the prior $ {\bf{\uptheta}}_i \sim N(\bf 0,I) $ .  The $ J \times 1$ vector {\bf{b}} holds the item intercepts $b_j$,  and $\varepsilon_{ij} \sim N(0,1) $ is a random measurement error. Fox provides more details on related Bayesian item response modeling (\cite{fox2010bayesian}). 

It should be noted that the augmented $z$'s are continuous response data sampled from categorical responses. A linear combination of slopes and abilities collapses to a univariate continuous propensity. This approach is mathematically convenient for several statistical techniques, and it also provides an opportunity to exploit some exotic new breakthroughs in random matrix theory as described at the end of this chapter.

\section{Estimation Approaches}
	
	Given the model described in Section~\ref{sec:MIRT}, its application necessitates an adequate formalism for estimation of item parameters given assessment data. One key takeaway from Equation~\ref{eq:p2pno} is that a linear combination of independently distributed latent traits (a vector of $\theta$'s), is analogous to a factor model (\cite{spearman1904general}). Further, a linear combination of normal random variables will yield a covariance matrix that can be factor analyzed, whether by a Karhunen-Lo\'{e}ve transformation, widely known as a principal component analysis, or a more general factor analytic formulation of the component factors. The influence of each factor in the probability of an examinee correctly responding to an item can be ascribed to the respective latent factor's discrimination parameter. 
	
	In a seminal paper, \textcite{BockAitkin1981} proposed an EM (expectation maximization) algorithm based on the work of \textcite{DempsterLairdRubin1977} and extended this procedure to multidimensional item-response data. As they noted, this EM algorithm has an alternative formulation according to the missing information principle. However, Bock and Aitken also realized that numerical quadrature becomes impractical for marginalizing across missing variables as the number of dimensions increases; this is informally reffered to as \textit{the curse of dimensionality}. 
	
	Until recently, estimation of factor coefficients, including loadings and thresholds, has been a major obstacle with categorical item responses, especially with high-dimensional data or a large number of variables. A major trend in addressing computational challenges is the implementation of stochastic approaches to EM (e.g. \cite{Albert1992}, \cite{MengSchilling1996}, \cite{zhao2008ml}, and \cite{cai2010high}) for the E-step. In this work, the goal is to contribute to this trajectory with the implementation of a new algorithm for IRT factor analysis based on the SAEM (stochastic approximation EM) algorithm of \textcite{delyon1999convergence}.
	
	The concept of missing data as introduced by \textcite{TannerWong1987} provided a conceptually important link between Monte Carlo methods and the EM algorithm. Using this bridge, \textcite{beguinglas2001} extended the gibbs sampling approach of  \textcite{Albert1992} to IRT factor models, while others have advanced stochastic versions of the Newton-Raphson algorithm (\cite{gu1998stochastic},  \cite{cai2010high}) in which Metropolis-Hastings sampling is combined with the Robbins-Monro procedure (\citeyear{robbins1951stochastic}) for establishing convergence.  
	
	Below, the EM algorithm is briefly described, followed by an outline of the general SAEM procedure. After providing some background on the foundational work of \textcite{MengSchilling1996}, an algorithm is presented for the estimation of coefficients in factor models for categorical data. Meng and Schilling proposed a gibbs sampling approach to the factor analysis of dichotomous variables based on sufficient statistics. Previously, \textcite{Rubin1982} had described a highly similar approach for the factor analysis of observed continuous variables. In the current work, several existing strategies are added to the procedure of Meng and Schilling for carrying out factor analysis of polytomous item responses. While none of these components are new, they are assembled into a novel procedure that is efficient for large data sets.

 
\section{The EM and SAEM Algorithms}

	Test data can be described as a set of random variables, conditionally dependent on examinees' knowledge and the items of the test. The mathematical model described above is an attempt to describe such data, where responses to items adhere to a probabilitistic function with parameters for item discrimination, difficulty, and the ability of the examinee. Given a set of observed item responses, the goal of the psychometrician is to estimate the value of the proposed parameters. In this model, the item parameters are artifacts of the test structure while the ability of the examinees are missing data also to be estimated. The model parameters and estimations of student ability should converge to true values as the likelihood function is maximized given the observed test data. This estimation problem is a multidimensional optimization problem. 
	  
	The EM algorithm as described by \textcite{DempsterLairdRubin1977} can be described as a generalized solution to this missing data problem. Given a set of observed data $y$, let 
\begin{equation} 
l\left( {\xi |y} \right) = \log f\left( {y|\, \xi } \right)
\end{equation}
\noindent be the observed data log likelihood, where $\xi$ is a set of fixed parameters to be estimated. In introducing a set of missing data  $\psi$, the EM algorithm obtains an estimate of $\xi$ on iteration \textit{t} by maximizing the conditional expected log likelihood, better known as the $\mathcal{Q}$ function 
\begin{equation}\label{eq2}
{\mathcal{Q}_t} = \mathcal{Q}\left( {\xi |\,{\xi ^{(t - 1)}}} \right) = \int {\log f_y\left( {\psi| \xi } \right)f\left( {\psi |y,{\,\xi ^{(t - 1)}}} \right)d\psi } 
\end{equation}
\noindent where $\log{f_y(\psi|\xi)}$ is the complete data likelihood. As \textcite{gupta2011theory} describe, $\psi$ is the data ``you wish you had'' (\textrm{p226}). In many situations, the maximization of $\mathcal{Q}_t$ is simpler than that of $l(\xi| \,y)$, and the EM algorithm proceeds iteratively in two steps at iteration \textit{t}:
\begin{enumerate}
\item E-step: Take the expectation $\mathcal{Q}_{t}$
\item M-step: Find  $\xi^{(t)} = \arg \underset{\xi}{\max}\, \mathcal{Q}_{(t)}$
\end{enumerate}

On iteration $t$ , $\mathcal{Q}_{t}$ is evaluated in the E-step and then maximized in the M-step. Among models of the exponential family, sufficient statistics exist for model parameters \parencite{pitman1936sufficient}. In turn, if a sufficient statistic exists for a parameter, then the MLE (maximum likelihood estimate) must be a function of it \parencite{gupta2011theory}. For exponential family distributions, \textcite{DempsterLairdRubin1977} showed the E-step has a particularly simple form of updating the sufficient statistic. Despite the advantage of monotone convergence, however, the EM algorithm can be very slow with numerical integration over the missing data. Efficiency diminishes rapidly as the number of dimensions increases. Further, the choice of initial values for parameters can result in convergence at local maxima in the hyperparameter space rather than the global maximum of the likelihood function.


In the SAEM approach, the E-step is replaced by a stochastic approximation calculated from multiple draws of the missing data, though a single draw can be used if the E-step is computationally intensive. Let $y$ be the observed data, then at iteration $t$ of SAEM:
\begin{enumerate}
\item Missing data. Obtain one draw from from the distribution  $f(\psi|\, \xi^{(t-1)},y)$.
\item S-step (stochastic approximation). Update $\mathcal{Q}_{t}=\mathcal{Q}_{t-1} + \gamma_{t}\,\lbrace\log{f_y(\psi|\, \xi)}-\mathcal{Q}_{t-1}\rbrace$.
\item M-step. $\arg \underset{\xi}{\max}\, \mathcal{Q}_{t}$.
\end{enumerate}
\noindent Here, $\gamma_t$ is the current iteration's value of the RM (Robbins-Monro) gain coefficient. Implementation of SAEM (as well as EM) is highly simplified when the complete likelihood belongs to a curved exponential family. Using this result in Step 2 of the SAEM algorithm, the vector of sufficient statistics $s(y,\psi)$ is computed at iteration $t$ and then updated according to
\begin{equation}  \label{eq:RMiter}
S^{(t)}=S^{(t-1)}+\gamma_{t} \,\lbrace s(y,\psi^{(t)} )- S^{(t-1)}\rbrace.
\end{equation}
\noindent In this equation, $s(y,\psi^{(t)} )$ is the sufficient statistic calculated on iteration \textit{t}. In the M-step, $\xi$ is updated as a function of $S$. \textcite{delyon1999convergence} showed the SAEM algorithm converges to the MLE under general conditions.

 
\section{IRT Factor Analysis as a Missing Data Problem}

The goal is to maximize the observed data log likelihood $l\left( {\xi |\,y} \right) = f\left( {y|\,\xi } \right)$ , but it is often easier to work with the $\mathcal{Q}$ function used in EM estimation as shown in Equation~\ref{eq2} . Given the current estimate of the fixed item parameters ${\xi ^{(t - 1)}}$ , the posterior distribution $f\left( {\psi |\,{\xi ^{(t - 1)}},y} \right)$ of the unobserved latent variable $\theta$  is generated, and parameters are updated by maximizing  ${\mathcal{Q}_t}\left( {\xi |\,{\xi ^{(t - 1)}}} \right)$. In the case of the IRT factor model, Equation~\ref{eq2} can be modified as
\begin{equation}
{\mathcal{Q}_t}\left( {\xi |{\xi ^{(t - 1)}}} \right) = \int {\log f_y\left( {z,\theta|\,\xi } \right)} \;f\left( {\theta |z,{\xi ^{(t - 1)}}} \right)f\left( {z|y,{\xi ^{(t - 1)}}} \right)d\theta dz.
\end{equation}
\noindent Recalling that persons are indexed by $i$ and items by $j$, the complete data log likelihood for the normal factor model can be expressed as
\begin{equation}
\label{eq9}
\log f_y\left( z,\theta |\,{\xi ^{(t - 1)}} \right) = \exp \left[ {-\frac{1}{2}\sum\nolimits_{i} { \left\{ {{\left[ {{{\bf{z}}_{i}} - \left( {{\bf{A}}{{\bf{\uptheta }}_{i}} - {\bf{b}}} \right)} \right]}^T}\left[ {{{\bf{z}}_{i}} - \left( {{\bf{A}}{{\bf{\uptheta }}_{i}} - {\bf{b}}} \right)} \right] + {\bf{\uptheta }}_{i}^T {\bf{\uptheta }}_{i} \right\} } } \right] .  
\end{equation}
\

The last term in Equation~\ref{eq9} represents the normal prior ${\bf{\uptheta }}\sim N\left( {{\bf{0}},{\bf{I}}} \right),$  which has the effect of fixing the latent scale for person parameters and the multivariate variance term $\bf{\Sigma}=\bf{I}$ constitutes an identification restriction. Calculations of model parameters require the definition of the following conditional expectations
\begin{eqnarray}
\label{eq10} \mathbb{E}[{\bf{\uptheta}}_i|\,{\bf{z}}_i] &=&  \beta ({{{\bf{z}}_i} + {\bf{b}}})
\\
\label{eq11} \mathbb{E}[{\bf{\uptheta}}_i{\bf{\uptheta}}_i^{T}|\textbf{z}_{i}]&=& (\textbf{I} - \textbf{A}^{T}\textbf{A})^{-1} +\beta \textbf ({{{\bf{z}}_i} + {\bf{b}}})({{{\bf{z}}_i} + {\bf{b}}})^{T}\beta^{T} \\
&=&   \textbf{I} -\beta \textbf{A} + \beta \textbf ({{{\bf{z}}_i} + {\bf{b}}})({{{\bf{z}}_i} + {\bf{b}}})^{T}\beta^{T} 
\\
\beta&=&(\textbf{I}+\textbf{A}^{T}\textbf{A})^{-1} \textbf{A}^{T}
\\ &=& \textbf{A}^{T}(\textbf{I} + \textbf{A}\textbf{A}^{T})^{-1}.
\end{eqnarray}

As shown by \textcite{beguinglas2001}, once the $z$ values are sampled from a truncated normal distribution, $\bf{\uptheta}$ can be sampled as
\begin{equation}
{\bf{\uptheta}}_i \sim N \big\{ \mathbb{E}[{\bf{\uptheta}}_i|\,{\bf{z}}_i],({\bf{I}} + {\bf{A}^{T}} {\bf{A})}^{-1} \big\}  \label{eq:thetaN}
\end{equation}

\noindent Maximizing the complete data likelihood with respect to $\bf{A}$ and $\bf{b}$ using expected values (as in standard EM factor analysis) results in
\begin{equation}\label{eq12}
{\bf{b}} = {n^{ - 1}}\left( {{\bf{A}}\sum\nolimits_i {{ \mathbb{E} [{\bf{\uptheta }}_i|\textbf{z}_i]}} \, - \sum\nolimits_i {{{\bf{z}}_i}} } \right)
\end{equation}
\begin{equation}\label{eq13}
{\bf{A}} = \big\{ {\sum\nolimits_i {\left( {{{\bf{z}}_i} + {\bf{b}}} \right)} \, \mathbb{E}[{\bf{\uptheta }}_i^T}|\textbf{z}_i] \big\} {\left( {\sum\nolimits_i {{ \mathbb{E}[{\bf{\uptheta }}_i}} {\bf{\uptheta }}_i^T|\textbf{z}_i]} \right)^{ - 1}}.
\end{equation}
\noindent Next, define the sufficient statistic ${\bf{S}}_1 ={n^{ - 1}}\sum\nolimits_i {{{\bf{z}}_i}}$ and substitute into Equation~\ref{eq12}. Then it can be shown after simplification
\begin{equation}\label{eq14}
{\bf{b}} = - {{\bf{S}}_1}.
\end{equation}

\noindent Now define the sufficient statistic
\begin{equation} \label{eq:sigmaZ}
 {{\bf{S}}_2} = {n^{ - 1}}\sum\nolimits_i {\left( {{{\bf{z}}_i} + {\bf{b}}} \right){{\left( {{{\bf{z}}_i} + {\bf{b}}} \right)}^T}}
\end{equation}
\noindent and substitute into Equation~\ref{eq13}. After tedious but straightforward simplification, the following is obtained
\begin{equation}\label{eq15}
{\bf{A}} =  {{\bf{S}}_2}{\left( {{\bf{I}} + {{\bf{A}}}{\bf{A}}^T} \right)^{ - 1}}{\bf{A}}.
\end{equation}

\noindent Rearranging and expressing ~\ref{eq15} in terms of eigenvalues and eigenvectors gives
\begin{eqnarray}
\label{eq16}{{\bf{S}}_2} - {\bf{I}} &=& {{\bf{A}}}{\bf{A}}^T = {\bf{VD}}{{\bf{V}}^T}
\\ \label{eq17} {\bf{A}} &=& {\bf{V}}{{\bf{D}}^{1/2}}.
\end{eqnarray}

For any given confirmatory analysis, only the first $Q$ eigenvalues and eigenvectors are retained. Parameters are identified provided that each of the first $Q$ eigenvalues in $\textbf{D}$ is greater than zero. \textcite{zhao2008ml} provide a more detailed discussion of this approach and several algorithms that may be useful for categorical data.  While the lower triangular restriction of \textcite{anderson1956statistical} is popular for the identification of $\textbf{A}$ (and also requires a set number of factors), \textcite{conti2014bayesian} point out that ``the choice and the ordering of the measurements at the top of the factor loading matrix (i.e., the lower triangle) is not innocuous'' and may create additional identification problems \textrm{(p32)}. The lower-triangle restriction is essentially a confirmatory rather than an exploratory approach, and much slower computationally than applying Equation~\ref{eq17}. Note that parameter identification is a separate issue from rotational indeterminacy, and rotation is generally required after estimation of factor coefficients for practical interpretation.

\section{SAEM Algorithm for Exploratory IRT Factor Analysis}   \label{sec:SAEMIRT}

Building on the previous work of \textcite{BockAitkin1981}, \textcite{takane1987relationship},  \textcite{Albert1992}, \textcite{MengSchilling1996}, \textcite{fox2003stochastic}, and \textcite{cai2010high}, the estimation procedure in this paper applies a variant of the SAEM method of \textcite{delyon1999convergence} as explained further by \textcite{kuhn2005maximum}. The algorithm below is given for polytomous items, of which the procedure for dichotomous items is a special case. For $K$ ordinal categories of an observed item response indexed by $k$, the normal ogive model is given as
\begin{equation}\label{eq20}
P\left( {{Y_{ij}} \le k|{\eta _{ijk}}} \right) = \Phi \left( {{\eta _{ijk}}} \right)
\end{equation}
where
\begin{equation}
{\eta _{ijk}} = {{\bf{A}}_j}{{\bf{\uptheta }}_i} - \left( {{b_j} + {\tau _{jk}}} \right)
\end{equation}
for $k=1,2,..,K-1$ with $P\left( {{Y_{ij}} \le K} \right) = 1.$ In Equation~\ref{eq20}, ${\tau _{jk}}$  is a decentered category threshold such that ${\Sigma _k}{\tau _{jk}} = 0$ and other parameters are described above. The resulting probability of a response to category $k$ is then given by
\begin{equation}\label{eq22}
P\left( {{Y_{ij}} = k|{\eta _{ijk}}} \right) = P\left( {{Y_{ij}} \le k|{\eta _{ijk}}} \right) - P\left( {{Y_{ij}} \le k - 1|{\eta _{ijk}}} \right),
\end{equation}
\noindent with the corresponding latent response process
\begin{eqnarray}
{x_{ijk}} &=& {{\bf{A}_j}}{\uptheta _i} - {b_j} - {\tau _{jk}} + {\varepsilon _{ijk}}\\
&=& {\eta _{ijk}} + {\varepsilon _{ijk}}.
\end{eqnarray}

\noindent This is a variant of the graded response model of \textcite{samejima1969estimation}.

To take advantage of the estimators in Equations~\ref{eq14} and ~\ref{eq17}, the estimation procedure can be represented as a series of S-steps (the stochastic version of the E-step) and M-steps based on the following functions:
\begin{center}
\begin{tabular}{l c l}
step S1: && $f\left( {\theta |\,z,A,b,\tau } \right)$ \\  
step S2: && $f\left( {x|A,b,\tau ,\theta ,y} \right)$ \\   
step S3: && $f\left( {z|A,b,\tau ,\theta ,y} \right)$ \\  
step M1: && $f\left( {b,\tau |\,x} \right)$ \\
step M2: && $f\left( {A|\,z} \right)$ 
\end{tabular} 
\end{center}

In the sequence above, three missing variables ($\theta$, $x$, and $z$) are sampled in steps S1, S2, and S3, and fixed item parameters are obtained by maximization in steps M1 and M2. For $K=2$, note that $x = z$ and there are no thresholds $\bf{\uptau}$; thus, step S2 is bypassed. For $K > 2$, $x$ is used to estimate item thresholds $\bf{\uptau}$ and $\bf{b}$, while $z$ is used to estimate factor loadings $\bf{A}$ and sample the latent person parameter ${\bf{\uptheta}}$. 

As shown below, $x$ carries information about item thresholds, while $z$ carries the information in the observed categorical response of an examinee. The steps of the algorithm are:
\begin{enumerate}
\item Sample $\bf{\uptheta}$.
\item Draw missing values $x$ for estimating item thresholds: Option propensities $x$ for ordinal item responses $1 \leq y\leq K$  are drawn as

\begin{equation}\label{eq24}
  {x_{ijk}}|{\eta _{ijk}} \sim
    \begin{cases}
      {{N_{\left( { - \infty ,0} \right)}}{\rm{ }}\left( {{\eta _{ijk}},1} \right)} & \text{if}\ {{y_{ij}} \le k} \\
      {{N_{\left( {0, + \infty } \right)}}{\rm{ }}\left( {{\eta _{ijk}},1} \right)} & {{\rm{if  }}{\, y_{ij}}\,{\rm{ >  }}\,k,}
    \end{cases}
\end{equation}

\noindent for $k=1,2,..., K-1$. Random values of $x$ are independently generated for each individual for each item from the truncated normal distributions in Equation~\ref{eq24}. Item intercepts $b_j$   and decentered thresholds $\tau_{jk}$ are obtained from sufficient statistics derived from the $x$ variables
\begin{eqnarray}
   {{{\bar x}_{ij\boldsymbol{\cdot}}}}&=&\sum\nolimits_k {{{ x}_{ijk}}} /(K-1)\\
{b_j} &=&  - \sum\nolimits_i {{{\bar x}_{ij\boldsymbol{\cdot}}}} /n\\
{\tau _{jk}} &=&  \sum\nolimits_i {\left( {{-x_{ijk}} - {b_j}} \right)} /n\,.
\end{eqnarray}
 \item  Draw missing values $z$ for estimating person parameters and factor loadings: For dichotomous items $K=2$, let ${z_{ij}} = {x_{ij}}$. For $K>2$, draw $z$ from the truncated normal distribution
\begin{equation}\label{eq28}
  {z_{ij}}|{\eta _{ijk}} \sim
    \begin{cases}
       {T{N_{\left( {{d_{jL}},\infty } \right)}}\left( {{{\eta '}_{ij}},1} \right)} &  {{\rm{ if }} \,{y_{ij}} = K-1}\\
      ...&\\
      {T{N_{\left( {{d_{j1}},{d_{j2}}} \right)}}\left( {{{\eta '}_{ij}},1} \right)} & {{\rm{if }}\, {y_{ij}} = 1}\\
    {T{N_{\left( { - \infty ,{d_{j1}}} \right)}}\left( {{{\eta '}_{ij}},1} \right)} &   {{\rm{if }}\, {y_{ij}} = 0}
    \end{cases}
\end{equation}\newline
\noindent where ${d_{jk}} = {b_j} + {\tau _{jk}}$  and  ${\eta '_{ij}} = {{\bf{A}}_j}{{\bf{\uptheta }}_i}$. In this sense, $z$ carries the ordinal information in the categorical item response.
\item Update sufficient statistics:
\begin{eqnarray}
{{\bf{\upmu }}_{jk}^{(t)}} = {\bf{\upmu }}_{jk}^{(t - 1)} + {\gamma _t}\left\{ {{{{\bf{\bar x}}}_{jk}} - {\bf{\upmu }}_{jk}^{(t - 1)}} \right\}
\\
{\bf{\Sigma }^{(t)}} = {\bf{\Sigma }}^{(t - 1)} + {\gamma _t}\lbrace {{\bf{S}}_2 - {\bf{\Sigma }}^{(t - 1)}} \rbrace
\end{eqnarray} 
\noindent where $\gamma_t$ is the current value of the RM gain coefficient. Note that $\gamma_t = \gamma_0 = 1$ until the MCMC (Markov chain monte carlo) chain exhibits stationary behavior \parencite{robbins1951stochastic}.
\item Obtain \{$\bf{b}$, $\bf{\uptau}$\} from  
${\bf{\upmu }}$, and ${\bf{A}} = {\bf{V}}{{\bf{D}}^{1/2}}$ from $\bf{\Sigma}$ .
\item Repeat Steps 1-5 until convergence.
\end{enumerate} 

In Step 4, the original E-step is carried out by averaging over stochastically generated missing data, and for this reason, it is sometimes referred to as an S-step. In the M-step, factor loadings and thresholds are obtained through maximization. The proposed algorithm shares a number of S-step features with the gibbs algorithm of \textcite{beguinglas2001}. In Step 3, convergence is defined relative to the sufficient statistics rather than values of fixed item parameters. For this purpose, the change in the trace of ${\bf{\Sigma}}$ between iterations was used. In Step 4, the updated sufficient statistics are used in Equations~\ref{eq14} and ~\ref{eq16}. In developing the operational code for this algorithm, it has been noted that the steps of the algorithm can be rearranged to ensure a more robust approach to the hyperspace of ergodicity, but keeping consistent with the literature in grouping the S-steps and M-steps, the negligible difference in the outcomes from this rearrangement will only be anecdotally noted in the following chapter.

To define a convergence criterion, a window size (say $W_C = 3$) is selected along with a tolerance constant. Iterations are terminated when the maximum covariance change for the trace of $\bf{\Sigma}$ is less than $\epsilon$ for $W_C$ iterations; $\epsilon$ typically falls in the range of [1e-4, 1e-3]. See \textcite{houts2015flexmirt} for an example of this convergence strategy with respect to individual parameters during MHRM (Metropolis-Hastings Robbins-Monro) iterations. For step size, let ${\gamma _t} = {\left( {{1 \mathord{\left/
 {\vphantom {1 t}} \right.
 \kern-\nulldelimiterspace} t}} \right)^\alpha },\;t \ge 0$, where ${\gamma _t} = {\left( {{1 \mathord{\left/
 {\vphantom {1 t}} \right.
 \kern-\nulldelimiterspace} t}} \right)^\alpha },\;t \ge 0$. A larger step size (say $\alpha  = 1$) may accelerate the rate of convergence, but result in a local maximum. A smaller step size $\alpha < 1$  may allow the sequence of estimators to approach the neighborhood of the global maximum, but will slow convergence. \textcite{jank2006implementing} provided a set of recommendations to monitor convergence to a global solution.

\section{Error Estimation Methods for Structural Parameters} \label{sec:ErrEsts}
 
 Given that the proposed algorithm has an analytic form for the likelihood function (Equation~\ref{eq9}), convergence to the parameters' maximum likelihood estimate is obtained and it is possible to simultaneously estimate the Fisher information matrix \parencite{delyon1999convergence}. The Fisher information matrix is a function of $\bf{\xi}$,
 
\begin{equation}\label{eq:info}
  I\left(\bf{\theta}\right) = -\partial^2 l\left(\bf{\xi}|\bf{y}\right),
\end{equation}
\noindent where the diferentials here and below are taken with respect to $\bf{\xi}$. Due to the form of the observed likelihood, $I(\bf{\theta})$ has no closed form solution. However, with the complete data likelihood estimation can be broken down into three pieces. For compact notation define $\bf{\varphi} = (\bf{\theta},\bf{z})$ and let $f_{t}$ be the log likelihood function  $l\left( {\bf{\xi}} | {\bf{y}}, \varphi^{(t)}\right)$ evaluated at ${\bf{\xi}}^{(t)}$  Based on the Louis missing information principle \parencite{louis1982finding}

\begin{equation}\label{eq:missinfo1}
  \partial^2 l\left(\bf{\xi}|\bf{y}\right)= \mathbb{E} \{\partial^2 f_{t}\} + \mathsf{Cov} \{\partial f_{t}\}
\end{equation}
\begin{equation}\label{eq:missinfo2}
 \mathsf{Cov} \{\partial f_{t}\} = \mathbb{E}\{\partial f_{t} \partial f^T_{t}\} + \mathbb{E} \{\partial f_{t}\} \mathbb{E} \{\partial f_{t}\}^T 
\end{equation}

\noindent which results in a representation of $I(\bf{\theta})$ as a combination of conditional expectations. After augmented values are drawn in Step 1 of the algorithm, the three differential components of the equations above are stochastically approximated as 
\begin{eqnarray}\label{eq:hesscomps}
D_{t+1} &=& D_t + \gamma_{t+1} \left[ \partial^2 f_{t+1} - D_t \right]
\\ G_{t+1} &=& G_t + \gamma_{t+1} \left[ \partial f_{t+1} \partial f_{t+1}^T - G_t \right]
\\ \Delta_{t+1} &=& \Delta_t + \gamma_{t+1} \left[ \partial f_{t+1} - \Delta_t \right]
\end{eqnarray}
\noindent The Hessian is then updated according to 
\begin{equation}\label{eq:hess}
 H_{t+1} = D_{t+1} + \left\{ G_{t+1} - \Delta_{t+1} \Delta_{t+1}^T \right\}. 
\end{equation}
\noindent For a given iteration, $\partial^2 f_{t}$ for the complete data log likelihood has a particularly simple form that is identical for all items. It should be noted, however, that  $\partial^2 f_{t}$  is only one component of the final Hessian in Equation~\ref{eq:hesscomps}. Under general regularity conditions, $-H_{t+1}$ converges to $-\partial^2 l\left(\bf{\xi}|\bf{y}\right)$ at a limiting estimate of $\bf{\xi}$, and the inverse of $-H_{t+1}$  converges to the asymptotic covariance of the estimators \parencite{delyon1999convergence}. As the chain converges, the RM procedure is applied and the Hessian is updated according to the equations above. This calculation happens during convergence with the inversion of the Hessian occuring \textit{at} convergence. In Section~\ref{sec:errors} this error will be referred to as the ICE (Iterative Converging Estimate). It is important to emphasize that the Hessian is available upon convergence and needs to be inverted just once. In contrast, the Hessian in the stochastic version of Newton-Raphson must be inverted in each iteration cycle.

 If the RM is engaged too aggressively ($\alpha=1$) and there has not occured some method similar to simulated annealing forcing the parameters into the region of maximum likelihood, it is possible that the inverted Hessian results in a negative eigenvalue. It is also possible in low sample simulations $H$ can be unstable in situations where many parameters are being estimated.  For this reason, a number of other approaches to standard error estimation will be explored. Stochastic approaches to error estimation can leverage properties of the MCMC chain using several variations both during and \textit{post}-convergence; knowledge of the converged parameters and abilities allows for draws from the estimated posterior conditional on the observed data and missing data via the Markov chain. 

Two variations use the Hessian described above and can be implemented at and after convergence. As the Hessian is defined as a function of the converged parameters $\hat{\bf{\xi}}$, in the first variation one draw of $\bf{\theta}$ from the posterior, using the conditional distribution in Equation~\ref{eq:thetaN} samples the convex hyperspace of the structural parameters and allows an estimate of the standard errors upon inversion of $-H$. This first variation will be referred to as the SPCE (Simple Post-Convergence Error). A second variation of a stochastic error approximation is based on the idea the RM procedure on the Markov chain of sufficient statistics yields parameter estimates much more quickly than traditional MCMC approximation methods. In this second variation, 2000 gibbs cycles are reinitialized starting at the converged estimates of the parameters and abilities. The RM procedure in Equations~\ref{eq:hesscomps} is applied while the chain evolves, but is not applied to the chain's sufficient statistics; the Hessian is iteratively updated following each draw, and then $-H$ is inverted. This second variation will be referred to as the IPCE (Iterative Post-Convergence Error).  

\begin{table}
\centering
\begin{tabular}{rp{10cm}} \hline
Error Acronym & Description \\ \hline
ICE & Iterative Converging Estimate; inversion of Hessian at convergence \\
SPCE\textsubscript{O} & Simple Post-Convergence Error; single sample of $\Theta$ \textit{post}-convergence, ogive form \\
SPCE\textsubscript{L} & Simple Post-Convergence Error; single sample of $\Theta$ \textit{post}-convergence, logistic form  \\
IPCE\textsubscript{O} & Iterative Post-Convergence Error; chain of gibbs cycles \textit{post}-convergence, ogive form  \\
IPCE\textsubscript{L} & Iterative Post-Convergence Error; chain of gibbs cycles \textit{post}-convergence, logistic form \\
CLT\textsubscript{F} & Asymptotic variance in MCMC central limit theorem during burn-in gibbs cycles; all chains \\
CLT\textsubscript{I} & Asymptotic variance in MCMC central limit theorem during burn-in gibbs cycles; chains independent within items \\
MCMC & Simple variance in the Markov chain with thinning parameter; burn-in gibbs cycles    \\
\end{tabular}
\caption[Descriptions of Error Estimations]{These error estimation methods will be referenced in later sections. For any estimation, the $\zeta$ iterations of \textit{burn} followed by Robbins-Monro \textit{squeeze} \{$B_T,\zeta-B_T$\} iterations of the SAEM-IRT algorithm used here is understood as a sequence of phases; initialization, burn-in, pseudo-annealing, Robbins-Monro, and convergence. Iterations \{$1, .8 B_T, .2 B_T,\zeta-B_T,W$\} comprise the lengths of the logical phases of gibbs conditioning. The 20\% of burn-in iterations prior to the psuedo-annealing phase are used in CLT and MCMC errors.} \label{tab:ErrorAcro}
\end{table} 

\begin{figure}
\centering
\includegraphics[width=5.5in]{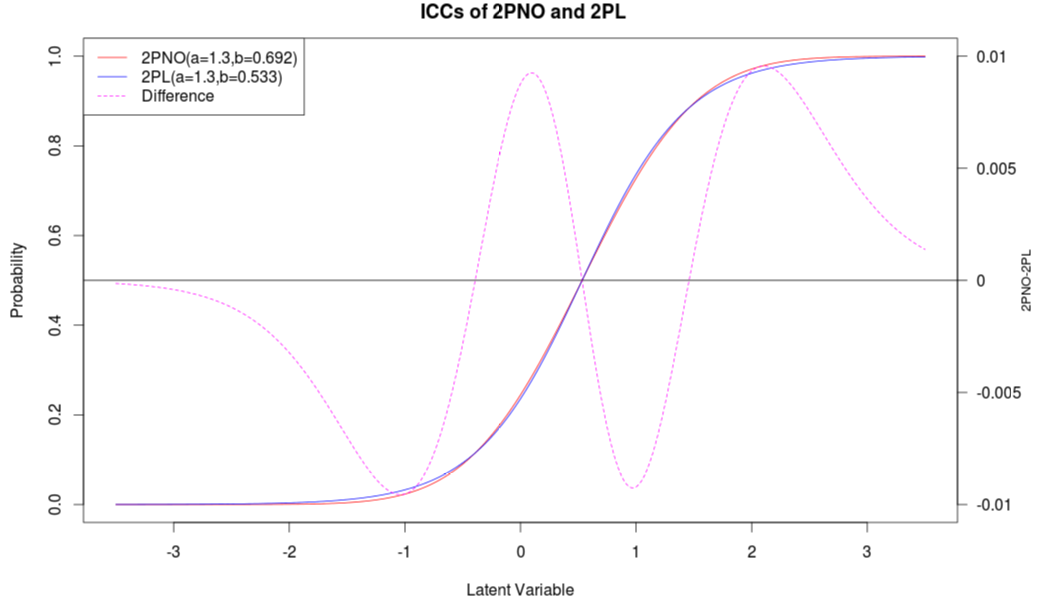}
\caption[Ogive and 2PL ICC Comparison]{The two parameterizations of item characteristic curves are nearly identical. The dotted line represents a magnified difference of the calculate probabilities, $P_{2PNO}(\theta)-P_{2PL}(\theta)$.}\label{fig:ogive2PL}
\end{figure}

The Hessian is a function of the likelihood which is a function of the ogive item response function, and the second derivatives of the ogive are complicated by the asymptotic behavior at the extrema. As aforementioned, the Hessian may have instabilities in the hyperspace as the MCMC chain oscillates, resulting in negative eigenvalues or infinite error terms after inversion. In addressing the convexity of the Hessian, a comparison of the 2PL and 2PNO is in order. A slope and intercept are sampled and shown in Figure~\ref{fig:ogive2PL}. Setting $a=\alpha=1.3$ and $b=.692$, and $\beta=b/a$, the two probability functions can be overlaid. Two points should be noted about the difference between these logistic and ogive item response functions: 1) even at the maximum, the difference in probability between the two curves is less than 1\%, and 2) the logistic curve approaches its asymptotic limits of 0 and 1 more gradually than the normal ogive. The latter point is computationally significant. When compared to the logistic response function, the instability in the Hessian is exacerbated by the asymptotic behavior at the tails of the normal ogive. For this reason, it is reasonable to attempt to calculate the error terms from the inverted Hessian in these three post-convergence error estimations using the logistic as an approximation to the ogive. This is accomplished by transforming the ogive's structural parameters; the scale of the slope is reduced by the scaling constant 1.7 with the intercept scaled down by the same factor and the discrimination parameter. The acronyms above (SPCE, IPCE) will be used in Section~\ref{sec:errors} in conjunction with a hyphenated suffix of O or L to indicate an ogive or logistic item response function is being used in the calculation of the Hessian.

While the cheapest\footnote{Computationally, any work that can be accomplished simultaneously with the convergence of the estimation algorithm is far less \textit{expensive} than initiating a `post-convergence' round of iterative calculations.} approach does not require post-convergence computations, the SPCE requires only a single draw of $\bf{\theta}$, ICE requires many draws of $\bf{\theta}$ as the chain converges, and IPCE is clearly the most expensive requiring a restart of the chain. All three variations apply to this implementation of the SAEM method in high dimensional exploratory factor analysis and should be compared as an analysis of generalizability of the properties of the MCMC chain given that it is converging or has already converged prior to their application. 

Other stochastic error estimation methods include the empirical estimation of the Markov chain's properties. The option is available to approximate standard errors using the chain itself, as well as using the asymptotic variance in the Markov chain central limit theorem. \textcite{geyer1992practical} explains that sums of adjacent pairs of autocovariances of a stationary, irreducible, reversible Markov chain are positive and have regularity properties that can exploited. This theory allows for the construction of \textit{adaptive window estimators} which should be constrained to the sequence of lags until the autocovariance of a single parameter's chain decays to the noise level. This method is noted to have a positive bias \parencite{geyer1992practical}, but the resulting standard error estimates may be practically useful within SAEM as it is applied here. In this method, the autocovariances of the structural parameters are measured and a lag is chosen to ensure an independent sample from which to calculate the chain's variance from each converged estimate. The asymptotic variance error estimate will be referred to as the `CLT' (Central Limit Theorem), and the errors derived from the chain itself will be noted as `MCMC' errors and will be calculated during multiple windows in the parameters' respective chains. Within the `CLT' approximation, two approaches are used; the $\mbox{CLT}_I$ will run the MCMC central limit theorem across the chains of structural parameters specific to the item (assuming ``mutual independence"), while the $\mbox{CLT}_F$ will be run on the full chain (all chains of all structural parameters).

\section{Factor Score Estimation Methods}

	Examinee Factor Scores can be computed via traditional techniques such as the MLE, EAP (Expected A Posteriori), and MAP (Modal A Posteriori) estimation using converged parameters. The MLE of an examinee's ability can be analytically calculated in this approach by taking derivatives of the likelihood function given the examinee's response pattern. It is also possible to numerically solve for the MLE using gradient descent such as the Newton-Raphson method or other optimization algorithms like coordinate descent (though grid-search algorithms become extremely expensive as dimensionality increases). The EAP is easily accomodated given the nature of this algorithm, both through calculations of the expectation of $\theta$ and sampling from the posterior as in  Equation~\ref{eq:thetaN}. In this work, $\theta$ uses a conjugate prior that is symmetric; while each calculation can be applied to the posterior at convergence, the analytic expecations of the MLE, EAP, and MAP are identical.  
		
This formulation of SAEM also allows exploration of iterative techniques for estimating factor scores. Recall that the gibbs draw for $\bf{\theta}$  in each iteration is given in  Equation~\ref{eq:thetaN}. At convergence, it is possible to run additional iterations for generating a chain for each vector of factor scores.  However, fixing $\bf{\xi} = \hat{\bf{\xi}}$ during iterations would generate estimates of $\hat{\theta_i}$ with too little variation; that is, the error in estimating  $\hat{\bf{\xi}}$  would not be reflected in the standard error of $\hat{\theta_i}$. Denote the MLE estimator of $\bf{\xi}$  at convergence as $\hat{\bf{\xi}_0}$.The following method is proposed to address this issue:

\begin{enumerate}
\item Setting  $\bf{\xi} = \hat{\bf{\xi}_0}$, run each of the steps of the algorithm with the gain constant $\gamma=1$. This obtains an MCMC value of $\hat{\bf{\xi}_1}$ on Step 4.
\item Run one additional draw of $\bf{\theta}$ with $\bf{\xi} = \hat{\bf{\xi}_1}$ to obtain a new estimate of $\theta_i$.
\item Repeat the two steps above $t$ times to obtain a chain of $\theta_i$. Apply standard methods to compute the mean and standard deviation of this chain.
\end{enumerate}

The latter statistics represent $\hat{\theta_i}$  and its standard error appropriately inflated by estimation error in $\bf{\xi}$. Note this is essentially using data augmentation for missing value imputation \parencite{little2002statistical} with one important exception: in this proposed method fixed parameters are reset to $\bf{\xi} = \hat{\bf{\xi}_0}$ on each iteration based on the rationale that in practice $\hat{\bf{\xi}_1}$ does not drift too far from the converged value $\hat{\bf{\xi}_0}$.

\section{Formulations of Dimensionality Using Random Matrix Theory} \label{sec:RMT}

The algorithm proposed offers an opportunity to utilize recent research in RMT (Random Matrix Theory). \textcite{wishart20generalized} first formalized random matrix ensembles consistent with the properties of covariance and correlation matrices of random variables. \textcite{wigner1955wcharacteristic} more than seventy years ago, proposed his theory that the interactions of atomic nuclei could be encapsulated by the formulation of a Hamiltonian composed of a high dimension random Hermitian matrix; in attempting to describe the energy levels of nuclei, Wigner detailed many properties of a few forms of matrix ensembles, including reduced forms of their eigenvalue distributions. Historically, Wigner's work encouraged a deeper interest in the properties of real-symmetric and Hermitian random matrices as these matrices hold significant applications in mathematical physics and probability theory. It also led to the formulation of free probability theory and the spectral theory of random matrices. The empirical spectral limit, noted the \textit{semicircle distribution}, of Wigner matrices was derived by Eugene Wigner. 

Applicable to the current work, \textcite{MarcenkoPastur1967} solved for the limiting distribution of eigenvalues of a subset of Wigner matrices, namely sample covariance matrices as defined by \textcite{wishart20generalized}. This algorithm exploits the properties of the random variables implicit to the augmented data \parencite{MengSchilling1996}; the augmented data matrix is an assembly of random gaussian variables conditional on items, examinees, and observed data. When centered and scaled, the sufficient statistics used to extract item discriminations is a GOE (gaussian orthogonal ensemble). The asymptotic properties of the covariance of GOEs and their eigenvalues becomes remarkably useful for the purpose of measuring the statistical significance of these item discriminations.

The augmented data sampled using Equation~\ref{eq28} gives rise to the covariance matrix defined in Equation~\ref{eq:sigmaZ}. The structure of the distribution of $\bf{\uptheta}$ in Equation~\ref{eq:thetaN} results in the covariance $\bf{S}_2$ satisfying the definition of a Wishart matrix. When applying a measuring instrument such as a psychological assessment, the algorithm in this work is modeled such that the convergence of parameter estimation uses empirical spectral analysis of the covariance matrix of augmented data. 

In this section, the notation of $J$ items and $N$ examinees play a significant mathematical role and will be used to appropriately connect the conceptual framework of the augmented data matrix of this algorithm to results derived from RMT. In the interest of parsimony, applications of this RMT will be restricted only to normalized Wishart matrices where $N \ge J$. Complex Hermitian matrices, unitary ensembles, and symplectic ensembles do not apply to this treatment of assessment data, but the reader can be directed to the work of Freeman Dyson and Madan Mehta for a thorough discussion on foundational RMT (\cite{dyson1962statistical1}; \citeyear{dyson1962statistical2}; \citeyear{dyson1962statistical3}; \citeyear{dyson1962brownian}; \cite{mehta1967random}).

Spectral analyses are commonly used for tests of equality between population covariances and a scaled identity matrix; i.e. if a set of \textit{J}-dimensional zero-mean gaussian random variables are sampled $N$ times, the population covariance (a $J \times J$ matrix) is structured as $\bf{\Sigma} = \sigma^2 \bf{I}_{J \times J}$. Denoting each draw of the gaussian vector as $\mathbi{v}_1, \mathbi{v}_2, ... \mathbi{v}_N$, the resulting real-valued normalized random Wishart matrix will be defined as
\begin{equation}\label{eq:Wishart}
\mathbf{S}_N = \frac{1}{N} \sum_i \mathbi{v}_i \mathbi{v}_i^T.
\end{equation}
\noindent Note that this formula only implies a transpose rather than the Hermitian conjugate typically used in the treatment of Wigner matrices. The eigenvalues of this matrix will be rank-ordered from largest to smallest and labeled as $\lambda_1 \ge \lambda_2 \ge ... \ge \lambda_J$. This matrix has $N$ degrees of freedom and an average trace 
\begin{equation} \label{eq:meanTrace}
\Lambda = \frac{1}{J}  \text{Tr}(\mathbf{S}_N) 
\end{equation}
\noindent which is equal to the average eigenvalue $\frac{1}{J} \sum_j \lambda_j$.

In this formulation, the resulting matrix adheres to the assumption of \textit{sphericity} of the covariance matrix, and is known as a real white Wishart matrix. In another interpretation, a spectral analysis of this matrix,   
\begin{equation} 
\text{det}(A-\lambda I) = 0
\end{equation}
\noindent where $A$ is a central Wishart distribution and $\Sigma$ is its covariance, will have a result that its $J$ dimensions are rotationally invariant and there is no statistically significant principal component that can be extracted from it. 

Stepping back for a more qualitative interpretation that applies to the current use case for the proposed algorithm, a $J$-item assessment with probabilities of success that are not dependent on a latent ability of an examinee would result in discrimination parameters that are not significantly different from zero; this implies that the covariance matrix of such an assessment's centered (zero-mean) augmented data obeys the sphericity assumptions of a random Wishart matrix. Stated differently, \textit{and this is the main idea}, \textbf{a null hypothesis can be defined such that no eigenvalue of the covariance matrix of zero-mean augmented data is significantly greater than the expectation of the largest eigenvalue of a random Wishart matrix of the same number of dimensions} ($J$) \textbf{and degrees of freedom} ($N$), $H_0 : \Sigma = I$. The Tracy-Widom distribution provides the means for us to produce such a test. Practically, if a psychometric assessment was designed to contain $J$ items that obey structural probability functions that are conditional on, say, three latent dimensions, the eigenanalysis of the augmented data covariance should result in three eigenvalues significantly greater than the expectation of the largest eigenvalue $\lambda_1$. In other words, an assessment designed to measure three latent factors is expected to result in a spectral analysis of $\bf{S}_2$ that will negate this null hypothesis given that three eigenvalues should be significantly greater than the distribution of the largest eigenvalue $\lambda_1$ given $N$ examinees and $J$ items of test data.

Given that the calculation of discrimination parameters in this application of SAEM are coming from a principal components analysis, this probabilistic treatment of a component's significance is important. Informally, most statistical coursework advocates that it is sufficient to cut off significant eigenvalues at an ``elbow'' in a scree plot, or to study principal components with eigenvalues greater than one. Worse, in many large scale psychometric analyses of real data, dimensionality may be dictated by test design rather than empirical calculations or statistically motivated justifications. The nature of the algorithm being proposed here offers an opportunity to study the signal-to-noise ratio arising from an assessment's measurement, removing intuition and ``rules of thumb'' altogether by leveraging the advances of RMT.

\subsection{Marchenko-Pastur Law} \label{sec:MP}

Before introducting the Tracy-Widom distribution, it is useful to discuss the result of \textcite{MarcenkoPastur1967}. The investigation of the eigenvalue distribution of GOEs was important for physicists who were interested in the energy levels of atomic nuclei. In reference to experimental observations, spacings between such levels could provide empirical proof that the mathematical formalism applies to nuclear scattering and other observable phenomena. The derivation of analytical forms for these solutions also allow for inferences about the mathematical and physical properties of the eigenvalues of ensembles of random variables; to this end, the trailblazers of RMT were able to extract many features of these covariance matrices, and these will prove useful for assessing the statistical properties of the sufficient statistic $\bf{S}_2$.

First, it is for clarification that specific terms are defined with respect to the simulations below. The augmented data matrix has dimension $N \times J$. Mapping assessment data to the concept of RMT, this augmented matrix is thought of as $J$ measurements or possible dimensions of the latent scores of $N$ examinees. When $J < N$, the eigenvalues of the resulting covariance matrix $\mathbf{S}_N$ as defined in Equation~\ref{eq:Wishart} prove to be confined to a sequence of values or states with non-zero positive upper and lower bounds called the \textit{empirical spectral distribution}.\footnote{Typically this result is written in one of two ways, (1) as a statement that denotes the counting of a number of distinct values, specifically $\mu_{\lambda}(A) = \frac{1}{N} \#\{J \le N; \lambda_i \in A\}$, where $A \subset \mathbb{R}$, or (2) as a point process representation, i.e. $\mu_{\lambda}(A) = \frac{1}{N} \sum_{i=1}^N \delta_{\lambda_i}$ where $\delta$ is a Dirac measure. When $J>N$ for this GOE, the distribution is still confined to a region $A \subset \mathbb{R}$ with a finite upper limit, but the lower bound is at zero with a probability mass at zero of $\frac{J-N}{J}$.} The probability space near the bounds of the eigenvalue distribution are called the \textit{spectral edges} with the region between the bounds labeled the \textit{bulk spectrum}. As the eigenvalues are confined to this space between the edges and have a specific number of values ($J$), the spacings or gaps between the finite positions of these values on the real line satisfy the definition of a DPP (determinantal point process). DPPs are probabilistic constructs that describe a fixed set of potential arrangements of states.

\textcite{wigner1955wcharacteristic} and \textcite{mehta1967random} derived terms demonstrating that the individual energy levels described by their Hamiltonian operators were sparsely spaced in the bulk, and became spaced further apart at the edge furthest from zero. To solve for the distribution of the spacing, a series of mathematical transforms must be performed, and the determinantal form invokes the use of a Fredholm determinant with an integral operator sine kernel to describe the gap probability distribution in bulk, while at the edge the solution involves the Airy kernel as it necessitates a vanishing boundary condition. For the work here, the main point of interest concerns the limiting spectral edge.

Before assessing the distribution of the largest eigenvalue of a random Wishart matrix, a pedagogical approach motivates first the introduction of its asymptotic limit. For a random GOE with $\mathbb{E}(\mathbf{S}_N) = \mathbf{I}$, \textcite{MarcenkoPastur1967} derived the formula for the distribution of $\lambda$ in the limit as $J \rightarrow  \infty$, $N \rightarrow \infty$, and $J/N \rightarrow c$. In contrast to Wigner's semicircle distribution, the form of the equation defining the asymptotic distribution is sometimes referred to as the quarter circle law when $c \le 1$.  In the derivation, the distribution of the point process spectra is written  
\begin{equation} \label{eq:meandev}
\mu = \frac{1}{J} \sum_{j=1}^J \delta_{\lambda_j} 
\end{equation}
\noindent where $\delta_{\lambda_j}$ are the count of the distinct eigenvalue spacings, and the bounds of this distribution have deterministic values in the asymptotic case. In physics this is known as a \textit{density of states}. The cumulative distribution function, or \textit{integrated density of states}, when described in analytical form is differentiable and gives rise to the limiting probability density known as the Marchenko-Pastur law,

\begin{equation} \label{eq:mplaw}
\frac{d \mu}{dx} = \frac{1}{2 \pi c x} \sqrt{(x - \ell_{-}) (\ell_{+} - x)}
\end{equation}
\noindent where $\ell_{\pm} = (1 \pm \sqrt{c})^2$. In the asymptotic limit, $\lambda_1 \rightarrow \ell_{+}$ and $\lambda_J \rightarrow \ell_{-}$. When $c > 1$, this implies $J > N$ and the covariance matrix becomes singular, resulting in $P-N$ eigenvalues at zero and a finite probability of non-zero eigenvalues described by Equation~\ref{eq:mplaw} with $\ell_{-} = 0$. 

\begin{figure}
\centering
\includegraphics[width=5.5in]{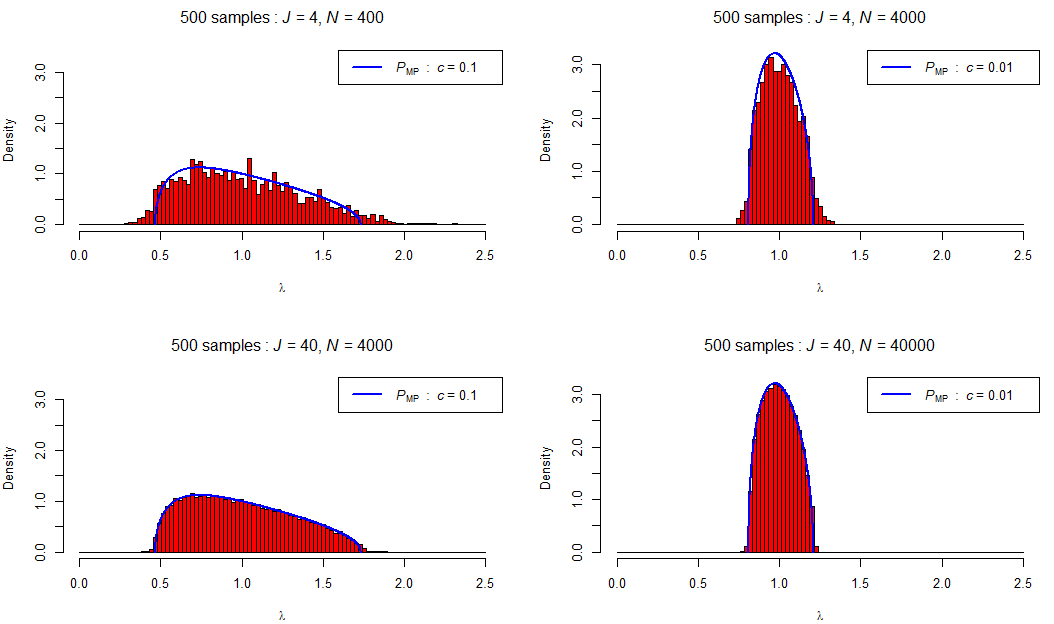}
\caption[Marchenko-Pastur Distribution]{
The histograms show the distributions of 500 samples of eigenvalues for four configurations of $J$ and $N$. The left panels produce the proportional constant $c=.01$ with the right panels set to $c=.1$. In proceeding from the top panels to the bottom, both $J$ and $N$ are multiplied by a factor of ten to demonstrate the empirical convergence of the eigenvalues to the Marchenko-Pastur probability distribution ($P_{MP}$), which is overlaid in blue.}\label{fig:evhists}
\end{figure}

To demonstrate the utility of the Marchenko-Pastur law, 500 GOE are sampled, and the eigenvalues of their covariance matrices can be histogrammed and scaled to a probability density. This exercise is shown in Figure~\ref{fig:evhists}. In the top row, a $J=4$ dimensional vector of i.i.d. (independent and identically distributed) gaussian random variables are drawn for $N=40$ and $N=400$ cases, and $J \times J$ covariance matrices are produced; the proportion $\frac{J}{N} = c = .01$ and .1, respectively, also produce the blue overlaid limiting distribution of the Marchenko-Pastur law. This is repeated 500 times and the eigenvalues are bucketed into 100 bins from 0 to 2.5. In the bottom row, the same exercise is repeated for 10 times as many random dimensions ($J=40$) and cases ($N = 400$ and 4000), covariances calculated, followed by an eigenanalysis; preserving the ratio $c$ in both configurations. The eigenvalues of 500 instances of these 100 extracted eigenvalues are more clearly confined to the shape of Equation~\ref{eq:mplaw} overlaid again in the lower histograms.


\subsection{Tracy-Widom Distribution} \label{sec:TW}

\begin{figure}
\centering
\includegraphics[width=5.5in]{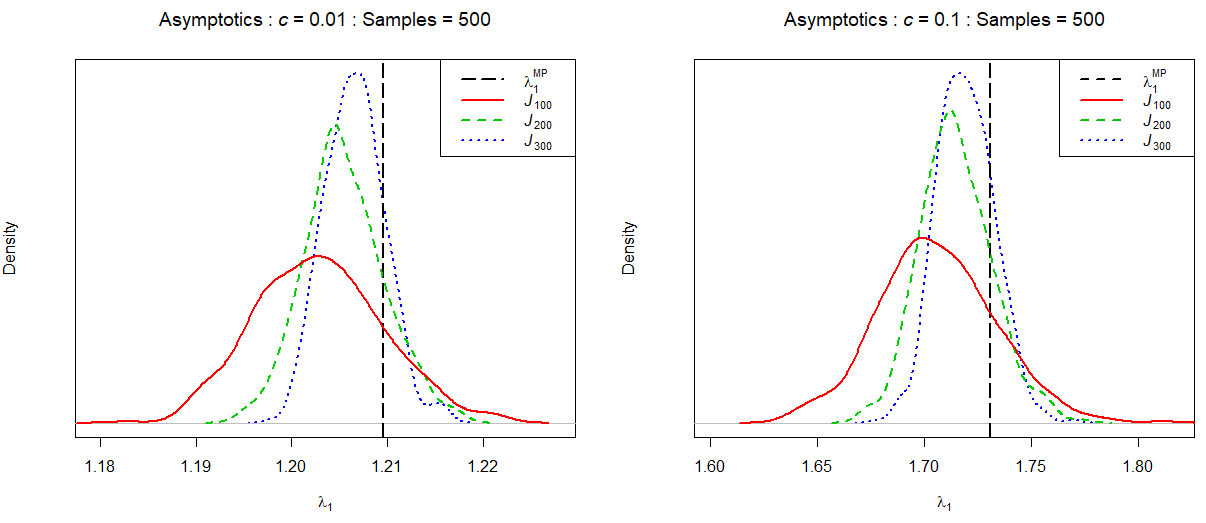}
\caption[Largest Eigenvalues of Gaussian Orthogonal Ensemble Covariances]{
The densities of the largest eigenvalue for several configurations of $J$ and $N$. The left panels produce the proportional constant $c=.01$ with the right panels set to $c=.1$. Three choices of items \{100, 200, 300\} demonstrate the convergence to the asymptotic limit ($\lambda_{MP} =\argmax P_{MP}$)  defined by the Marchenko-Pastur law.}\label{fig:evlimits}
\end{figure}

For the statistician looking to measure a ``signal,'' interest lies in the nature of the distribution of $\lambda_1$ when spectral analyses are performed. As it is impossible that measurements are performed under the conditions $J \rightarrow  \infty$ and $N \rightarrow \infty$, practical use of this it is important for researchers to understand the probability that one or more eigenvalues constitute a significant \textit{signal} within an eigenanalysis. To visualize the behavior of $\lambda_1$, and taking the same proportions of items to examinees ($c=.01, .1$) chosen in the four plots from Figure~\ref{fig:evhists}, ten increasing sets of item and examinee configurations are sampled 500 times and the largest eigenvalue from each sample is shown in the distributions plotted in Figure~\ref{fig:evlimits}. These plots provide direct observation of the limiting distribution of the largest eigenvalue of a GOE.
\noindent This problem was solved by \textcite{tracy1996orthogonal} and is thoroughly explored within a later article by \textcite{johnstone2001distribution}. The TW (Tracy-Widom) distribution is here introduced summarizing a small section of the latter work which studies its application to the covariance derived from \textit{rectangular} random gaussian matrices. The covariance of this $N \times J$ GOE is Wishart $W_J (I,N)$ of $J$ dimensions and $N$ degrees of freendom; again, $J$ and $N$ continue to be used here because this is a direct representation of the augmented data simulated in this application of SAEM.

First, centering and scaling constants are calculated,
\begin{eqnarray} \label{eq:TWcenterscale}
\mu_{J} & = & \left( \sqrt{N-1} + \sqrt{J} \right)^2, \\
\sigma_{J} & = & \left( \sqrt{N-1} + \sqrt{J} \right) \left(\frac{1}{\sqrt{N-1}} + \frac{1}{\sqrt{J}} \right)^{1/3}.
\end{eqnarray}
\noindent There are three \textit{orders} of the Tracy-Widom law, but only the first order is applicable for this use case. This distribution function is written as
\begin{equation} \label{eq:TW}
F_1(s) = \exp \left\{ -\frac{1}{2} \int_s^\infty q(x) + (xs)q^2(x)dx \right\},     s \in \mathbb{R}
\end{equation}
\noindent and the limiting law of the largest eigenvalue of a $J \times J$ gaussian symmetric matrix  is
\begin{equation} \label{eq:TWdist}
\frac{\lambda - \mu_{J}}{\sigma_{J}} \underrightarrow{\mathcal{D}} W_1 \sim F_1.
\end{equation}

\noindent This invokes the use of the aforementioned Airy function, Ai($x$), which is a solution to the nonlinear Painlev\'{e} II differential equation. 
\begin{eqnarray} \label{eq:painleve}
q''(x) & = & xq(x) + 2q^3(x), \\
q(x) & \sim & \text{Ai}(x)  \text{   as  } x \rightarrow +\infty.
\end{eqnarray}

While the full derivation is beyond the scope, a few remarks should be made. The joint density of the Wishart matrix is transformed through the clever use of a Vandermonde determinant identity and a Laguerre expansion that \textcite{mehta1967random} shows can be rewritten as a determinant of the Airy kernel. A Fredholm determinant identity is then applied and the result yields the Tracy-Widom distribution. The Painlev\'{e} forms are a series of nonlinear differential equations studied at the turn of the twentieth century with movable singularities at the poles of their complex form. The Airy function, as a solution to the form in \ref{eq:painleve}, has oscillatory behavior up to a turning point at which there is an exponential decay; the solution at this pole gives rise to the centering and scaling factors in Equations~\ref{eq:TWcenterscale}  \parencite{johnstone2001distribution}. The TW distribution is a statement of the variation at this point and this Airy function is reflective of the properties of the eigenvalues in the bulk as it approaches the limiting edge of the maximum eigenvalue. 
\begin{figure}
\centering
\includegraphics[width=5.5in]{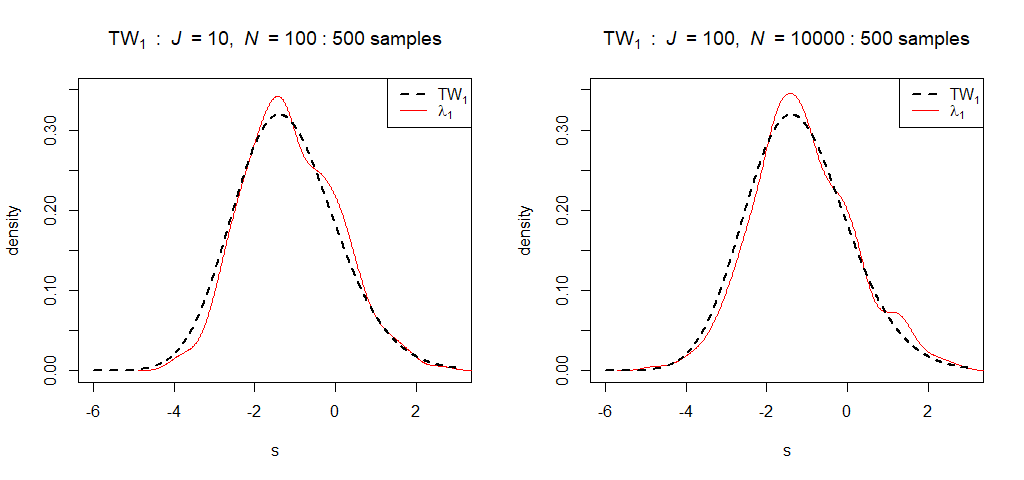}
\caption[Largest Eigenvalues and the Tracy-Widom Distribution]{
500 samples of the largest eigenvalues of GOEs sampled with the configurations of $J=10, N=100$ and $J=100,N=10000$ are shown. After centering and scaling according to Equations~\ref{eq:TWcenterscale}, the densities are well described the Tracy-Widom distribution.}\label{fig:twdens}
\end{figure}

The variability of the largest eigenvalue is well-described by the TW distribution. In continuing with the configurations used above, another 500 samples of $\lambda_1$ are sampled, centered, scaled, and plotted with the TW distribution $F_{\beta=1}$ overlaid in Figure~\ref{fig:twdens}. As it is successfully being used in signal processing applications (\cite{wax1985detection};  \cite{bianchi2009performance}), portfolio management (\cite{laloux2000random}; \cite{avellaneda2010statistical}), and quantum information processing \parencite{zyczkowski2001induced}, this utility allows for a statistical treatment of the psychometric measuring instrument's assessment of examinees' abilities. For the simulations in chapter 3, an exposition of the performance of the SAEM algorithm and the Tracy-Widom test will be delineated.

%

%




\chapter{Simulation Studies} \label{ch:ss}

To properly describe the performance of the algorithm, it is necessary to build incrementally, starting from simple foundations of a one-dimensional dichotomous response assessment up to a large-scale polytomous test of many dimensions. The recovery of generated parameters is pertinent for a thorough discussion of the capabilities of SAEM as it converges to a reconstruction of the simulated test design. This includes the estimation of latent factors as they compare to the simulated abilities. The validity of the mechanism of the algorithm requires that a researcher can make inferences from well understood quantitative outputs, and a thorough examination of the properties of SAEM's operation, estimates, and their respective errors are to follow. The algorithm will also be applied to three instances of real response data; each is multidimensional and will assist in the application of the Tracy-Widom test described in section~\ref{sec:TW}.

\section{Simulations, Estimations, and Diagnostics}

The goal of this study is intended to provide proof of concept, applying the above algorithm to conditions that provide insight into its utility as well as limitations. Several Monte Carlo approaches to multivariate normal probit algorithms have been advanced in the past two decades (\cite{MengSchilling1996}; \cite{chib1998analysis}; \cite{patz1999straightforward}; \cite{song2005multivariate}). The probit link function exhibits behavior at the tails that is less tractable than the logit link as the ogive CDF more rapidly approaches the lower and upper asymptotes of large absolute values of $\theta$. Computation of the derivatives for the Newton-Raphson cycles in traditional EM estimation are also computationally intensive. 

Given the variation in the draws of  $\theta$, the updates to the gradient and Hessian in \ref{eq:hesscomps} and \ref{eq:hess} may be inflated by orders of magnitude that are not typical of the convex region of the hyperspace where the true value of the parameters are located. However, calculating standard errors with post-convergence chains should provide at least a partial solution to this source of instability. To examine this issue, several approaches to SE (standard errors) calculation are to be explored. To establish a baseline for comparison, the RMSE is easily obtained for all configurations of tests, examinees, and dimensionality of latent factors. Any post-convergence method for obtaining standard errors are calculated by restarting the chain for 1000 iterations with converged estimates of random and structural parameters keeping $\gamma = 1$. Thinning will be applied using autocovariance diagnostics. 

Specific to this research, key improvements are expected to be made in computational efficiency. Exploratory factor analyses on large scale assessment take a significant investment in time and are rarely evaluated, as the dimensionality of these exams are typically prescribed \textit{a priori}. Further, as latent factors increase in dimensionality, traditional methods become cumbersome; thus, computational time becomes a key metric for simulations in this research. 

\subsection{Simulation Conditions}
In Table~\ref{tab:sims} conditions for simulations are expressed in terms of numbers of items, ordinal categories, examinees, and dimensions. As this code was developed to process the current algorithm, it does not depend on other IRT packages or software; thus statistical effects of multiple parameters on small and large assessments are to be investigated and clocked. The code is programmed using \texttt{R} \parencite{teamr} and is made publicly available, along with the simulations and their fits at \textcite{GeisGit2019}. In gauging performance, standard questions are to be answered around ability estimates, parameter estimates, rotational indeterminacy, convergence criteria, and computational speed. Significance of dimensionality will also be explored given the results of RMT in Section~\ref{sec:RMT}.

To study the accuracy of reconstruction of coefficients and their respective standard errors, fifty replications of the first five simulation conditions are performed. Condition 1 and Condition 2 are base case dichotomous univariate assessments, without and with a guessing parameter, respectively. Condition 3 enhances the item information with 4-category ordinal polytomous response data. In these first three conditions, there are 100 items and 5,000 examinees. Conditions 4 and 5 aim to explore the differences between bifactor and subscale items within a small multidimensional assessment of the same sample size as the univariate benchmarks. In these conditions, the exam is 4-category ordinal polytomous and there are 30 items and 5,000 examinees. 

\begin{table}
\centering
\begin{tabular}{lcrrrl}  
Condition & Guessing & Items ($J$) & Options ($K$) & Examinees ($N$) & Dimensions ($Q$) \\ \hline
1 (50) & No & 100 & 2 & 5,000 & 1  \\ 
2 (50) & Yes & 100 & 2 & 5,000 & 1  \\ 
3 (50) & No & 100 & 4 & 5,000 & 1  \\ 
4 (50) & No & 30 & 4 & 5,000 & 3 (bifactor)  \\ 
5 (50) & No & 30 & 4 & 5,000 & 3 (subscale)   \\ 
6 (5) & No & 100 & 4 & 10,000 & 5 (bifactor)   \\ 
7 (5) & No & 100 & 4 & 10,000 & 5 (subscale)   \\ 
8 (5) & No & 100 & 4 & 100,000 & 10 (bifactor)   \\ 
9 (5) & No & 100 & 4 & 100,000 & 10 (subscale)  \\ \hline
\end{tabular}
\caption[Simulation Conditions]{The simulation conditions that have been generated and estimated for this work. The number of replications is noted in parentheses within the Condition column.} \label{tab:sims}
\end{table} 

Conditions 6-9 probe the computational effects for large scale assessments as well as the rotational complexity using the converged estimates of factor loadings. In Conditions 6 and 7, five dimensions and 4-category polytomous items will be simulated in bifactor and subscale structure using 100 items and 10,000 examinees. In Conditions 8 and 9, the dimensions are doubled and the number of examinees scales to 100,000. 

\subsection{Parameter Distributions}
\begin{figure}
\centering
\includegraphics[width=5.5in]{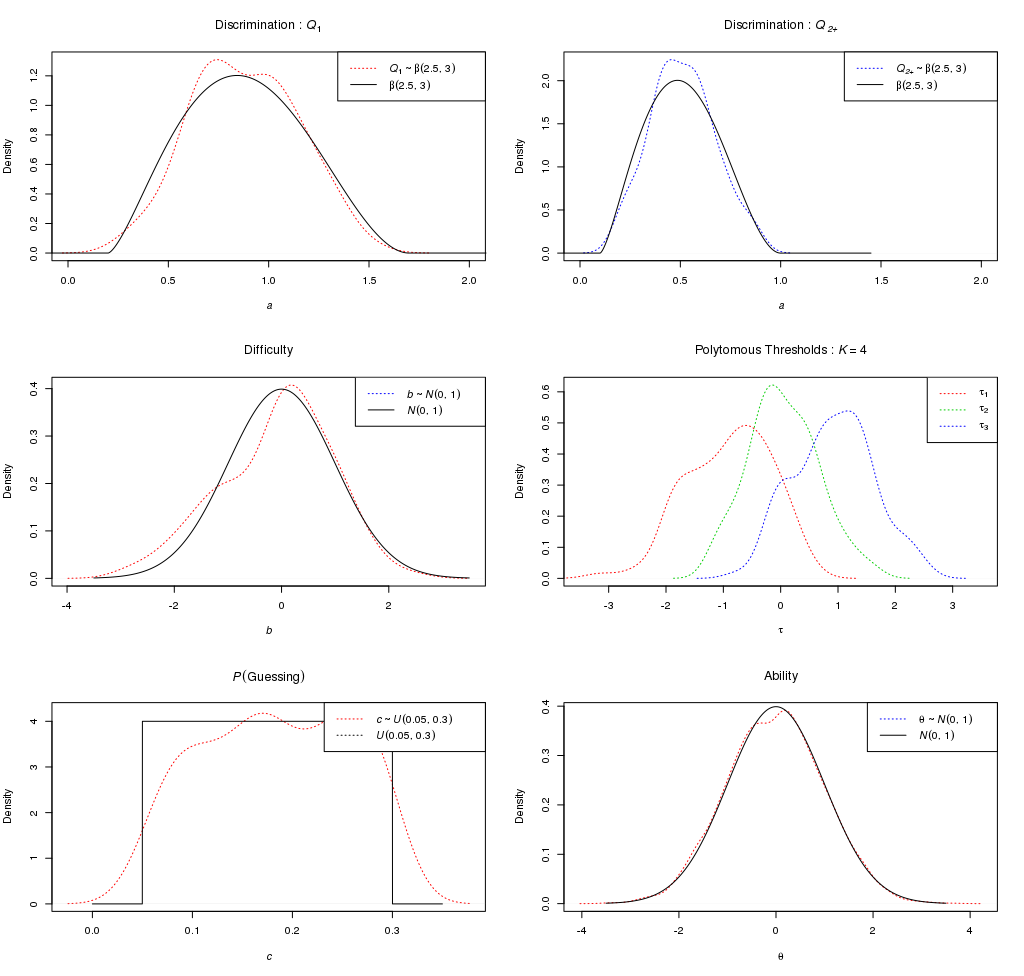}
\caption[Distribution of Simulated Parameters]{Simulated parameter densities are shown above along with their source distribution overlaid with one exception: the polytomous thresholds consist of $K-1$ draws from $N(0,1)$ and ordered from smallest to largest. All item parameters are drawn 100 times, reflecting the variation of the majority of the simulation conditions. The abilities are sampled 5000 times from the normal distribution. The upper left and right panels represent the distributions of discriminations for the primary dimension of each item in the simulations and the more constrained magnitude of the second dimension in bifactor conditions, respectively. The middle left and right panels are the difficulties and polytomous thresholds. The bottom left plot shows the density of guessing parameters. The bottom right is a visualization of the normal latent factor, our \textit{bell-curve}.}\label{fig:pardists}
\end{figure}

The parameters of the simulation conditions necessitate distributional parameterizations that are visualized in Figure~\ref{fig:pardists}. In the 1D simulations the slope or discrimination parameters were drawn from a 4-parameter beta, i.e., $B_4 (2.5, 3, .2, 1.7)$.  In the subscale multidimensional simulations, the loadings from all dimensions in A are also sampled from $B_4 (2.5, 3, .2, 1.7)$. In the bifactor multidimensional simulations, the loadings from the first dimension in A are sampled from $B_4 (2.5, 3, .2, 1.7)$ while the higher-dimensional items load on a second latent factor with slopes sampled from a more constrained $B_4 (2.5, 3, .1, .9)$ to constrain individual item reliability. All but two latent factors are set to zero in the bifactor simulations, while subscale conditions load on only one dimension for each item. The lower magnitude second-factor discriminations in the higher dimensions are intended to create a challenging situation for the bifactor MIRT condition.

The $b$ parameters are sampled from a N(0,1) to assess the tails of the hyperspace. If the assessment items imply a partial credit model, the $K-1$ thresholds are drawn from a $N(0,1)$ distribution and sorted such that $\tau_1  < \tau_2 < \tau_3$. The guessing parameter in Condition 2 is drawn from a uniform distribution, $c \sim U(.05,.3)$. The abilities of the examinees were sampled from a $N(0,1)$ and $Q$-dimensional $MVN(\mathbf{0},\mathbf{I})$. In Figure~\ref{fig:pardists}, the distributions are overlaid on top of a sample of 100 parameters from each item-specific structural parameter distribution. The final bottom right plot shows 5000 draws of abilities from a single normal latent factor.

\subsection{Pseudo-Annealing and Activation of Robbins-Monro Iterations}

The stochastic approximation updates were initiated once several iterations of burn-in were completed for each simulation condition. For simulations with only 5000 examinees, a longer burn-in period was set so that the parameters were expected to be traversing the convex hyperspace of the MCMC chain. For every condition above, the exponent in the gain constant of Equation~\ref{eq:RMiter} was set to 1. Following the discussion at the end of Section~\ref{sec:SAEMIRT}, the larger step size, $\alpha  = 1$, was used to maintain the highest rate of convergence. An enhancement akin to simulated annealing was also implemented in a small window of $\kappa$ iterations where $W_i = 1, 2, ... , W_{\kappa}$. This to prevent the activation of the Robbins-Monro gain constant from forcing a parameter's estimate from prematurely converging to a local maximum in the hyperspace.

\begin{figure}
\centering
\includegraphics[width=5.5in]{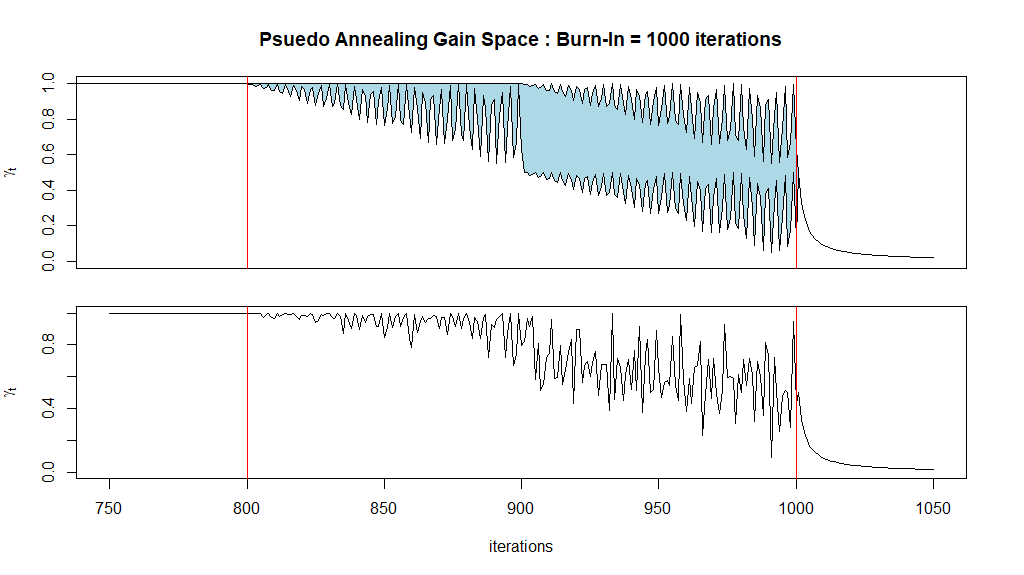}
\caption[Stochastic Pseudo-Annealing Applied Before Robbins-Monro Iterations]{The gain constant from Equation~\ref{eq:RMiter}, when engaged, decreases rapidly when $\alpha = 1$. The exponential decrease to the right of iteration 1000 shows this \textit{simple} Robbins-Monro decay of the gain constant. To mitigate the potential for the hyperparameters to get stuck in a local maximum of the likelihood function, this oscillatory stochastic regime (seen in iteration 800 to 1000) is developed to use stochastic noise to allow a gradual but random leakage of statistical information during the final 20\% of burn-in iterations of the MCMC chain.}\label{fig:RMwinplot}
\end{figure}

Before the RM gain constant is engaged, this window of $W_{\kappa} = .2*B_T$ or 20\% of the burn-in iterations is defined for stochastically leaking statistical information into adjacent iterations, breaking the axiomatic rule of a Markov process. In this window, the gain constant is effectively throttled such that the amount of statistical information retained from iteration to iteration is small, oscillating, and dynamic. Several modifications to the imposed variation in the upper and lower bounds of stochasticity may be allowed. As shown in Figure~\ref{fig:RMwinplot}, the amplitude of the oscillatory function increases linearly as the window evolves, thus allowing random fractional volumes of statistical information to be retained from iteration to iteration until the Robbins-Monro sequence in Equation~\ref{eq:RMiter} is fully engaged starting from $\gamma_t = \frac{1}{2}, \frac{1}{3}, ...$ until convergence. This pseudo-annealing technique is visualized in Figure~\ref{fig:RMwinplot}; it shows a plot of the possible values for $\gamma_t$ between the start and end of this window, before the RM gain constant is turned on in its deterministic geometric sequence. Effectively, $\gamma$ is sampled from a $U\left( 1-\frac{W_i}{W_{\kappa}}\cos^2(W_i) , 1 \right)$ in the first half of the window, and the second half of iterations in the window are to be sampled from $U\left( \frac{1}{2} - \frac{W_i}{W_{\kappa}}\cos^2(W_i), 1 - \frac{W_i}{W_{\kappa}}\cos^2(W_i)  \right)$. For a hypothetical MCMC chain with $B_T=1000$ burn-in iterations, Figure~\ref{fig:RMwinplot} shows the gain constant's state space visualized along with a random sequence of $\gamma$ being sampled from that space defined in the upper plot. The vertical lines at iteration 800 and 1000 are the edges of this pseudo-annealing window. Note again that the first value of $\gamma_t$ at iteration 1000 is $\frac{1}{2}$ so as not to reinstate a Markovian draw.

\begin{figure}
\centering
\includegraphics[width=5.5in]{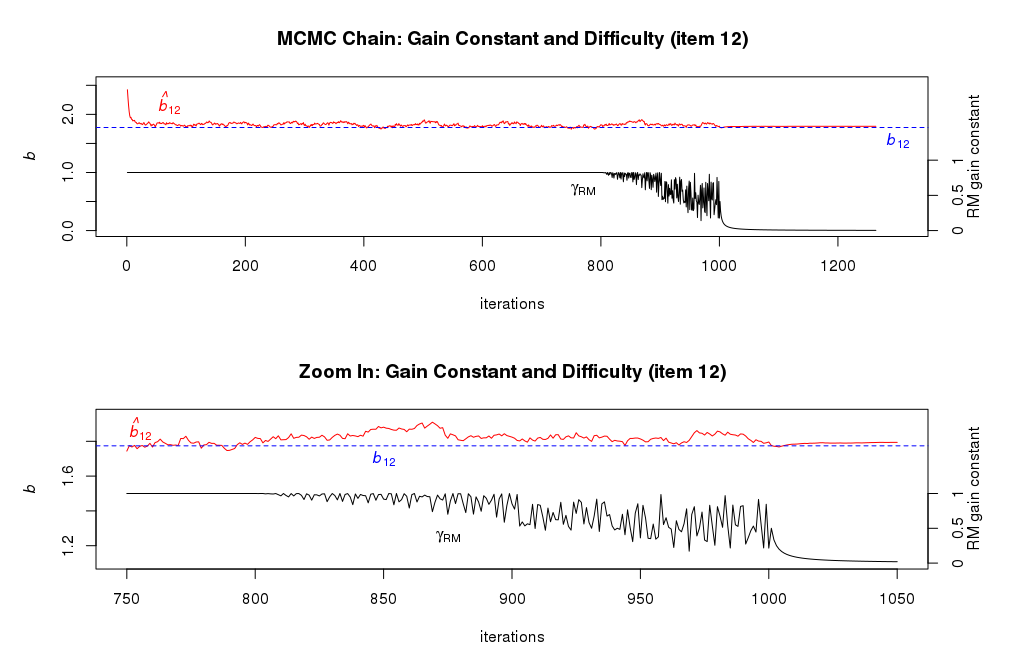}
\caption[Gain Constant and MCMC Chain Visualized from a Univariate SAEM Estimation]{The gain constant recorded from a simulation in Condition 1 is shown here (bottom line) along with a sequence of draws of a difficulty parameter for a single item (upper line) as well as the parameter's true value (dotted line). The choice of the item parameter plotted is for convenient visualization  purposes only. In the upper plot is the entire sequence of iterations of the MCMC chain, from the start until convergence. The lower plot shows the same three lines centered on the pseudo-annealing window. It is clearly visible that the variation in the chain is significantly suppressed once the Robbins-Monro updates are initialized at iteration 1000.}\label{fig:RMexample}
\end{figure}

In Figure~\ref{fig:RMexample}, a parameter was selected from the first replication of the first simulation condition in Table~\ref{tab:sims}. The difficulty parameter for item 12 was chosen for the convenience of visualizing its MCMC chain simultaneously with the gain constant's sequence of recorded values. At the beginning of the burn-in of the MCMC chain, the difficulty parameter quickly converges towards the true value shown as the dotted line. As the gain constant enters the pseudo-annealing window, a large amount of the variation seen in the parameter's chain remains intact as only a very small amount of the statistical information may pass from one iteration to the next when the gain constant remains close to one. Immediately after iteration 1000, the Robbins-Monro sequence begins and the variation in the chain quickly diminishes.

\section{Simulations of Probit Model Convergence of SAEM in One Dimension}

The performance of the SAEM algorithm is first benchmarked for one dimension of latent factor. Condition 1 from Table~\ref{tab:sims} contains 100 dichotomous items and 5000 examinees without a guessing parameter. Condition 2 contains the same configuration, but with guessing allowed for each item. Condition 3 differs from Condition 1 in that it is now a polytomous 4 category test. Once Condition 1 is assessed, it should be expected that Condition 2's inclusion of a guessing parameter will increase the RMSE for the reconstruction of the slope and intercept parameters, and Condition 3 should decrease the RMSE of the slopes but increase the RMSE of the first order intercept. It should also be expected that Condition 3 recovers examinee abilities better than Condition 1, and Condition 2 will underperform Conditions 1 and 3 in its estimation of examinee abilities. 

\subsection{Simulation and Estimation of the 2PNO}

\begin{figure}
\centering
\includegraphics[width=5.5in]{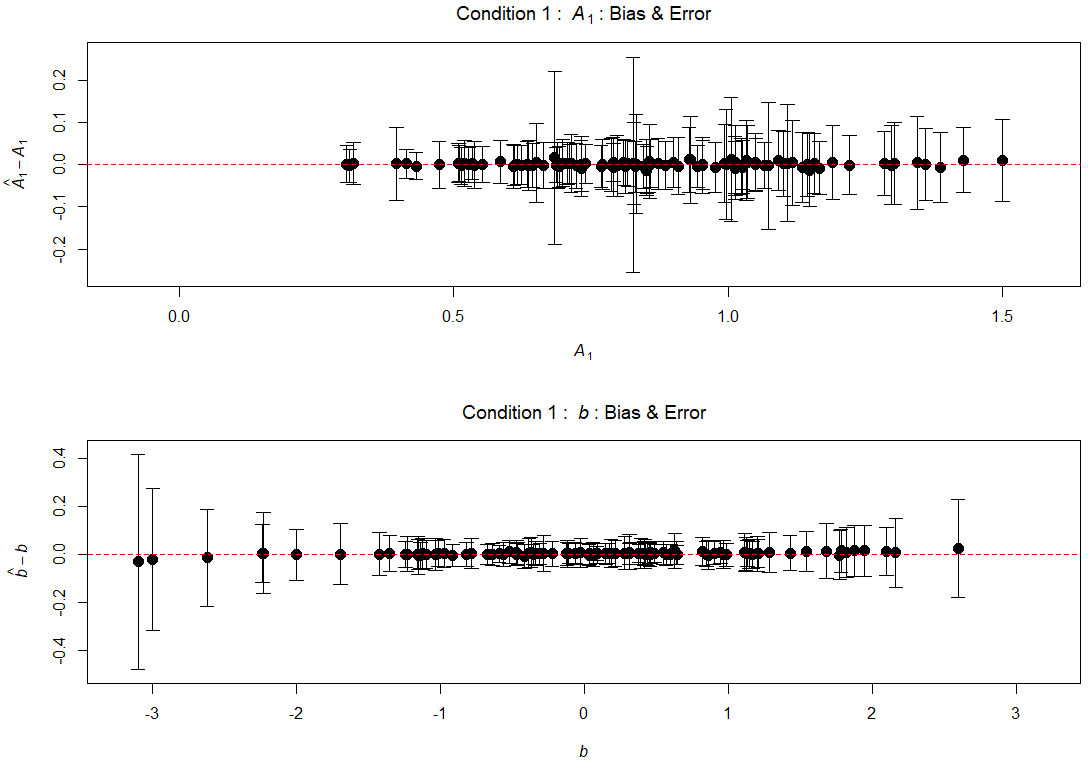}
\caption[2PNO Univariate Structural Parameter Estimation]{Condition 1 shows the results of 50 replications of parameter estimation using the SAEM algorithm coded in \textcite{GeisGit2019}. The means of all 50 estimates minus the actual values are plotted along with error bars representing a 95\% confidence interval calculated from the RMSE. Note that axes are not on the same scale. The results are in Table~\ref{tab:S1}.}\label{fig:S1}
\end{figure}

The plot of the results of the most parsimonious simulation condition are shown in Figure~\ref{fig:S1}. The generated values of slopes and intercepts are on the horizontal axis with the mean and variance of their 50 replications of residuals plotted on the vertical axis. An OLS (ordinary least squares) simple regression of the bias on the generated value was also fit for diagnostic information. Bias and RMSE are listed in Table~\ref{tab:S1}. The only semblance of a systematic error is seen in the OLS fit of the intercept; in the region of very low difficulty there is a bias of approximately -.1\%, while in the region of high difficulty there is an expected positive bias of approximately +.3\%; for simplicity, only the mean estimates of bias are shown in the table.
\begin{table}
\centering
\begin{tabular}{cS[table-format=1.4]S[table-format=1.4]}  
Paramater & $\overline{\mbox{bias}}$ & $\mbox{RMSE}$ \\ \hline
$A_1$ & .0002 & .0052  \\ 
$b$ & .0042 & .0068  \\  \hline
\end{tabular}
\caption[Diagnostics of Bias in Condition 1]{Over 50 replications, Condition 1 showed no bias in the slope's estimates nor as a function of the generated value. The intercept showed no bias when calculating the mean and its standard deviation, but there was a very small but significant systematic bias in the intercept as a function of the generated value (approximately .004 at $b=0$ and .014 at $b=2.5$). A $p$-value less than .001 is denoted by $^{***}$.} \label{tab:S1}
\end{table} 
  
\begin{figure}
\centering
\includegraphics[width=5.5in]{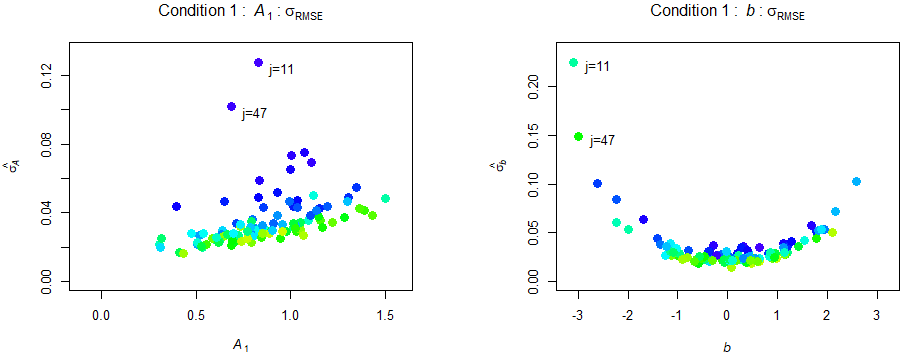}
\caption[2PNO RMSE Diagnostics for Condition 1]{Visualization of the RMSE for each of the structural parameters is useful for understanding the statistical information available in the parameter space. The RMSE for the slopes and intercepts of each of the 100 items is plotted against its generated value in the upper left and right plots, respectively. Points in the slopes plot are shaded as the item difficulties approach large relative absolute values. In the upper right intercepts plot, the points are shaded darker as slopes increase in value. Items 11 and 47 are singled out to demonstrate the effects of an extreme intercept on the RMSE of both the slope and intercept. }\label{fig:S1_RMSE}
\end{figure}

In Figure~\ref{fig:S1}, the scale of the $y$-axis demonstrates the very small RMSE resulting from this simulation. The few slopes with sizable RMSE can be seen in Figure~\ref{fig:S1_RMSE}; the items with large RMSE in the slope are sampling the statistical information from the tails of the prior distribution of abilities. Further, the large RMSEs in the intercepts are from these same \textit{long-tail} items $J = \{11,47\}$ of extremely low difficulty. There is not only a general increase in the RMSE of the slope as the generated slope increases, but there is a steeper increase in the uncertainty of the slope as the intercept extends into the tails of the ability of our examinees, and this is to be expected because there is far less available statistical information from the examinee pool in this region. in Figure~\ref{fig:S1_RMSE}, the right-hand plot also shows that higher values of slope influence the RMSEs of the intercepts; the shading of the points are darker as the slope increases.

\begin{figure}
\centering
\includegraphics[width=5in]{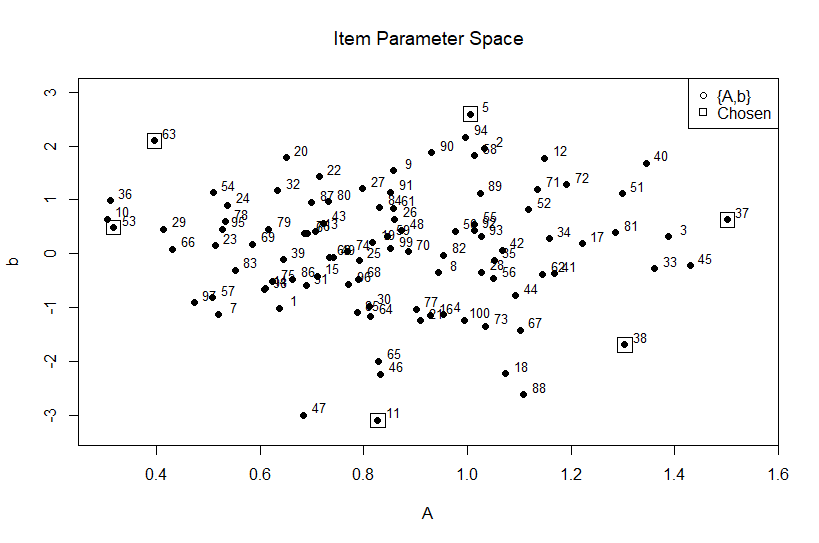}
\caption[Structural Item Parameters in Condition 1]{The item parameter space of all 100 items of Condition 1. Surrounded with a square are the 6 items chosen to visualize the chains and their drift in Figures~\ref{fig:S1_MCMC} and \ref{fig:S1_MCMC_2D}.}\label{fig:itemPars}
\end{figure}

The behavior of the MCMC chain can also be instructive via a visualization of the parameter estimates for each iteration as one or more replications evolve through the gibbs cycles. Six items were chosen for investigation based on their location in their generated parameter space as seen in Figure~\ref{fig:itemPars}; the six items were strategically chosen as they land on the edges of the joint distribution of structural parameters: two items are of central difficulty but high and low slopes (37 and 53), two items have slope nearly 1 but very high and low difficulties (5 and 11), and two items are at the corners of this parameter space (38 and 63). 

\begin{figure}
\centering
\includegraphics[width=5.5in]{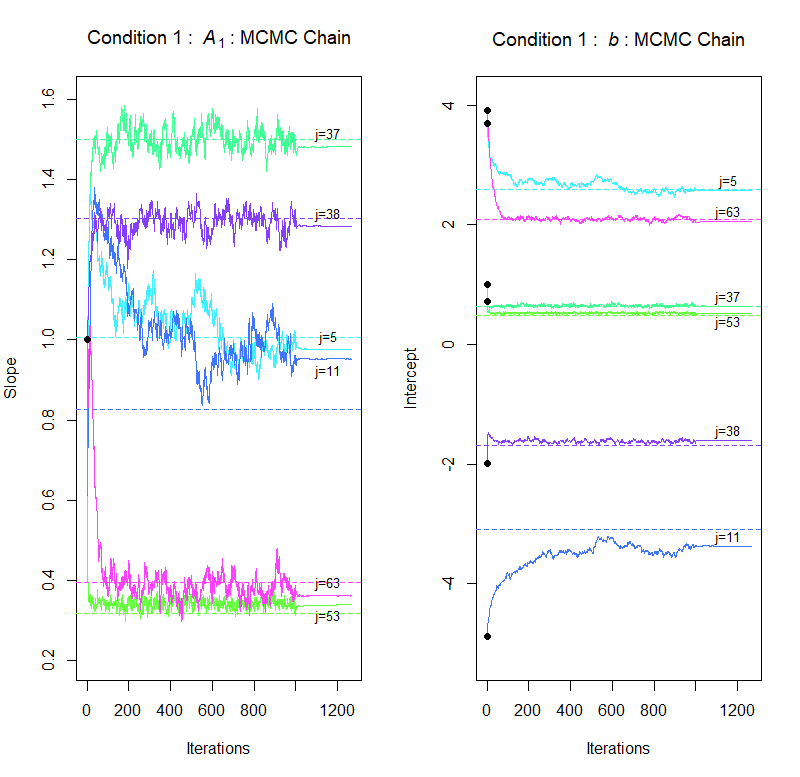}
\caption[2PNO MCMC Visualization for Condition 1]{Six items were chosen due to their visual separation in the parameter space, and the MCMC chain for each pair of slope and intercept are plotted and labeled. The points at Iteration 1 are also shown as the initial values where the slopes are set to 1 and the difficulties of each item are estimated from the simulated responses. The dotted line represents the generated parameter and the line with stochasticity shows every estimate at each iteration in the estimation. Note that the pseudo-annealing window is activated at iteration 800, and the Robbins-Monro is engaged at iteration 1000; this window is shown with dotted vertical lines. It is notable how the slope of item 11 shows the strongest deviation from its generated value as it is paired with a large absolute value of the intercept.}\label{fig:S1_MCMC}
\end{figure}

In Figure~\ref{fig:S1_MCMC} the MCMC chains of the structural parameters of these six items from the first replication of Condition 1 are shown. In the first iteration the slopes are fixed to 1 and the intercepts are estimated by converting the probability of a correct value into an estimate of the difficulty, e.g. if only 10\% of the examinees correctly answered an item its initial estimate for the intercept would be near the value $X$ for $P(\theta>X) = .10$ given $\theta \sim N(0,1)$; the solid points in the figure at iteration $t=1$ are those initialized values. Also important for the interpretation of the plots, the dotted lines show the point when the pseudo-annealing window is activated at iteration 800 and the Robbins-Monro is engaged at iteration 1000; the extreme drop in the variation of the chain is clearly recognized after iteration 1000.

The chain of the slopes in items 5 and 11 stand out from the other chains as they traverse a large range of values from their initial burn-in phase of estimates as the difficulty parameters are at the extremes of the bell-curve of abilities of our examinees. For these items, each change in direction in the intercept's chain has a notable autocovariance with its paired slope and there is very little demonstrable stability until about iteration 600. Further, the slope of item 11 has a large error in the positive direction at its convergence; the horizontal dotted line at $A=.827$ being its generated value. The variability in the chains of the slopes of items 53 and 63 are instructive as they demonstrate the effect of a central difficulty as opposed to a large absolute value of difficulty. 

\begin{figure}
\centering
\includegraphics[width=5.5in]{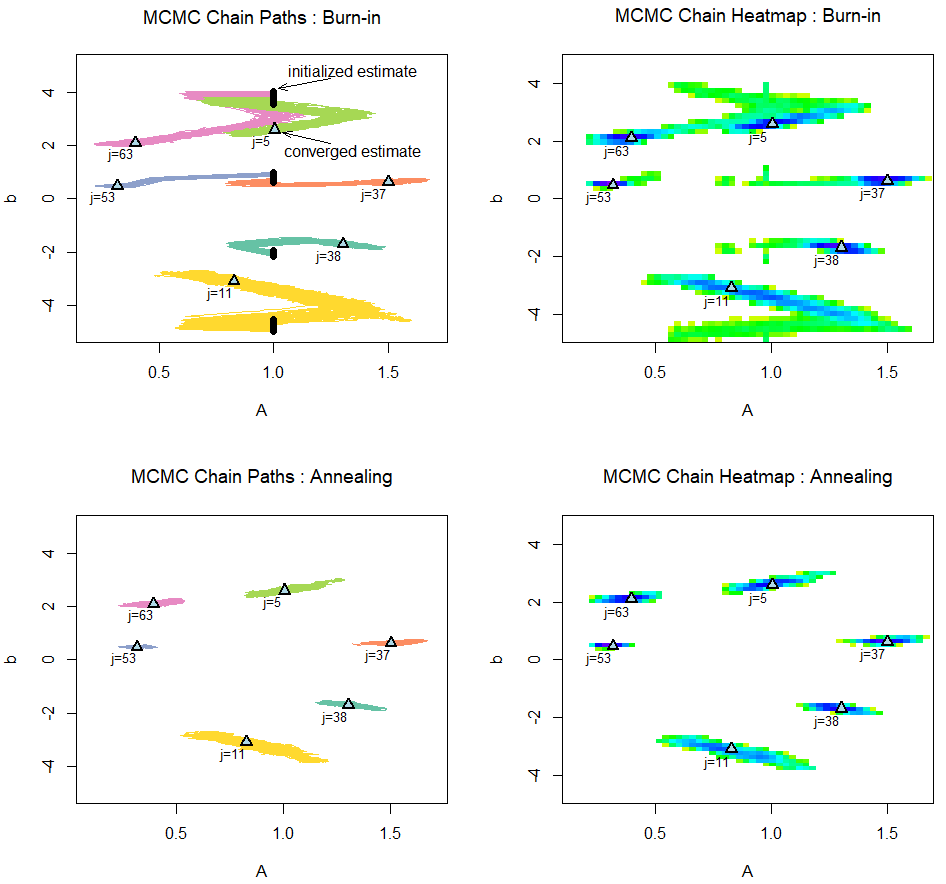}
\caption[2PNO MCMC 2D Visualization]{The same six items from Figure~\ref{fig:S1_MCMC} are also visualized in their joint parameter space, and the MCMC chain of all 50 replications for each item's pair of slope and intercept are plotted and labeled along with their starting points (dark circle), with lines showing the MCMC path to their true value (six triangular points). The plots on top show the burnin phase of the gibbs cycles, i.e. the gain constant $\gamma_t = 1$. In the bottom plots the stochasticity within the annealing iterations are shown to be quite stable. The plots on the left show the paths while the plots on the right are a heatmap, darker where the iterative estimates are well localized.}\label{fig:S1_MCMC_2D}
\end{figure}

In Figure~\ref{fig:S1_MCMC_2D}, the iterations of the estimates for the same six items from Figure~\ref{fig:S1_MCMC} are shown so as to demonstrate the progressive movement towards the region of highest likelihood. The parameters begin at the initialized points, and in the burn-in iterations, progress from the initialized values into the region of highest likelihood. When the pseudo-annealing window is engaged, the parameters are already in the vicinity of their true value, plotted as a triangular point and labeled for each of the six items of interest. The plots on the right are heatmaps with the darkest regions indicating strong localization of the estimates; it can be seen that these parameter estimates are highly localized in the region of highest likelihood before the Robbins-Monro sequence is engaged.

 \bgroup
\def\arraystretch{1.3}
 \begin{table}
 \centering
\begin{tabular}{ccc} 
Item &  Cov$(\hat{A}_{r},\hat{b}_{r})_{t=1000}$ & Cov$(\hat{A}_{r},\hat{b}_{r})_{\mbox{Converged}}$ \\  \hline
53 & $\left[ \begin{array}{rr} .0008 & .0001 \\ .0001 & .0004 \end{array} \right]$ &  $\left[ \begin{array}{rr} .0006 & .0001 \\ .0001 & .0004 \end{array} \right]$ \\
37 & $\left[ \begin{array}{rr} .0033 & .0010 \\ .0010 & .0014 \end{array} \right]$ &  $\left[ \begin{array}{rr} .0024 & .0007 \\ .0007 & .0013 \end{array} \right]$ \\  
5 & $\left[ \begin{array}{rr}   .0058 & .0070 \\ .0070 & .0113 \end{array} \right]$ &  $\left[ \begin{array}{rr} .0054 & .0068 \\ .0068 & .0105 \end{array} \right]$ \\  
11 & $\left[ \begin{array}{rr} .0164 & -.0262 \\ -.0262 & .0511 \end{array} \right]$ &  $\left[ \begin{array}{rr} .0162 & -.0265 \\ -.0265 & .0505 \end{array} \right]$ \\  
38 & $\left[ \begin{array}{rr} .0027 & -.0030 \\ -.0030 & .0050 \end{array} \right]$ &  $\left[ \begin{array}{rr} .0024 & -.0026 \\ -.0026 & .0040 \end{array} \right]$ \\  
63 & $\left[ \begin{array}{rr} .0021 & .0012 \\ .0012 & .0029 \end{array} \right]$ &  $\left[ \begin{array}{rr} .0019 & .0013 \\ .0013 & .0025 \end{array} \right]$ \\  \hline
\end{tabular}
\caption[Variation of Estimates in the Iterations of the Robbins-Monro]{The variance of the 50 replications of estimates at the end of the annealing window and at their converged values are produced and shown in this table. The subscript of $r$ denotes that the parameter estimate data used in the covariance $(\hat{A}_{r},\hat{b}_{r})$ are from each replication.} \label{tab:RMDrift}
\end{table} 
\egroup

To assess whether the Robbins-Monro iterations have some structural drift towards or away from the generated structural parameters, \textit{snapshots} of variances of the 50 replications of estimated parameters at different points in the MCMC chain can be produced. From this effort, it is notable that the variation of the estimates about the true values did not decrease over the span of the annealing window. In contrast, during the RM iterations each diagonal element of the variance among the 50 replications decreased from the initialization to the convergence of the algorithm. These covariances are shown in Table~\ref{tab:RMDrift}. After observing the parameter drift and this decrease in the covariance of the converged parameters, it is evident that the RM method favors a direction that nudges the stochastic estimates towards a converged estimate that approaches the generated parameters.

\subsection{Error Estimation of the 2PNO} \label{sec:errors}

As important as the model’s estimation of the parameters, is the model’s ability to explicate the precision of those values. SAEM poses a distinct challenge in using traditional diagnostics as the computational problem it attempts to solve precludes the use of many of the accepted diagnostic criteria associated with convergent MCMC chains; the Robbins-Monro algorithm accelerates convergence to bypass typical time and memory requirements of MCMC. Error estimation in MCMC methods at first glance seems simple as the samples of the chain are confined to the hyperspace of the parameters and seem to reflect the variability of the estimates themselves, but any application of the CLT (central limit theorem) under the assumptions of MCMC requires that the chain be stationary which is not easy to verify \parencite{geyer1992practical}. This difficulty is more pronounced as stochastic approximation is implemented since the chain is `'squeezed'' before errors are estimated in the traditional fashion. Consequently, standard error estimation is as challenging with the SAEM approach as the accuracy of the parameters themselves.  

\begin{figure}
\centering
\includegraphics[width=5.5in]{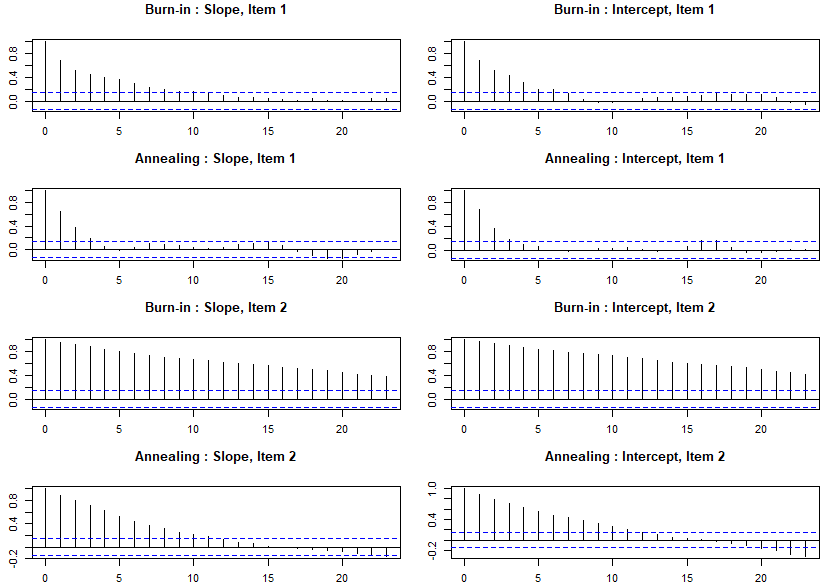}
\caption[Autocovariances for Items 1 and 2]{The autocovariance plots for the slopes and intercepts of the first two items during the final 200 iterations of the burn-in and the 200 iterations of the pseudo-annealling window. }\label{fig:ACF2pnoAnn}
\end{figure}

In order to confidently work with the variation within the MCMC chains, autocovariance should be properly analyzed to infer a lag for trimming. As shown in Figure~\ref{fig:ACF2pnoAnn}, the autocovariances of the structural parameters for the first two items are shown during the 20\% of iterations of burn-in, as well as during the 200 iterations of the pseudo-annealing window. The autocovariances are better behaved during the annealing window as the number of lags to confidently extract independent draws have substantially decreased. Looking across the items, the thinning of the chain of the slopes allowed a modulus of 8, while the thinning of the intercepts will use a modulus of 5. This choice of trimming is used for the MCMC errors. The CLT calculations of the MCMC chains incorporate the autocovariances and cross-correlations of the chains, thus eliminating the need for trimming.

\begin{figure}
\centering
\includegraphics[width=5.5in]{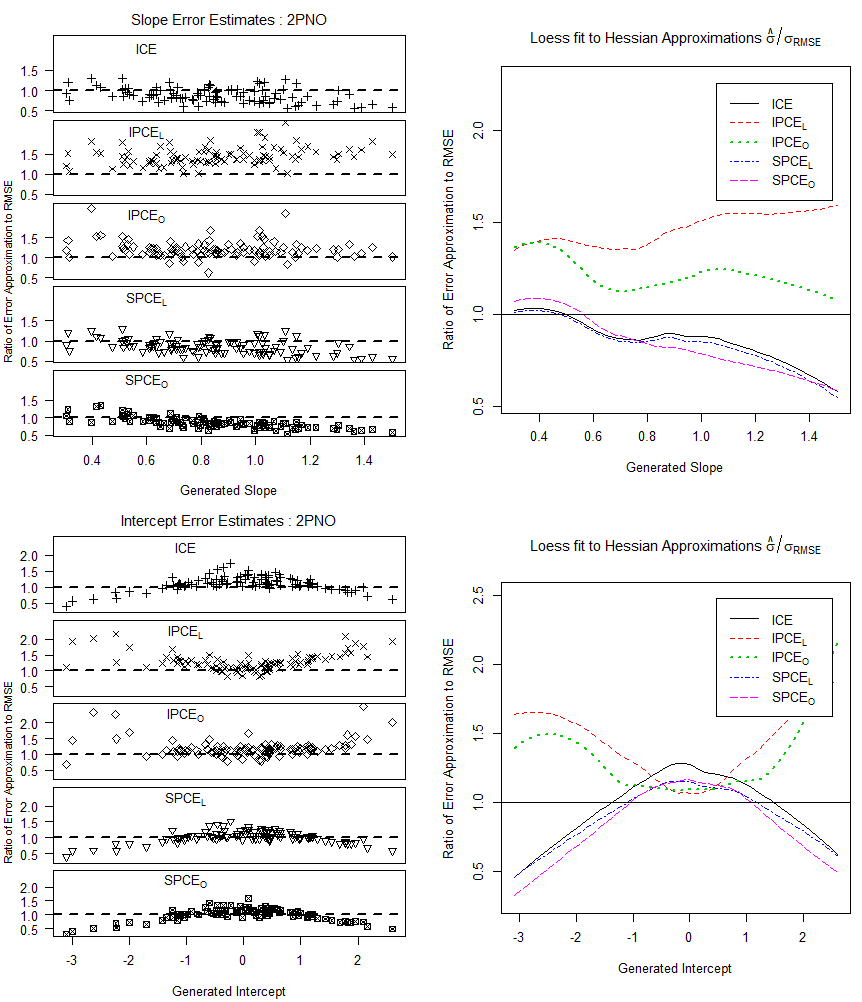}
\caption[Error Estimates from the Hessian for the 2PNO]{The error estimates derived from the Hessian of the methods described in Table~\ref{tab:ErrorAcro}.}\label{fig:Error_Hess}
\end{figure}

In Section~\ref{sec:ErrEsts} the error estimation methods were introduced. The parameters' RMSEs over the 50 replications are the \textit{gold standard} and will serve as the reference value for all error estimates; each error calculation shown in Figure~\ref{fig:Error_Hess} are plotted as ratios to the RMSEs. The estimates of the calculated errors shown in the figure are simple averages of the calculated error estimates for each structural parameter from all 50 replications. The five plots of slope errors and intercept errors are calculated using the five separate estimates of the Hessian. The ICE (Iterative Converging Error) is calculated during convergence where the RM gain constant is applied to the Louis missing information equations given in the Equations in \ref{eq:hesscomps}; it approximates the errors relatively well at low slopes and near the center of the intercept distribution, but then overestimates large slope errors and underestimates extreme intercept errors. The SPCE (Simple Post-Convergence Error) was calculated using the 2PNO and 2PL formulations as discussed in the previous chapter. The SPCE\textsubscript{L} best approximates the RMSE of the intercepts when compared to all five Louis missing information methods in that the errors have the closest ratio to 1. The SPCE\textsubscript{O} is very close to its 2PL approximation and the ICE for both slopes and intercepts. The IPCE (Iterative Post-Convergence Error) was also calculated using both the 2PNO and 2PL formulations. For a conservative measure, the IPCE\textsubscript{O} could be used for either the slopes or intercepts as it does not underestimate either parameter, though the overestimation of intercept errors far from the mean are quite large approaching a factor of two.

Unfortunately, the IPCE errors require the reinitialization of the chain after convergence, thus making it the most expensive of the Hessian approximations. As mentioned in Section~\ref{sec:ErrEsts}, the Louis missing information Hessian calculations have the potential to produce instabilities resulting in negative errors \textit{if} the gibbs cycles sample the tails or edges of the hyperspace of the parameters. It is not visualized on the plots of errors, but several of the post-convergence (IPCE) error estimates resulted in a negative error calculation for the parameters of some items. In more than 50\% of the replications, specifically items 5, 11, 47, and 88, the instabilities of the tails of the convex hyperspace resulted in negative terms; the source of the instabilities can be seen in Figure~\ref{fig:itemPars} and are clearly associated with extremes of difficulty, or high magnitudes of intercept. The ICE estimation showed some negative errors, but was much more stable as only 1 or 2 replications for nearly each item resulted in a negative term. The SPCE calculations did not result in a negative error, and this is explained by the calculation taking place at the converged estimates using only a single sample of the abilities; thus this sample occurred at the point of maximum convexity. For the practical use of these error calculations, it is useful to quote the results of the ICE and SPCE\textsubscript{O} as they are inexpensive to calculate, not requiring the reinitialization of the MCMC chain. 

\begin{figure}
\centering
\includegraphics[width=5.5in]{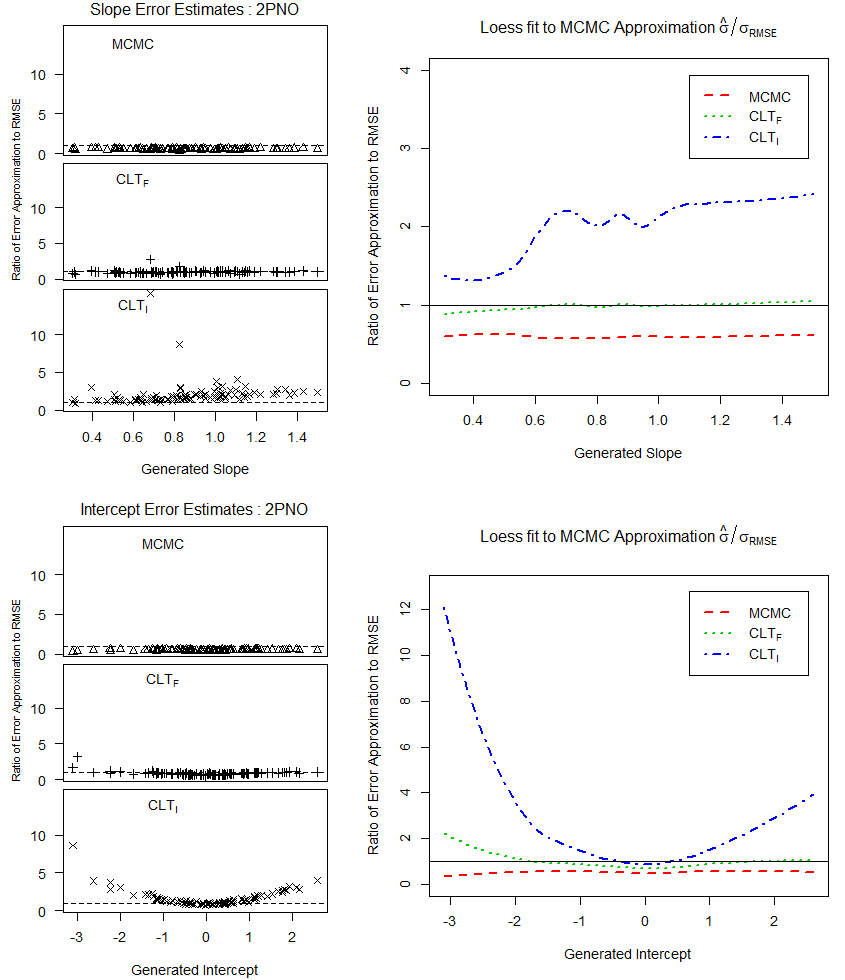}
\caption[Error Estimates from the MCMC Chains for the 2PNO]{The figure shows the error estimates derived from the MCMC Chains during the 20\% of burn-in iterations prior to the pseudo-annealing phase. Each method is as described in Table~\ref{tab:ErrorAcro}; $\mbox{CLT}_I$ and $\mbox{CLT}_F$ are standard deviations derived from the variance of the final 20\% of cycles prior to the pseudo-annealing phase, where the chains of structural parameters are trimmed and taken within items independently, and the full chains of all structural parameters are taken together, respectively. The estimates labeled as the MCMC estimates are the independent standard deviations of each chain of its respective structural parameter.} \label{fig:Error_MCMC}
\end{figure}

In Figure~\ref{fig:Error_MCMC}, the three plots specific to the slopes and intercept to the left of both summary plots on the right, are derived from the MCMC chains. \textcite{geyer1992practical} explains the asymptotic variance in the central limit theorem and these estimates are calculated during the burn-in phase (iterations 600-800 in condition 1; 20\% of the burn-in phase, before the annealing iterations are engaged). The MCMC errors are the standard deviations of the chains each structural parameter, thinned using the inspection of the autocovariances as described above. The CLT\textsubscript{F} is derived from the covariance of all of the structural parameters' chains in totality. The CLT\textsubscript{I} is derived from the covariance of the chains of each item's structural parameters, thus obeying the assumptions of mutual independence.

It must be noted that the variances of the chains of each parameter significantly underestimate the errors of their parameter; this suggests that a significant proportion of variation is shared between chains, especially as the mutually independent item errors are estimated to be extremely large when the intercept parameters approach the tails of their prior. When comparing the errors that are derived from all chains (CLT\textsubscript{F}) to the errors derived from the chains of each item (CLT\textsubscript{I}), the ``mutually independent'' errors significantly overestimate the errors. The covariation of structural parameters between items may play a significant role in the Bayesian approach used here. Remarkably, the diagonal elements of the covariance of the entire MCMC chain performs better than every other estimate shown in Figures~\ref{fig:Error_Hess} and \ref{fig:Error_MCMC}.

\subsection{Factor Score Estimation}

\begin{figure}
\centering
\includegraphics[width=5.5in]{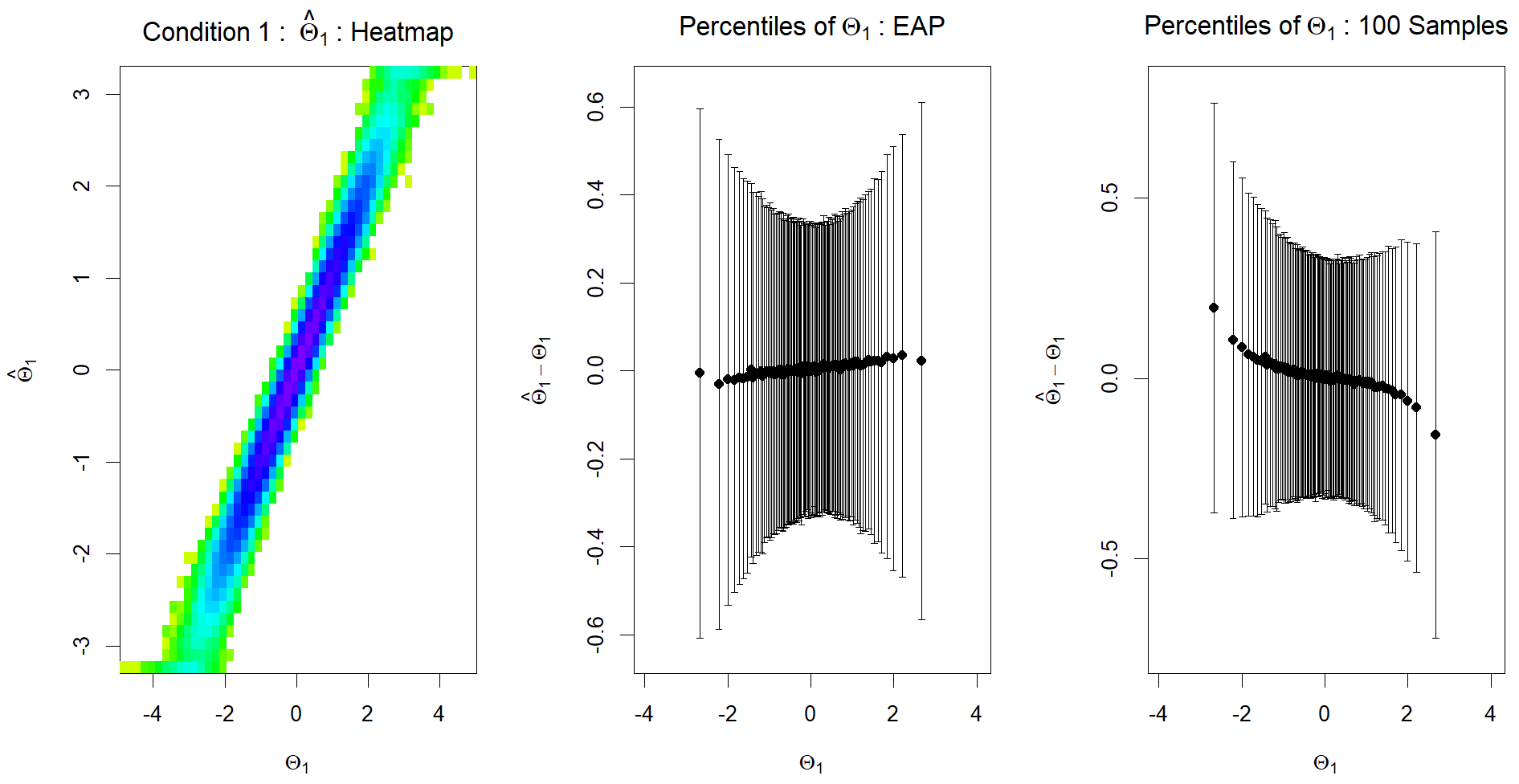}
\caption[Estimates of Examinee Abilities in Condition 1]{A heatmap of the 50 replications of 5000 examinees' abilities is shown in the first plot. The following two plots show the bias and 95\% confidence intervals of the factor estimates of 100 percentiles of the generated abilities. In the middle plot, the EAP is shown. In the right plot, samples of ability are taken using a single post-convergence gibbs cycle starting from the converged estimates of the parameters; this is done 100 times and a mean ability for each examinee is calculated along with its 95\% confidence interval for each of the 100 percentiles of generated abilities.}\label{fig:S1_Theta}
\end{figure}

With symmetric priors, the EAP, MAP, and MLE of the factor estimates are equal. An alternative of restarting the MCMC chain from converged parameters and sampling $\theta$ 100 times is also performed; the mean of these 100 samples is used as the alternative factor estimate. In Figure~\ref{fig:S1_Theta}, the heatmap of EAP estimates are first shown against the generated abilities. In the middle and right plots, the EAP and the alternative measure are plotted for each percentile of generated ability along with a 95\% confidence interval. 

In the EAP, there is a slight bias in underestimating and overestimating the abilities of those examinees from the lower and upper regions of the prior distribution, respectively. Starting from converged estimates, the sampled $\theta$ estimates show a reversal in the relationship of the EAP and a more significant bias as the generated ability approaches the tails. In the center quintile, the mean of the estimates is significantly non-zero at a value of .003 for both methods, showing positive bias but at only at 1\% of the 95\% confidence interval which is approximately .33 in the center.

\subsection{Simulation of the 3PNO}

It is trivial to expand the approach of the 2PNO to guessing in the manner described by \textcite{beguinglas2001}. The model for the 3PNO involves a binary classification where we assign the students to either knowing or not knowing the correct response. This model is designed such that a student who knows the correct response is expected to have 100\% probability for answering correctly and 0\% chance to answer incorrectly. If the student does not know the correct response, there is a finite probability that the resulting choice is correct or incorrect. In relation to our augmented data, the following assignments are:

\begin{eqnarray}
Y_{ij}=1 & \rightarrow &  W_{ij}=1 \mbox{  with probability  } \Phi (\eta_{ij}) \\
Y_{ij}=1 & \rightarrow & W_{ij}=0 \mbox{  with probability  } g_{j}(1-\Phi (\eta_{ij})) \\
Y_{ij}=0 & \rightarrow & W_{ij}=0.
\end{eqnarray}

\noindent Next, instead of choosing $z$ from the side of the truncated normal determined by $Y_{ij}$, the determination is now a result of the stochastic variable coded as the examinee's knowledge of the answer, $W_{ij}$. The itemized choices for assigned $W_{ij}$ become the intermediate step before the augmented $z$ is drawn. The value for $g_{j}$ becomes another sufficient statistic calculated at the end of each iteration; it is defined as a sum conditional on examinees' not knowing the correct answer, $g_{j} = \sum_{i|W_{ij}=0} \frac{Y_{ij}}{N_{W_j=0}}$. When the Robbins-Monro is invoked, this guessing parameter converges along with the other parameters.

\begin{figure}
\centering
\includegraphics[width=5.5in]{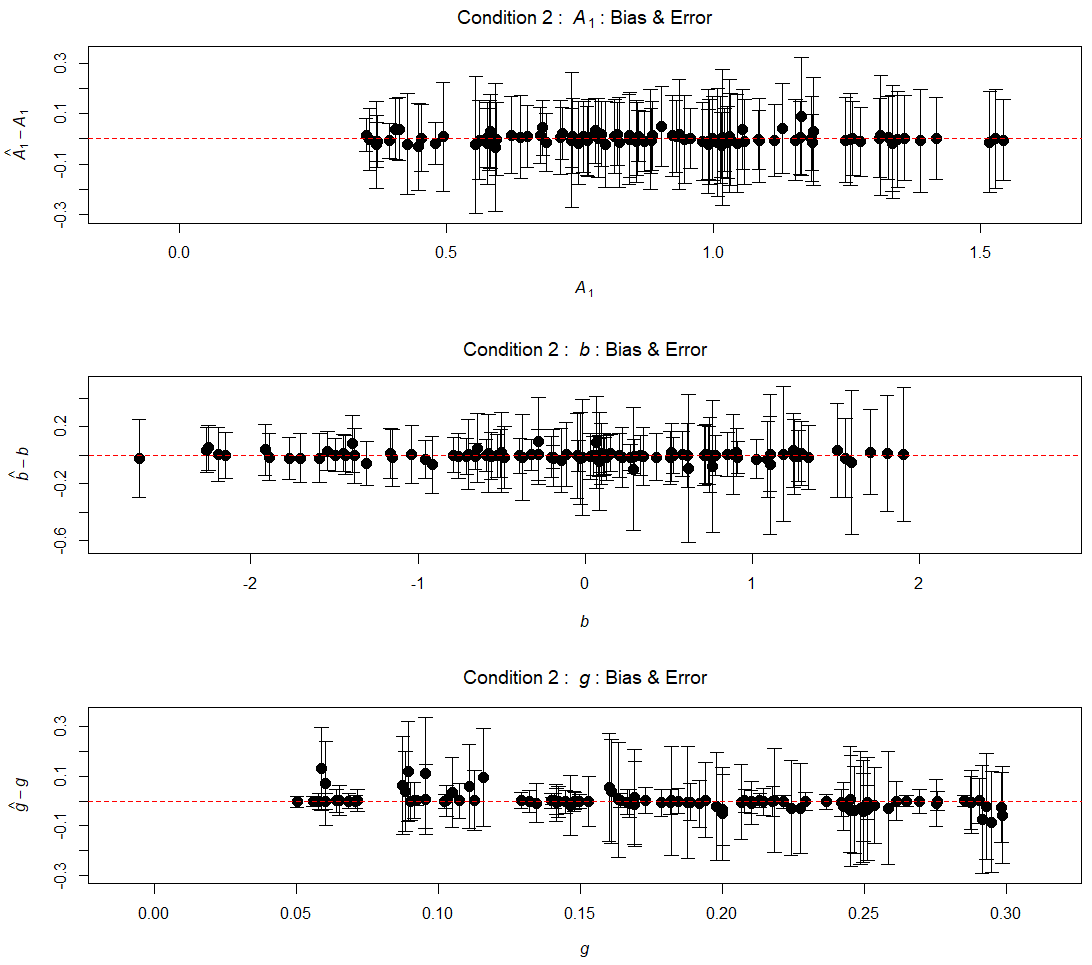}
\caption[3PNO Univariate Structural Parameter Estimation]{This figure shows the results of 50 replications of parameter estimation of Condition 2 using the SAEM algorithm coded in \textcite{GeisGit2019}. All 50 structural parameter estimates minus the actual values are plotted along with error bars representing a 95\% confidence interval calculated from the RMSE. Mean estimates for detection of bias results are in Table~\ref{tab:S2}. Note that axes are not on the same scale.  Relationships between errors and simulated values are shown in Figures~\ref{fig:S2_sigmaAb} and \ref{fig:S2_sigmaG}.}\label{fig:S2}
\end{figure}

 \begin{figure}
\centering
\includegraphics[width=5.5in]{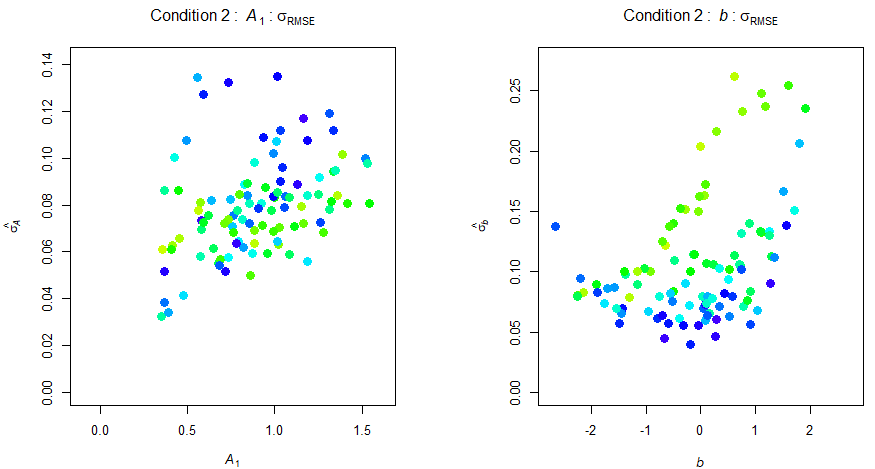}
\caption[3PNO Slope and Intercept RMSE Diagnostics]{The RMSE for the slopes and intercepts of each of the 100 items is plotted against its generated value in the left and right plots, respectively. In the left plot of slopes, points are shaded as the item difficulties approach large relative absolute values to demonstrate the effects of an extreme intercept on the RMSE of the slope. In the right plot of intercepts, points are shaded as the slopes or factor loadings increase in value.}\label{fig:S2_sigmaAb}
\end{figure}

 \begin{figure}
\centering
\includegraphics[width=5.5in]{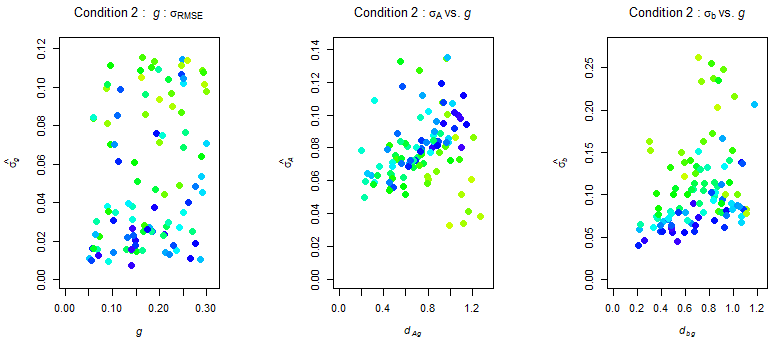}
\caption[3PNO RMSE Diagnostics with Guessing]{The RMSE for the guessing parameter is first shown against its simulated value with the points shaded as the generated slopes are increased. In the middle and right plots, the RMSEs of the slope and intercept are shown against a scaled two-dimensional Euclidean distance; $d_{Ag} =  \norm{\{|A-\bar{A}|,|g|\}}_{L_{2}}$ and $d_{bg} =  \norm{\{|b|,|g|\}}_{L_{2}}$. In the slopes plot, the points are shaded as the slope gets larger. In the right plot of the RMSEs of the intercept, the points are also shaded darker as the generated slopes increase.}\label{fig:S2_sigmaG}
\end{figure}

\begin{table}
\centering
\begin{tabular}{crr}  
Paramater & $\overline{\mbox{bias}}$ & $\mbox{RMSE}$ \\ \hline
$A_1$ & .0020 & .0187 \\ 
$b$ & -.0032 & .0302 \\ 
$g$ & .0002 & .0326  \\ \hline  
\end{tabular}
\caption[Diagnostics of Bias in Condition 2]{Over 50 replications, the mean of parameter estimates' residuals is shown along with RMSE. } \label{tab:S2}
\end{table} 

Figure~\ref{fig:S2} shows the RMSEs of the structural parameters and the mean of the 50 replications of estimates with the generated parameters subtracted. The guessing parameter shows a significant bias with the large RMSE estimates being overestimated and underestimated at smaller and larger guessing values, respectively. In Figures~\ref{fig:S2_sigmaAb} and \ref{fig:S2_sigmaG}, the RMSEs are further inspected against the values of the other generated structural parameters. A few relationships are notable. The greater the generated slope the more accurate the estimate of the intercept, which is consistent with the results of the 2PNO, but this effect becomes much more pronounced as the guessing parameter increases. The intercepts' errors also increase as the item difficulty increases in general, which is an unexpected result, though the relationship can also be seen in \textcite{beguinglas2001} where the errors of their intercept parameters were largest at high difficulty. The slopes RMSE increases much more quickly at large absolute values of the intercept and large values of guessing. The errors of the guessing parameter are significantly worse as the slope decreases but were less affected by the generated intercept. 

%

Error estimates of the 3PNO are best approximated by the analytical results of the Hessian approximations. Unlike the 2PNO, the CLT approximations of the MCMC chains heavily overestimate the errors of the structural parameters, regardles of whether the items are treated independently or all chains are considered in the covariance of the hyperspace of the gibbs cycles. The SPCE method underestimates the errors, especially at large difficulty as the guessing parameter strongly influences both the discrimination and difficulty parameters when these parameters are less than one or less than zero, respectively. 

Another approach that is very practical, especially in this Bayesian framework, is to restart the chain at converged estimates and investigate the Bayesian posterior densities of the structural parameters. Assuming the simulations have converged to a reasonable model, the chains of the stationary process can be compared to the generated parameter and its 95\% confidence interval of the RMSE. When restarting the first replication of this simulation at converged parameters and continuing the gibbs process for 2000 iterations, the slopes were within the bounds of twice the RMSEs in 94.3\% of the iterations, intercepts were within their 95\% confidence interval in 91.6\% of all samples, and the 90.9\% of the posterior distribution of sampled guessing parameters were within the bounds of twice their RMSEs. In this run, item 86 ($a = 1.33$, $b = .424$, $g = .276$) showed the worst performance in the posterior with the slope and intercept in the 95\% confidence interval defined by their RMSEs 18.2\% and 28.8\% of the iterations, respectively. Of the 100 items, 51 had greater than 95\% of all three parameters' posterior samples coming from within their respective 95\% confidence intervals over the 2000 iterations. 

The factor estimates produced are quite consistent with the 2PNO factor estimates in the previous subsection. The differences between the two estimates provided by the EAP and sampling method were negligible. In both conditions, the EAP outperforms the implementation of the fixed parameter draws as the tails are well-behaved.

\subsection{Simulation of the Polytomous 2PNO}

In moving to the polytomous condition, the intercepts are split. Further, the guessing parameter is removed as polytomous scoring is typically applied in items for which guessing is expected to be a phenomenological rarity, i.e. the probability, $P(Y_{ij}>0|W_{ij}=0)$ is very close to zero. Four ordinal polytomous categories were chosen with all other parameters remaining the same. 

\begin{figure}
\centering
\includegraphics[width=5.5in]{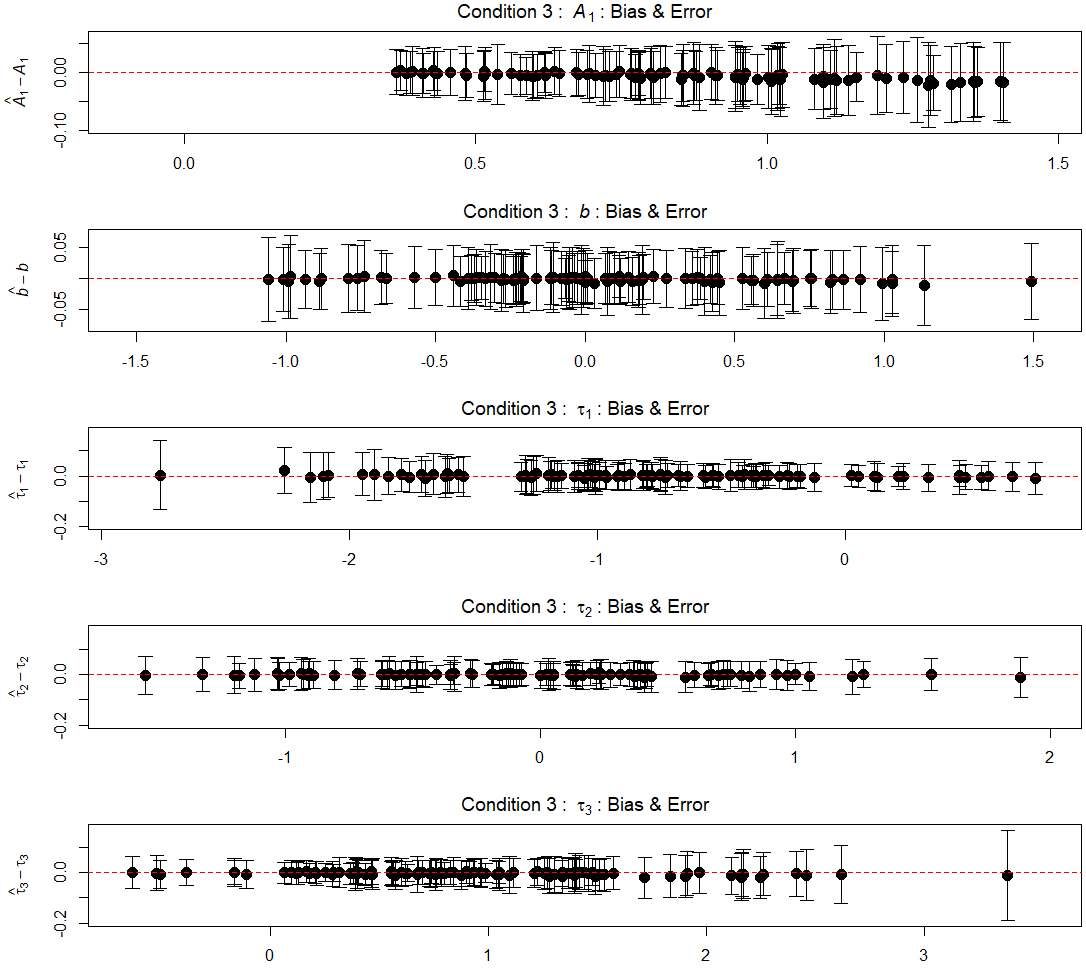}
\caption[Polytomous 2PNO Univariate Structural Parameter Estimation]{Condition 3 shows the results of 50 replications of parameter estimation using the SAEM algorithm coded in \textcite{GeisGit2019}. All 50 structural parameter estimates minus the actual values are plotted along with error bars representing a 95\% confidence interval calculated from the RMSE. Note that axes are not on the same scale. Mean estimates for detection of bias and the results are in Table~\ref{tab:S3}.}\label{fig:S3}
\end{figure}

\begin{table}
\centering
\begin{tabular}{cS[table-format=1.4]S[table-format=1.4]} 
Paramater & $\overline{\mbox{bias}}$ & $\mbox{RMSE}$ \\ \hline
$A_1$ & -.0063 & .0057 \\  
$b$ & -.0016 & .0030  \\ 
$\tau_1$ & .0008 & .0046  \\  
$\tau_2$ & -.0015 & .0034  \\ 
$\tau_3$ & -.0041 & .0054  \\   \hline
\end{tabular}
\caption[Diagnostics of Bias in Condition 3]{Over 50 replications, Condition 3 showed statistically significant bias in multiple parameters, but the deviation from zero is negligibly small in comparison to the values of the parameters.} \label{tab:S3}
\end{table} 

\begin{figure}
\centering
\includegraphics[width=5.5in]{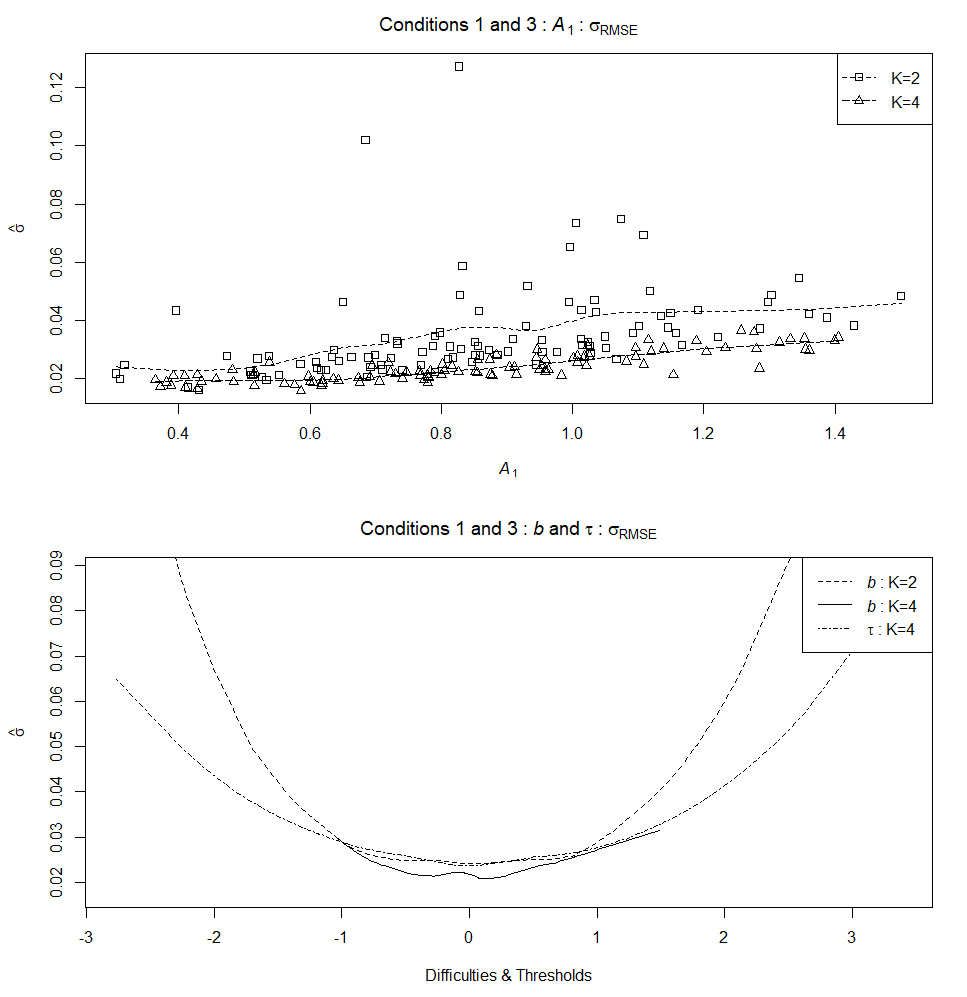}
\caption[Comparison of RMSEs of Dichotomous and Polytomous 2PNO]{Here the RMSEs of Condition 1 and Condition 3 are superimosed across all 100 item slopes and difficulty parameters. There is a significant reduction in the uncertainty of the converged parameters in Condition 3. The points are not overlaid in the bottom plot as the visualization of the differences near the center become invisible.}\label{fig:CompS1S3}
\end{figure}

In Figure~\ref{fig:S3}, the 95\% confidence intervals from the RMSEs are overlaid on the mean difference of estimates from true values for 50 replications. It is notable that extreme errors in the slope no longer seem affected by large intercepts. This may occur as the three thresholds separating the four categories are drawn from the same normal distribution and the average of these three values is the average difficulty by default, which has an expected standard deviation of $\frac{1}{\sqrt{3}}$; thus, the greater stability of the slope RMSEs. In Figure~\ref{fig:CompS1S3}, the RMSEs of condition 1 are overlaid on those of condition 3. The polytomous nature of this simulation demonstrates that the RMSEs are considerably better across all structural parameters.

The error estimates are well approximated by the MCMC CLT approach during the burn-in iterations in this polytomous simulation. The mutual independence assumption (CLT\textsubscript{I}) causes errors to be slightly underestimated as opposed to relative overestimation of errors in the 2PNO, but the full covariance of the MCMC chains gives estimates of errors of the structural parameters that are quite consistent with the RMSEs as with the 2PNO. The Bayesian draws were also performed on a single replication prior to convergence; the chains of 53 items having all five structural parameters ($a, b, \tau_1, \tau_2, \tau_3$) pulled from their 95\% confidence intervals at least 95\% of the time, and 91 of the items having at least four of the parameters drawn from their confidence intervals more than 95\% of the time.

The factor score estimates, again, showed strong consistency with the 2PNO and 3PNO. The EAP estimates also outperform the sampled estimates, specifically at the tails as in previous cases. It is worth noting that the bias in the sampled estimates has an inverse relationship to the bias of the EAP at the tails. Were the researcher forced to choose, the results from all three univariate conditions demonstrate that the EAP outperforms the stochastic estimate of the latent factors. For both methods of factor estimation, the uncertainty of the abilities are reduced with the increased granularity of the response categories; the 95\% confidence intervals are about 80\% of the size of the results of the dichotomous 2PNO from Condition 1 (in the center bucket of abilities, $\mbox{RMSE}_{\hat{\theta}=0} = .25$ as compared to .33).


\subsection{Tests of Multidimensionality for Univariate Simulations} \label{sec:2pnoDims}

In keeping with the goal of \textit{exploratory} factor analysis, each simulation condition should be run with alternate choices of dimensionality. With the 2PNO, the dimensions can only be increased. Each replication was fit setting the number of factors to two and five. The eigenvalues of the matrix $\bf{S}_2$ from Equation~\ref{eq:sigmaZ} are to be studied at convergence to address relationships that indicate the dimensionality of the underlying response data. 

\begin{figure}
\centering
\includegraphics[width=5.5in]{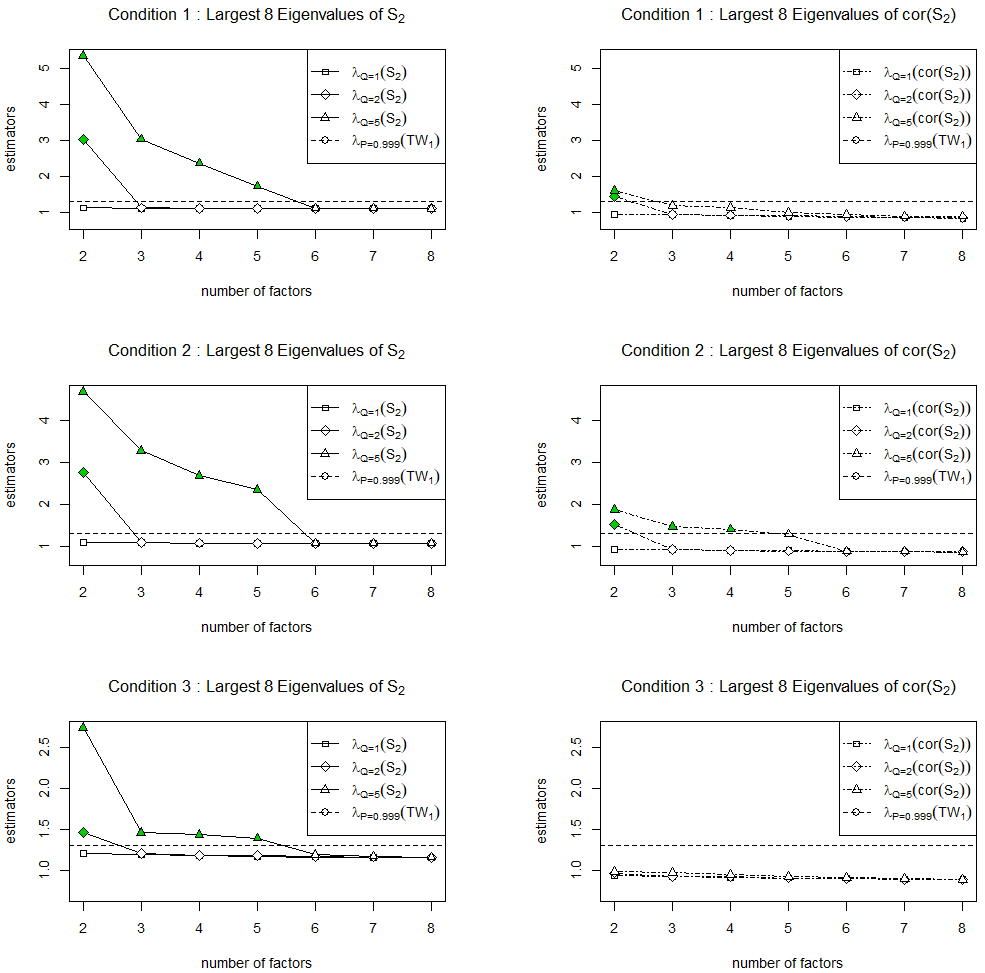}
\caption[Largest Eigenvalues for Alternate Specifications of Dimensionality in Univariate Conditions]{In these plots, the largest eight eigenvalues, not including the first, are shown for alternate specifications of the fit to the first replication from the univariate simulation conditions. The square, triangle, and diamond points are the results from the specification of one, two, and five dimensions. The point is solid when it is greater than the TW prediction at a p-value of .001\%. The covariance of $\bf{S}_2$ is also transformed to a correlation and subjected to an eigenanalysis; those points are shown connected with a dotted line.}\label{fig:S123EV}
\end{figure}

In Figure~\ref{fig:S123EV}, the top seven eigenvalues for three separate fits of the first replication of response data are shown; the second plot is a zoom of the second through seventh eigenvalues as it is known this is a one-dimensional fit, and the focus is on the effect of these alternate configurations on the second eigenvalue. The squares, triangles, and diamonds are fits using a configuration of one, two, and five dimensions. The horizontal dotted line is the overlay of the maximum eigenvalue expected from the Tracy-Widom distribution when $\frac{J}{N} = .02$ with a p-value of .1\% ($\lambda_{P=.999}(TW_1)$); the points are solid when greater than this value. Solid lines connect the eigenvalues of $\bf{S}_2$ and dotted lines connect the values of the eigenvalues of $\bf{S}_2$ converted to a correlation matrix. It is obvious from these plots that the use of the Tracy-Widom test explained in Section~\ref{sec:TW} must be supplemented with other heuristics.

In the first plot, it is clear that the first eigenvalue explains the dominant amount of variation described by the augmented data, the mean of the eigenvector's loadings is also significantly non-zero ($p$-value $< 1 \times 10^{-16}$). Based on \textit{rules of thumb} and the law of parsimony, most would venture to agree that the second eigenvalue is extremely weak in $Q=2$ and $Q=5$ converged fits. For the $Q=1$ configuration, only the first eigenvalue is significant according to the TW test on both the covariance and correlation matrices. For the $Q=2$ configuration, the second eigenvalue survives the TW test, while the mean of the second eigenvector's loadings is not statistically different from zero. In the $Q=5$ configuration, the same properties hold true on the second thru fifth eigenvalues and vectors until the sufficient statistic is transformed to a correlation matrix; then, only the second eigenvalue survives the TW test. 

While it is common to argue away the artifact as noise, it may be the case that the correct choice of factors results in a strong divergence from the TW distribution or higher moments about the transformed values $\mu_{J}$ and $\sigma_{J}$ from Equation~\ref{eq:TWcenterscale}; the amount of statistical information, in this case from the first eigenvalue, pushes down the values of the other 99 components of the 100 $\times$ 100 $\bf{S}_2$ sufficient statistic. Thus, it is worth noting other empirical methods employed to justify statistical decisions in factor analysis for the univariate case. 


In studying the univariate latent factors, it is important to also review a similar analysis for conditions 2 and 3. The converged estimates show nearly identical results for the polytomous configuration of condition 3, and a slightly noisier third, fourth, and fifth eigenvalue in the five-dimensional analysis of the 3PNO of condition 2. In general, there is very little difference from the conclusions about dimensionality derived from the eigenanalysis of the 2PNO. As the simulations move into higher dimensions, it will be informative to see the behavior of the TW test as the number of fit dimensions are decreased to a configuration of lower dimensionality than the generated parameters.

\subsection{Review of Heuristics on Factor Selection}

Two tests, in particular, have been suggested to be used in evaluating whether it is useful to pursue a factor analysis. First, \textcite{bartlett1951} proposed that a correlation matrix is distributed as chi square if it were not different than an identity matrix; dubbed the ``sphericity test'', the null hypothesis of the Bartlett test states the residual correlations are not significantly different from zero. This is a statistical test that is useful for pedagogy, but the tetrachoric or pearson correlation of bernoulli response patterns with items hinging on a particular knowledge domain will show significant residual correlations, and the response data from these simulations result are significant at a $p$-value less than one in the number of subatomic particles in the universe.

The KMO (Kaiser-Meyer-Olkin) test was described in \textcite{dziuban1974correlation} and is used to indicate suitability for factor analytic methods. The formula ranges in value from 0 to 1 with values greater than .5 considered essential for attempting a factor analysis, but preferably greater than .8. This statistic is a proportion of common variance among variables (or in this case, items) to total variance. The calculation of this criterion statistic on the correlations of $\bf{S}_2$ from the univariate simulations above results in values greater than .99 for each replication of each simulation condition.

Having motivated sufficiency for employing factor analysis, several methods have been advanced to motivate the deductive decisioning for retaining the number of factors, eigenvectors, or principal components derived from factor analytic methods. In Chapter~\ref{ch:theory}, the scree plot elbow is covered as a qualitative approach. A more robust attempt at an estimator for this elbow method comes first from \textcite{onatski2010determining}. An estimator $\delta$ is chosen such that the researcher must compare the adjacent eigenvalue difference at suspected threshold to a prediction ($\delta$) from an OLS regression of the next four eigenvalues. In relation to the eigenanalysis above, all of the values shown to be significant in Figure~\ref{fig:S123EV} survive this test as well. 

\begin{figure}
\centering
\includegraphics[width=5.5in]{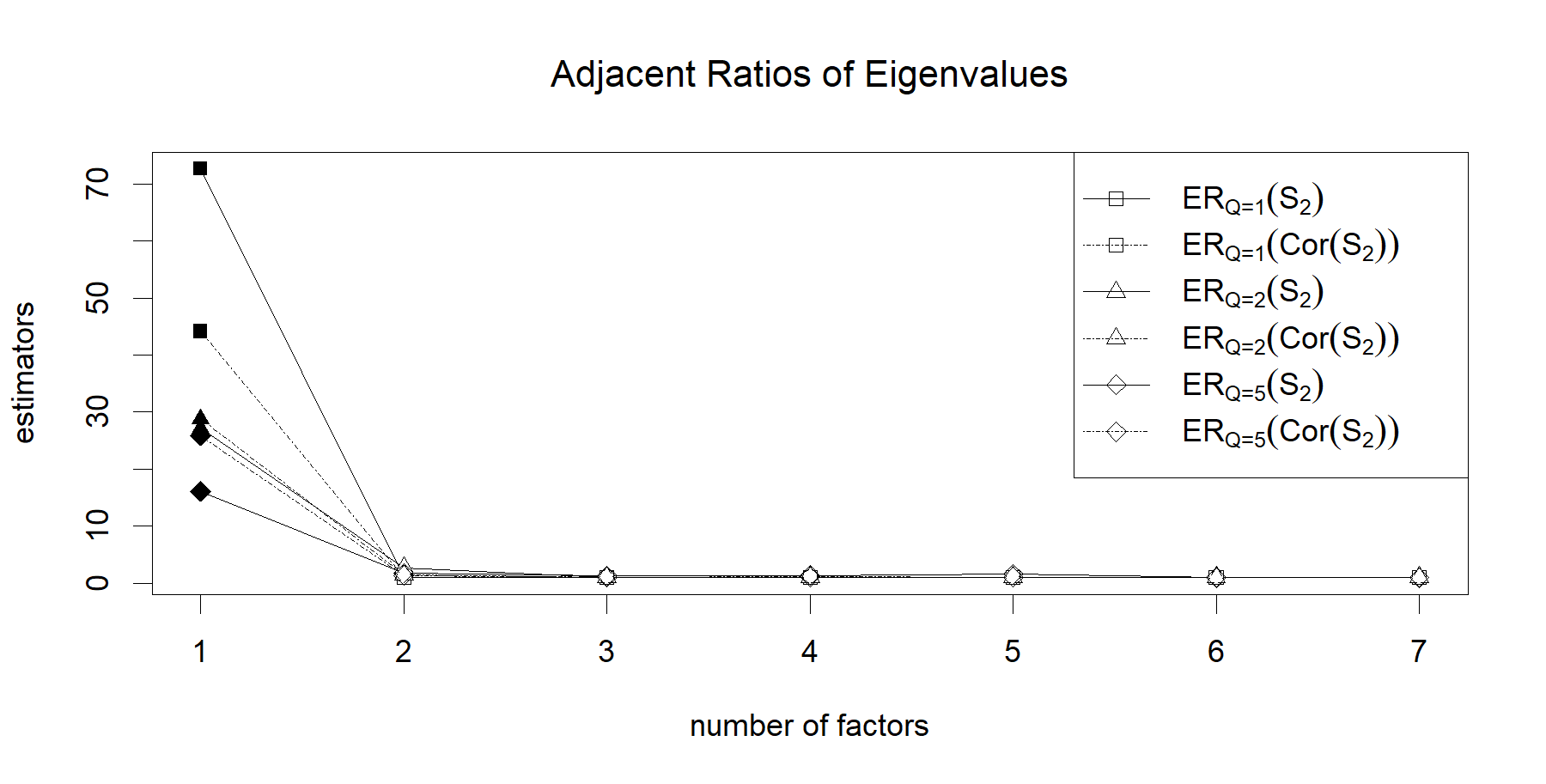}
\caption[Ratios of Largest Eigenvalues for Alternate Specifications of Dimensionality]{In these plots, the ratios of the largest seven eigenvalues to their lower adjacent value are shown for alternate specifications of the fit to the first replication from the first simulation condition. The square, triangle, and diamond points are the results from the specification of one, two, and five dimensions. The point is solid when it is the greatest ratio. The covariance of $\bf{S}_2$ is also transformed to a correlation and subjected to an eigenanalysis; the ratios of the eigenvalues from these transformations are shown connected with a dotted line.}\label{fig:S1ER}
\end{figure}

In \textcite{ahn2013eigenvalue}, the authors supplement the estimators put forward by Onatski by choosing the maximum ratio of the tested eigenvalue and its adjacent lower eigenvalue. This is performed in Figure~\ref{fig:S1ER}, and gives plenty of confidence in the choice of one dimension. 

Two more measures used for assessing eigenvalue thresholds are the empirical Kaiser criterion (\cite{Kaiser1958}; \cite{braeken2017empirical}) and the ``parallel test'' (\cite{horn1965rationale}; \cite{hayton2004factor}). Using the EKC (empirical Kaiser criterion), a reference value is calculated using the limit of the Marchenko-Pastur law, and the eigenvalues from the correlation matrix are compared to these reference values. For the parallel test, several samples of data of the same number of variables ($J$) and degrees are freedom ($N$) are randomly drawn, correlation matrices are derived, and eigenvalues are extracted and rank-ordered for comparison to the real data. It is noteworthy that both methods use random matrix theory to motivate their applicability to our problem. 

In reference to the simulations above, the EKC does not provide a silver bullet. When configured for one dimension, the EKC applied to the correlation of the augmented data matrix yields one dimension. When increasing the dimensions configured for estimation, the EKC increases the number of retained factors. For the estimation of two dimensions, the EKC returns two dimensions; in the five dimension configuration, the EKC dictates that we retain four.

One last qualitative ``rule of thumb'' worth mentioning is the explanation of variance as derived from the eigenvalues. The law of parsimony states the goal to explain the largest amount of variance by the least number of factors. The sum of the eigenvalues can be thought of as 100\% of the variation in multidimensional data. As factor analysis is a form of dimension reduction, the simplest approach is the choice of retaining any factor with an eigenvalue greater than one; this implies the factor explains more than a single variable and is known as the Kaiser-Guttman criterion (\cite{guttman1954some}; \cite{kaiser1960application}). Monte carlo studies (\cite{zwick1986comparison}; \cite{cliff1988eigenvalues}) have demonstrated that this rule can under and overestimate the number of factors, and some have gone as far as to say that it is the most inaccurate approach available \parencite{velicer1990component}.

From a psychometric perspective, the factors are meant to enable interpretation of the statistical information garnered from items used to measure the sample population's ability distribution. In the univariate simulations tested above, the first factor explained 47\% at most and 40\% at the least. When focusing on the second component and the results of the eigenanalyses of the correlation matrix, the largest amount of variation explained is 1.9\% and comes from the 5-dimensional fit of the 3PNO of condition 2. The second component of the eigenalysis of the covariance matrix comes from the 5-dimensional configuration of condition 1 and shows up as 2.7\%. In practice, it is unlikely that the explained variation derived from these configurations would motivate the researcher to attempt more than a second dimension of loadings; under this case, a confirmatory analysis of two dimensions would give a maximum of 1.65\% of the variation explained by the second dimension of the covariance matrix of condition 1. A 2-dimensional configuration of condition 2 and 3 gives 1.51\% and .97\%, respectively.

In Chapter~\ref{ch:em}, it will be seen that one or more of the methods described here have been tried on real response data from assessments such as the Force Concept Inventory.

\section{Simulations of Probit Model Convergence of SAEM in Multiple Dimensions}

The next four conditions are different than the previous conditions in that the loadings are now spread across several dimensions. In conditions 4 and 5, the number of items are decreased to assess convergence. In conditions 6 and 7, the number of items are equivalent to the number in the univariate conditions. In these multidimenstional simulations, two primary test conditions are explored: bifactor and subscale item loading.

For condition 4, there are 30 items and the bifactor loadings are imposed such that the first 10 items only load on the first latent factor, while items 11-30 load on the first factor and another factor. The second and third factors load on items 11-20 and 21-30, respectively, along with the first factor. In condition 5, it is a subscale configuration; the first factor does not load on items 11-30 as it does in condition 4. In the bifactor configurations of conditions 4, 6, and 8, the value of the slope of the second factor from each bifactor item are all sampled from $B_4 (2.5, 3, .1, .9)$ rather than the first factor's distribution of $B_4 (2.5, 3, .2, 1.7)$, thus decreasing the statistical information in the upper dimensions.

In conditions 6 and 7, there are response patterns for 100 items loading on five dimensions in bifactor and subscale confiurations generated for 10,000 examinees. In condition 6, loadings are in a bifactor structure with the first factor loading on all 100 items and the other 4 factors loading on 20 items each. In condition 7, loadings are subscale and each of the five factors are exclusively loaded on 20 items. In conditions 8, the simulation is scaled to 10 dimensions and 100,000 examinees with the same bifactor and subscale structure imposed such that condition 8 has one factor loading on all 100 items and the other nine dimensions loading on 10 items each. Condition 9 is the subscale configuration of condition 8; 10 items load on each of the 10 dimensions.

\subsection{Structural Parameter Estimates for Conditions 4 and 5}

\begin{figure}
\centering
\includegraphics[width=5.5in]{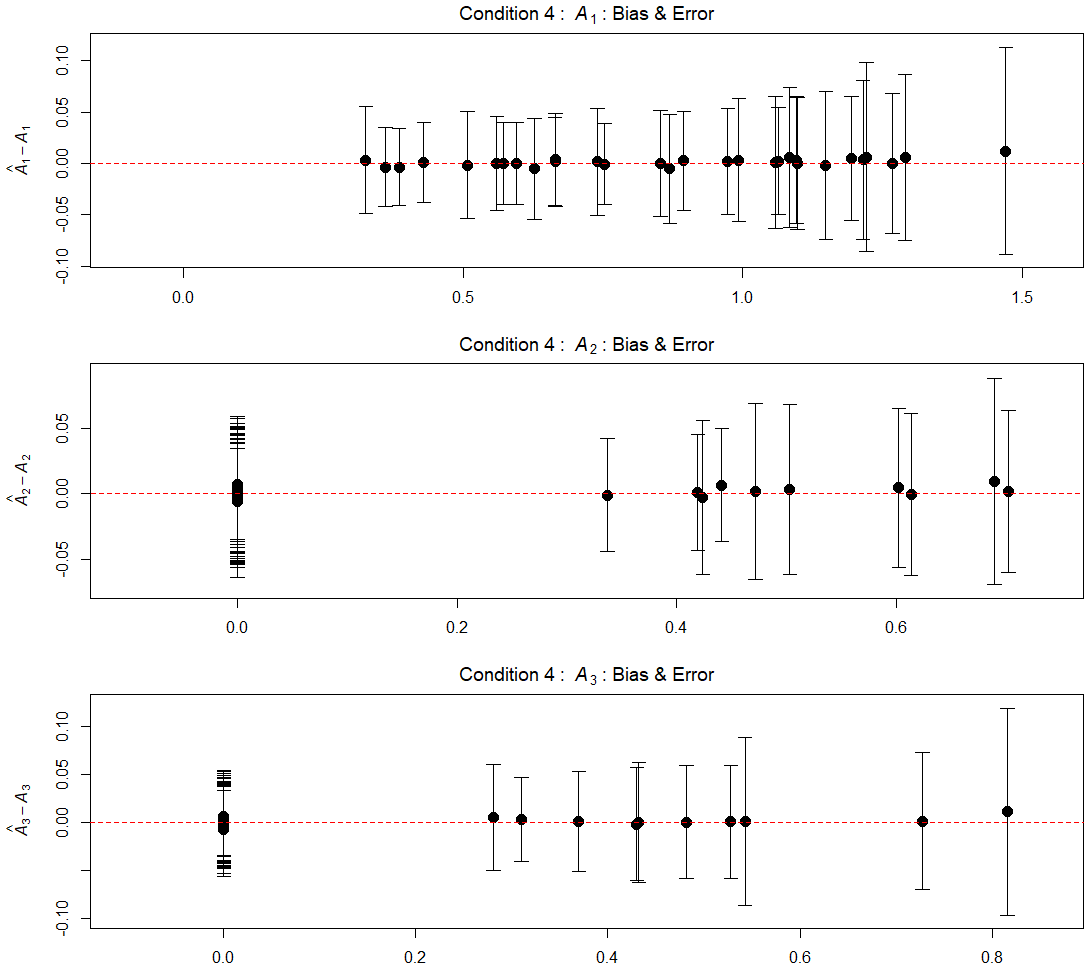}
\caption[Polytomous Multivariate Slope Parameter Estimation in Condition 4]{The results of 50 replications of parameter estimation using the SAEM algorithm coded in \textcite{GeisGit2019} show the three bifactor dimensions after a target rotation. All 50 structural parameter estimates minus the actual values are plotted along with error bars representing a 95\% confidence interval calculated from the RMSE. Note that axes are not on the same scale. Mean estimates for detection of bias are in Table~\ref{tab:S45}.}\label{fig:S4a}
\end{figure}

\begin{figure}
\centering
\includegraphics[width=5.5in]{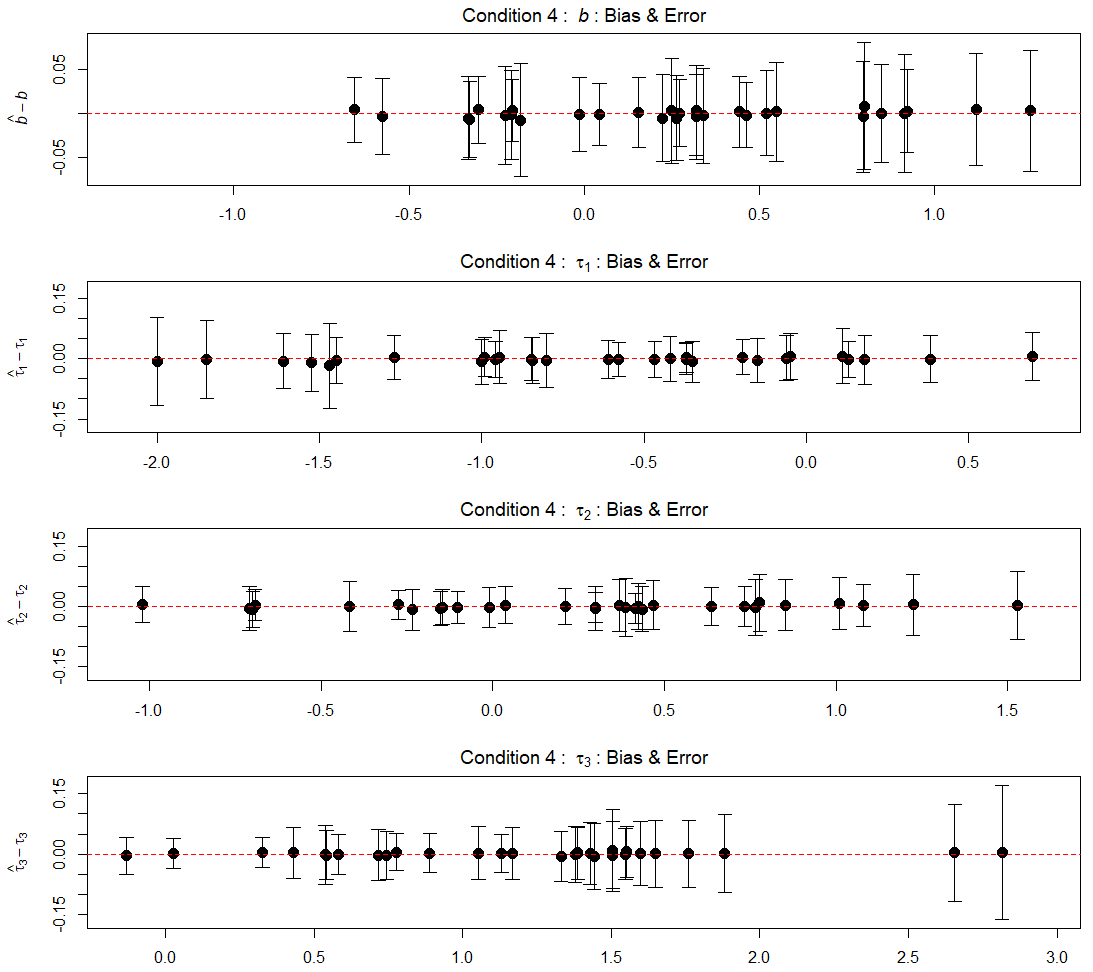}
\caption[Polytomous Intercept Parameter Estimation in Condition 4]{The results of 50 replications of parameter estimation using the SAEM algorithm coded in \textcite{GeisGit2019} show the intercepts and thresholds of the polytomous structure. All 50 structural parameter estimates minus the actual values are plotted along with error bars representing a 95\% confidence interval calculated from the RMSE. Note that axes are not on the same scale. Mean estimates for detection of bias results are in Table~\ref{tab:S45}.}\label{fig:S4b}
\end{figure}

The benchmark plots demonstrating the RMSE and reconstruction accuracy of the 50 replications are shown in Figure~\ref{fig:S4a} and \ref{fig:S4b} along with the mean estimates of bias in Table~\ref{tab:S45} to demonstrate quantitative assessments of these structural parameters. Following the target rotation (see Section~\ref{sec:MVRot}), the reconstruction of the generated loadings is accurate with very little relative bias. Though it is not listed, the first order OLS coefficient from a simple regression of bias on the simulated paramater in the first factor demonstrates statistically significant bias, but this effect is negligibly small with respect to the absolute value of the loadings. 

\begin{table}
\centering
\begin{tabular}{lcrr}  
Condition & Paramater & $\overline{\mbox{bias}}$ & $\mbox{RMSE}$  \\ \hline
4 (Bifactor) & $A_1$ & .0014 & .0036 \\  
4 (Bifactor) & $A_2$ & .0007 & .0033  \\  
4 (Bifactor) & $A_3$ & .0010 & .0037  \\  
4 (Bifactor) & $b$ & -.0004 & .0038  \\  
4 (Bifactor) & $\tau_1$ & -.0018 & .0052  \\ 
4 (Bifactor) & $\tau_2$ & -.0003 & .0043  \\  
4 (Bifactor) & $\tau_3$ & .0010 & .0035  \\    \hline
5 (Subscale) & $A_1$ & .0005 & .0029 \\ 
5 (Subscale) & $A_2$ & .0011 & .0041  \\  
5 (Subscale) & $A_3$ & .0007 & .0034  \\  
5 (Subscale) & $b$ & -.0002 & .0037  \\  
5 (Subscale) & $\tau_1$ & -.0012 & .0064  \\ 
5 (Subscale) & $\tau_2$ & -.0003 & .0035  \\  
5 (Subscale) & $\tau_3$ & .0010 & .0045 \\ \hline  
\end{tabular}
\caption[Diagnostics of Bias and RMSE in Conditions 4 and 5]{Over 50 replications, condition 4 showed statistically significant bias in the first slope, but the deviation from zero is negligibly small in comparison to the values of the parameters. Condition 5 showed statistically significant bias in two parameters, but the deviation from zero is negligibly small in comparison to the values of the parameters.} \label{tab:S45}
\end{table} 

\begin{figure}
\centering
\includegraphics[width=5.5in]{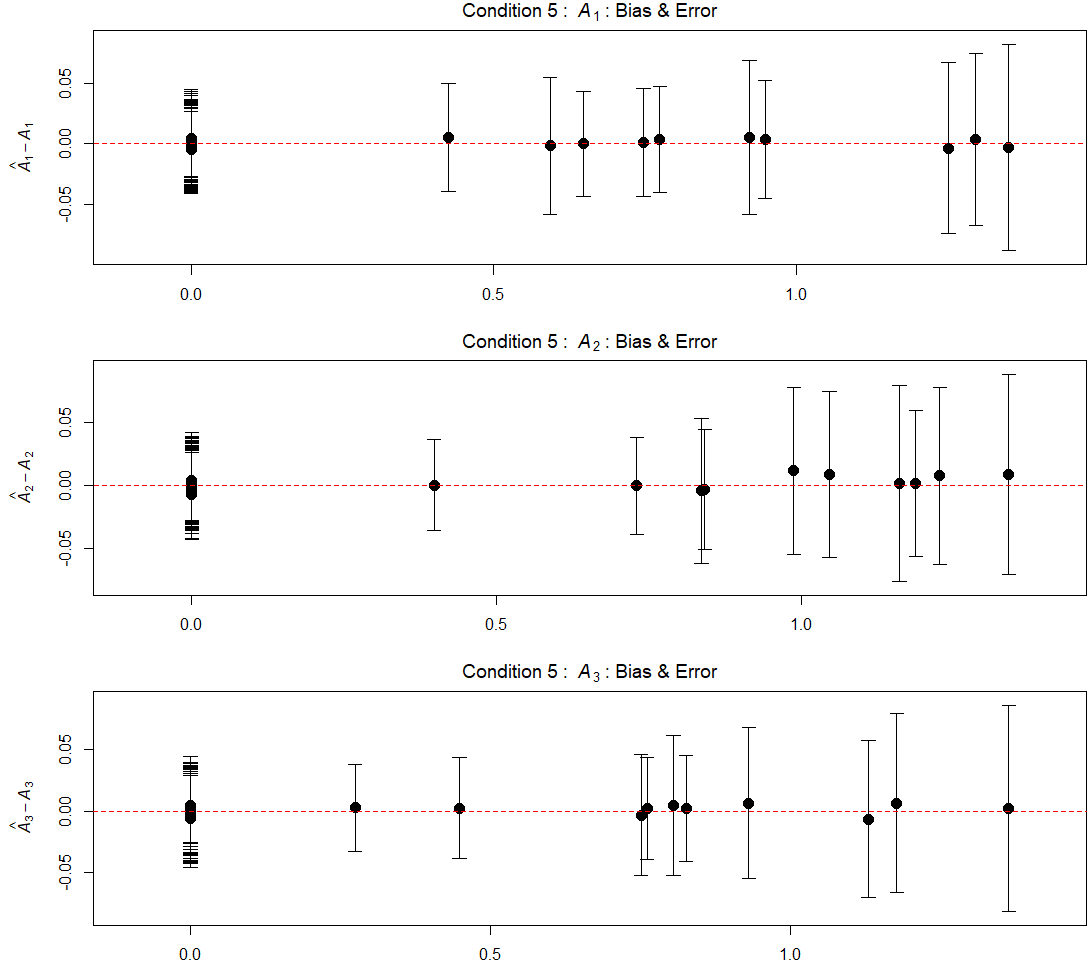}
\caption[Polytomous Multivariate Slope Parameter Estimation in Condition 5]{The results of 50 replications of parameter estimation using the SAEM algorithm coded in \textcite{GeisGit2019} show the three subscale dimensions after a target rotation. All 50 structural parameter estimates minus the actual values are plotted along with error bars representing a 95\% confidence interval calculated from the RMSE. Note that axes are not on the same scale. Mean estimates for detection of bias are in Table~\ref{tab:S45}.}\label{fig:S5a}
\end{figure}

\begin{figure}
\centering
\includegraphics[width=5.5in]{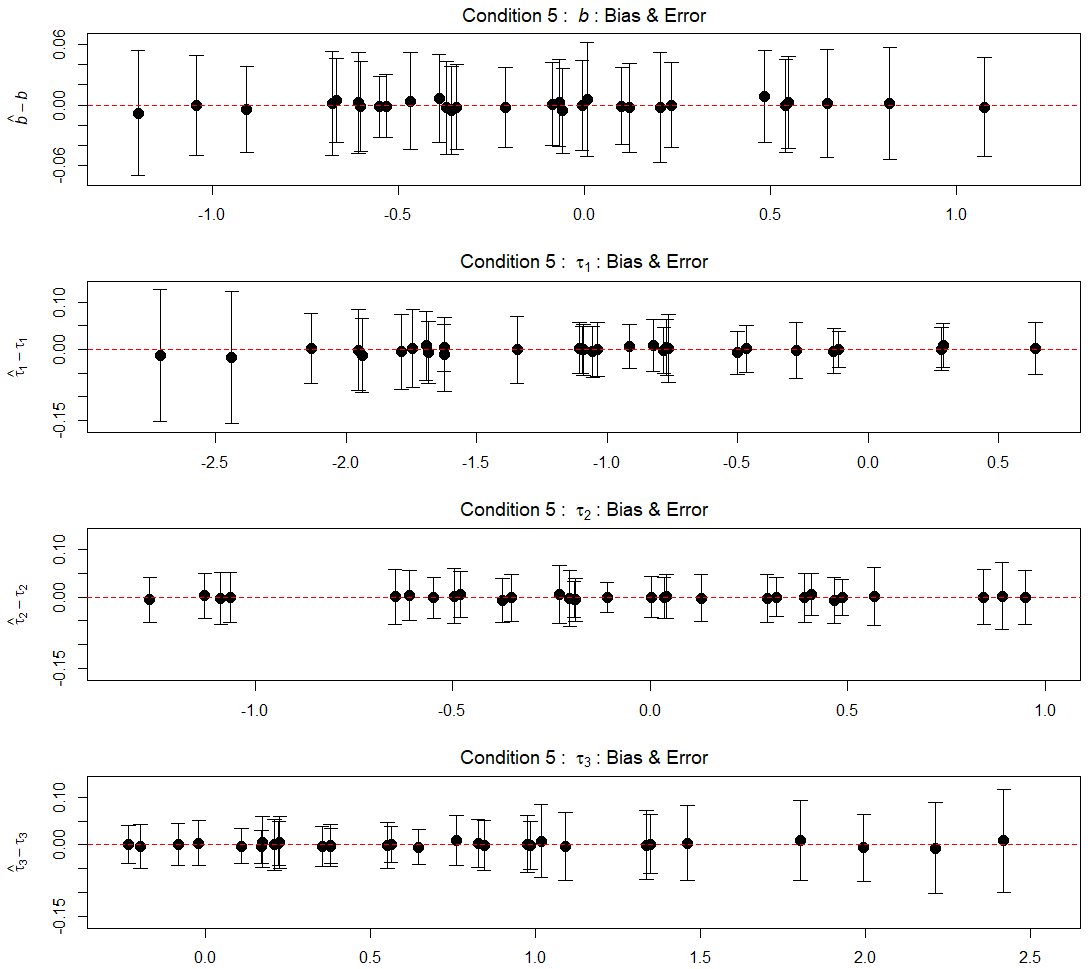}
\caption[Polytomous Intercept Parameter Estimation in Condition 5]{The results of 50 replications of parameter estimation using the SAEM algorithm coded in \textcite{GeisGit2019} show the intercepts and thresholds of the polytomous structure. All 50 structural parameter estimates minus the actual values are plotted along with error bars representing a 95\% confidence interval calculated from the RMSE. Note that axes are not on the same scale. Mean estimates for detection of bias results are in Table~\ref{tab:S45}.}\label{fig:S5b}
\end{figure}

For condition 5, the benchmark plots demonstrating the RMSE and reconstruction accuracy of the 50 replications are shown in Figure~\ref{fig:S5a} and \ref{fig:S5b} are in Table~\ref{tab:S45}. In extracting the final rotation required to target the generated loadings, the factor estimates can also be rotated to extract the examinee abilities.

Error estimates are not straightforward with multidimensional slope parameters; two problems arise in this approach, (1) is the rotational indeterminacy of the factors, and (2) the power iteration algorithm for eigenanalysis. Both issues can be resolved by forcing a target rotation in each iteration. This is not very pragmatic during estimation, but after convergence a reinitialization of the MCMC chains at converged parameters is shown to perform well. 

For condition 4, 1000 iterations with target rotations were run for the first replication after convergence. For 25 of the 30 items, at least 5 parameters were drawn from their 95\% confidence interval more than 95\% of their respective samples. Two items showed only three parameters satisfying the 95\% benchmark. Item 1 ($\mathbf{A} = [.36, 0, 0], b = .91, \mathbf{\tau} = [-.37, .30, 2.81]$) has a discrimination that is very small at .36, as well as a very high $\tau_3$ parameter; overall, the four difficulty parameters failed to achieve the 95\% even though the mode of each posterior was within .2 of the generated values. Item 23 ($\mathbf{A} = [.56, .50, 0], b = .27, \mathbf{\tau} = [-.47, -.10, 1.39]$) also reveals a similar difficulty with small discriminations with misses in the non-zero discriminations and the highest and lowest $\tau$ parameters, though the mode of the posterior distributions are no more than .2 from the generated values for all of the parameters.

The same exercise was run on the first replication of condition 5. Overall, 25 items showed five or more of the seven parameters satisfying the 95\% benchmark. Item 14 ($\mathbf{A} = [0, 1.23, 0], b = -.07, \mathbf{\tau} = [-.47, .04, .22]$) failed for four parameters, including all three slopes. Despite this miss, the modes of the posteriors were less than .1 from the generated value. The calculated RMSEs of the first and third slopes of item 14, given all 50 replications, is less than .02 which makes for a small target.

\subsection{Notes on Factor Analysis and Factor Rotation Methods} \label{sec:MVRot}

The reader may find several thorough treatments of factor analysis (\cite{Muthen1978}; \cite{gorsuch1988exploratory}; \cite{floyd1995factor}; \cite{osborne2008best}) and rotational considerations in factor analysis (\cite{fabrigar1999evaluating}; \cite{schmitt2011current}; \cite{osborne2015rotating}) in the literature. For the purpose of the analyses in this research, a few important details should be reviewed. Key to the interpretation of the converged estimates applied within the context of factor analysis are the goals of the research. In this effort, loadings and eigenvectors extracted from $\mathbf{S}_2$ require rotations for different purposes. For these simulation conditions, the goal is to recover the simulated slopes. With real response data, the goals may include inferences about the information within psychological or knowledge domains measured via an assessment. 

\begin{figure}
\centering
\includegraphics[width=5.5in]{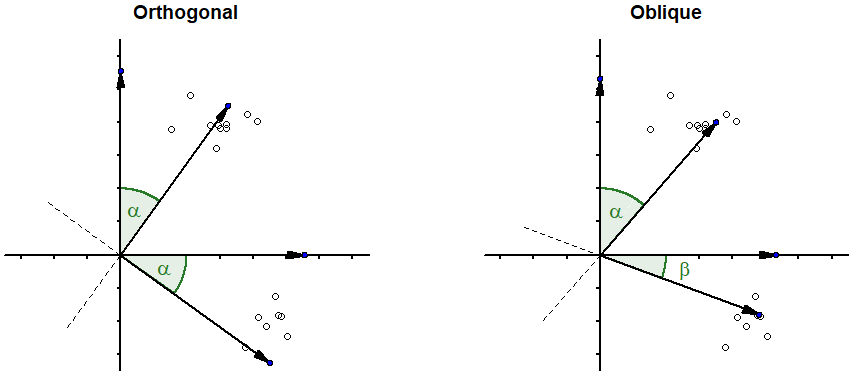}
\caption[Orthogonal and Oblique Rotations]{The figure on the left demonstrates an orthogonal rotation with fictitious loadings in two dimensions, while the figure on the right shows an oblique rotation.}\label{fig:Rotations}
\end{figure}

With respect to the simulation studies in this chapter, the loadings were sampled independently for each dimension. This implies \textit{orthogonality} of the dimensions, and the priors and variance calculations used within the gibbs cycles conditional on the examinee population are consistent with this orthogonal treatment. In the \texttt{GPArotation} package \parencite{bernaards2015package}, orthogonal and oblique rotation functions are available for evaluation of a targeted rotation. Orthogonal rotations, sometimes called \textit{rigid rotations}, are intuitive axis-angle rotations in a euclidean space where the vectors describing the orientation of each factor have a dot product of zero; statistically, the covariance of input factor loadings do not change under orthogonal rotations. Within an oblique rotation, the axes of the reference frame are not constrained to remain perpendicular (see Figure~\ref{fig:Rotations}.

Target rotations require a designation of the elements that should be constrained to zero, all other unspecified parameters are expected to be non-zero. To reconstruct the orientation of the vectors of factor loadings extracted from the simulations in this chapter, the \texttt{GPArotation::targetQ} function \parencite{bernaards2015package} is employed where the target matrix is defined to be unknown in each instance that an item loading is generated. This is the oblique form of the target rotation and thus allows for tests of systematic shifts in the non-zero loadings; an orthogonal target rotation is rigid and may conceal slight shifts in the factor space.

In condition 4, there are 50 loadings and thus the target matrix is defined such that the second dimension has items \{1-10,21-30\} defined to be zero, with the third dimension specifying items 1-20 also set to zero; all other loadings are initialized as \texttt{NA}. The \texttt{targetT} function will not favor positive or negative loadings, thus it is also possible that the loadings have a reversed sign. This rotational indeterminacy is a common issue in factor analytic approaches to multidimensional data. 

As the number of dimensions increases, the difficulty of recovering the original slopes increases and necessitates a novel approach for allowing multiple initializations of the loading matrix that is derived from the eigenanalysis of $\mathbf{S}_2$. For the purposes of this dissertation, the nuances of high dimensional rotations are beyond the scope of the current analysis. To control for the difficulties of high dimensional rotations, the code in \textcite{GeisGit2019} includes a routine that samples 8 random angles and 8 random pairs of dimensions to impose 8 random two-dimensional rotations of the input loading matrix. If the researcher passes the simulated loadings object into the function call of estimation, the rotation process will retain the output with the smallest difference from the target loading matrix.  

In the following chapter, several functions of the \texttt{GPArotation} package can be utilized in exploratory approaches to real response data. Three types of rotations will be employed on real response data, varimax, infomax, promax, and oblimin rotations. The varimax rotation optimizes for the smallest number of elements with the largest loadings; it is named appropriately as it maximizes the sum of the variances of the rotated squared loadings \parencite{Kaiser1958}. The infomax rotation is named as it maximizes entropy; statistically this is similar to a varimax but it meant to maximize the statistical information of the data \parencite{bell1995information}. The promax rotation is the oblique version of the varimax in that it allows for correlations in factors if it increases the sum of the variances of the rotated squared loadings. The oblimin rotation is an oblique rotation that allows for factors to correlate in order to allow a ``simpler'' structure, i.e. larger loadings on fewer factors for each item \parencite{katz1975primary}.

\subsection{Factor Estimates of Conditions 4 and 5}

Several methods are available for factor estimation once the structural parameters of the items have converged. In the case of real response data treated with MIRT and factor analysis methods, confirmatory analysis allows for a target rotation of the extracted loadings as discussed in \ref{sec:MVRot}. In the exploratory case, the researcher would first invoke a choice of the number of dimensions, followed by a rotation that enables the contextual inferences implied by the items of the exam; the researcher may employ varimax, infomax, oblimin, or any other variation of roatation that enables the interpretation of the item structure. Examinee abilities would be estimated after the rotation of item loadings is decided.

Using an MLE or EAP of the final structural parameters requires a multidimensional grid of likelihood calculations conditional on the response pattern of the examinee. This type of search, while more robust in the univariate cases as shown in previous sections, is computationally expensive, and becomes exponentially more complex as the number of dimensions increases. For this reason, only the 100 bayesian samples will be employed for the multidimensional cases simulated in conditions 4 thru 7. Were the MLE or EAP calculated, the researcher has the choice to do the grid search before or after the loading matrix was rotated to a chosen interpretable orientation; running the ability estimation prior to rotation would involve a more complex calculation of errors as those, too, would require proper treatment in the rotation. 

\begin{figure}
\centering
\includegraphics[width=5.5in]{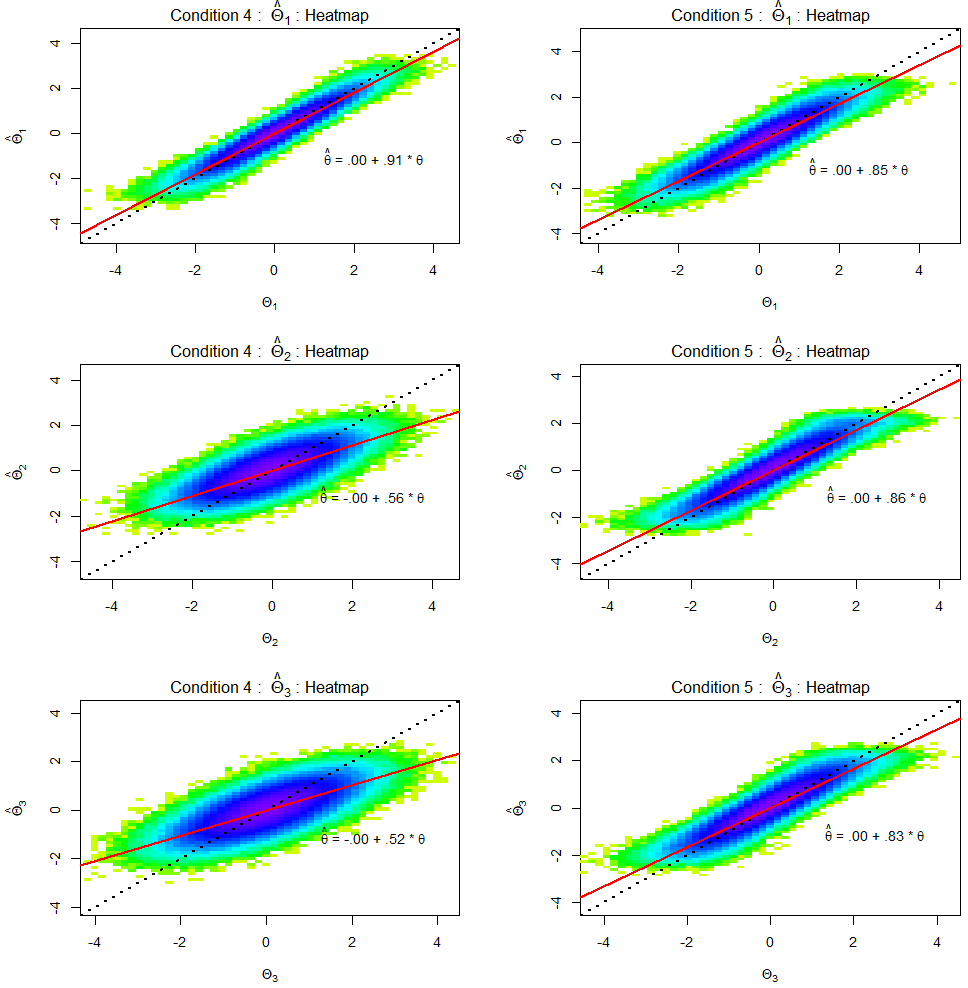}
\caption[Estimates of Examinee Abilities in Conditions 4 and 5]{Heatmaps and residuals of the 50 replications of 5000 examinees' abilities is shown for all three dimensions of the multivariate bifactor polytomous configuration. Estimates of ability come from samples taken using a single post-convergence gibbs cycle starting from the converged estimates of the parameters; this is done 100 times and a mean ability for each examinee is calculated. The solid line represents a simple regression of the predicted to the simulated value. The dotted line shows a slope of 1.}\label{fig:ThetaS45}
\end{figure}

For the multidimensional cases shown here, only the bayesian factor estimates are shown. In the plots shown in Figure~\ref{fig:ThetaS45}, 100 multivariate samples of latent factors are taken at converged structural parameters and averaged, then compared to the generated values. Note that the uncertainties of the estimates are substantially wider than the univariate cases as there are only 30 items. For condition 4, the loadings on the first factor occur in all 30 items, with 10 items loading on the second and third factors at a substantially smaller value of the slope as the higher dimensions are all sampled from $B_4 (2.5, 3, .1, .9)$ rather than the first factor's distributions of $B_4 (2.5, 3, .2, 1.7)$. In the subscale configuration of condition 5, all three dimensions are sampled from the wider Beta distribution.


It is striking how the bayesian method of sampling the latent factors is influenced by the central tendency of the prior, considerably overestimating low abilities and underestimating high abilities in all conditions. As the statistical information available increases, for example the 30 items loading on the first factor in condition 4, the bias in the ability estimates is decreased; the first dimension of ability in condition 4 is more accurately approximated than any other latent factor from either condition 4 or 5. It is also notable how the increased range of the factor loadings in condition 5 significantly increases the statistical information available for factor reconstruction; the discriminations come from a distribution allowing much higher values, further improving the estimation of the examinee abilities even though there are only 10 items loaded on each dimension in the subscale configuration.

\subsection{Tests of Multidimensionality for Multivariate Simulations} \label{sec:MVDims_S4S5}

\begin{figure}
\centering
\includegraphics[width=5.5in]{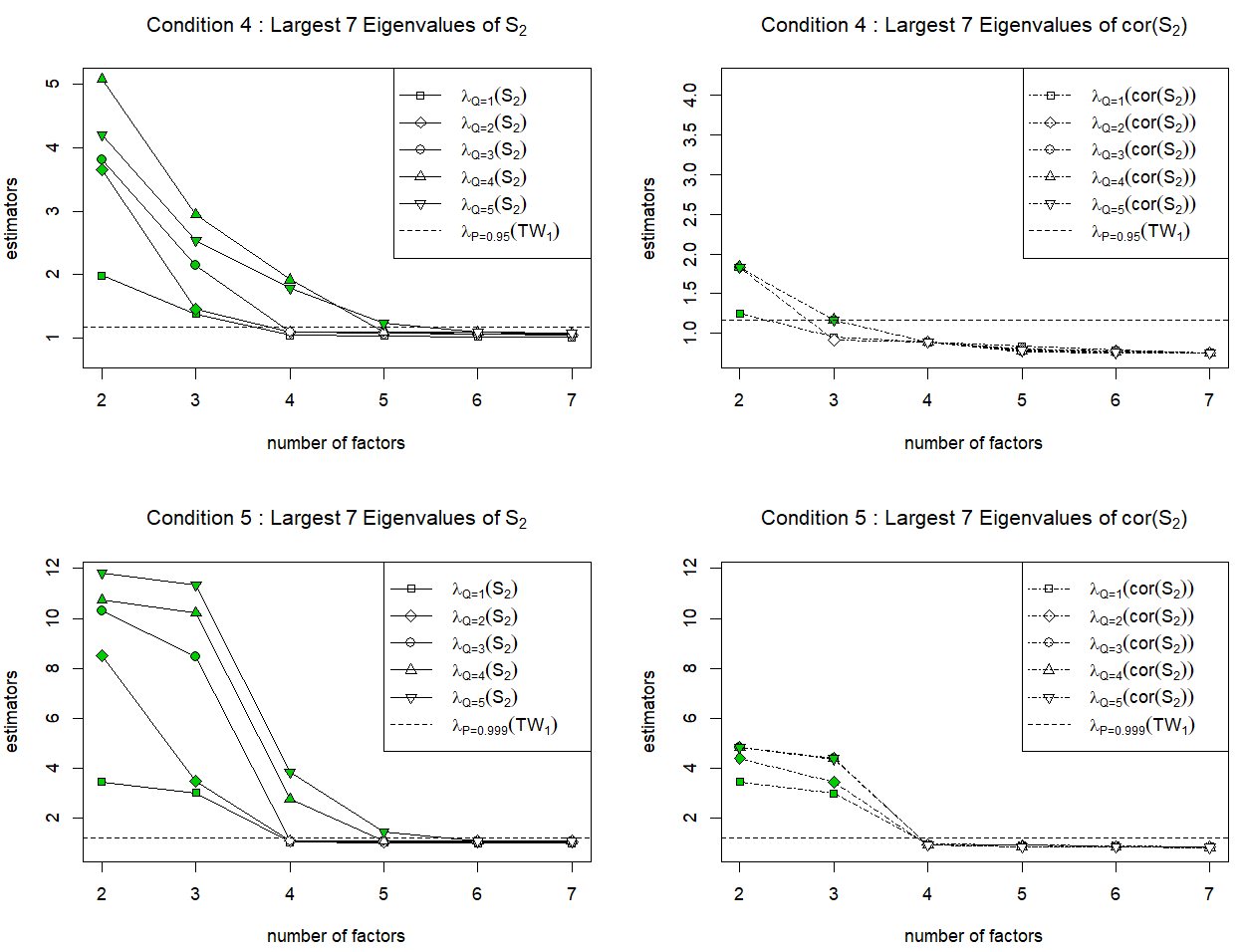}
\caption[Dimensional Sensitivity Analysis of Conditions 4 and 5]{The eigenvalues arising from the augmented data's covariance matrix at convergence is plotted for five configurations of dimensions imposed on the SAEM algorithm. The true number of dimensions of the generated item loadings is three. In the upper left plot, the natural log of the eigenvalues is plotted to enable visualization. The eigenvalues of $\mathbf{S}_2$ are shown with a solid line, while those of the correlation matrix of $\mathbf{S}_2$ are shown with a dotted line. When the point is significant at a \textit{p}-value of .001 (the horizontal dotted line), the point is shaded. }\label{fig:EVTestsS4S5}
\end{figure}

In testing the Tracy-Widom distribution's application for the multivariate simulations, the same analyses from Section~\ref{sec:2pnoDims} are performed on fits performed using one, two, three, four, and five dimensions; the true value of three dimensions is altered to assess the performance of the eigenvalue tests. No rotations are performed, just a simple eigenanalysis of the covariance matrix $\mathbf{S}_2$. The results of the analysis of the first replication of simulated data are shown in Figure~\ref{fig:EVTestsS4S5}.

At first look, condition 5 (subscale) shows a clear spike in the eigenvalue ratio heuristic seen in the lower right plot; for condition 5, the same sampling mechanism was applied to all dimensions, each loading independently on 10 items of this 30 item assessment. It may be the case that the statistical information derived from these three independent and orthogonal dimensions of item loadings allow for a strong signal using this ratio. Notable evidence for the applicability of the Tracy-Widom test also is evident in condition 5; the significance of the first three eigenvalues for the one, two, and three dimensional configurations of the estimation algorithm is visible and provides a strong, consistent signal in the correlation matrix regardless of the dimensionality imposed on the SAEM iterations. It is worth noting the augmented data matrix retains a signal of the second and third dimensions when the algorithm is constrained to retain only the first eigenvalue in each gibbs cycle, and further, the correlation matrix only shows three significant eigenvalues when five dimensions are retained at each gibbs iteration. 

A tighter analysis is required of the bifactor configuration (condition 4) and this should be expected as all 30 of the items most strongly loaded on the first dimension with the second and third dimensions being applied with a lower discrimination on a third of the item set; a lower discrimination directly correlates to less statistical information. In the upper left plot, the logarithm of the eigenvalues is employed to enable better granularity in the visibility on the graph. Similar results to condition 5 are noticeable but with some notable differences. First, the eigenanalysis of the covariances of the one, two, and three dimensional cases show three significant eigenvalues; the only ambiguity to this rule is the insignificance of the third eigenvalue when running the eigenanalysis of the correlation matrix in the one and two dimensional configurations. Again, in both the four and five dimensional configurations, the fourth and fifth eigenvalues of the correlation matrix are insignificant although the eigenvalues of the covariance are significant. It is also clear that the ratio-tests are not useful for these estimations.

\subsection{Structural Parameter Estimates for Conditions 6 and 7}

For the next two conditions, the bifactor and subscale configurations are imposed on 10000 examinees and 100 items using five dimensions to demonstrate the decrease in the RMSE of all structural parameters as the statistical information available is at the minimum, doubled. Only the first dimension in the bifactor structure loads on all 100 items, and the number of items assigned to every other latent factor in both the bifactor and subscale configurations is 20. Again, in the bifactor configuration the second loading is sampled from a  As opposed to the previous conditions, only five replications are performed. The tests performed on conditions 4 and 5 are repeated.

\begin{figure}
\centering
\includegraphics[width=5.5in]{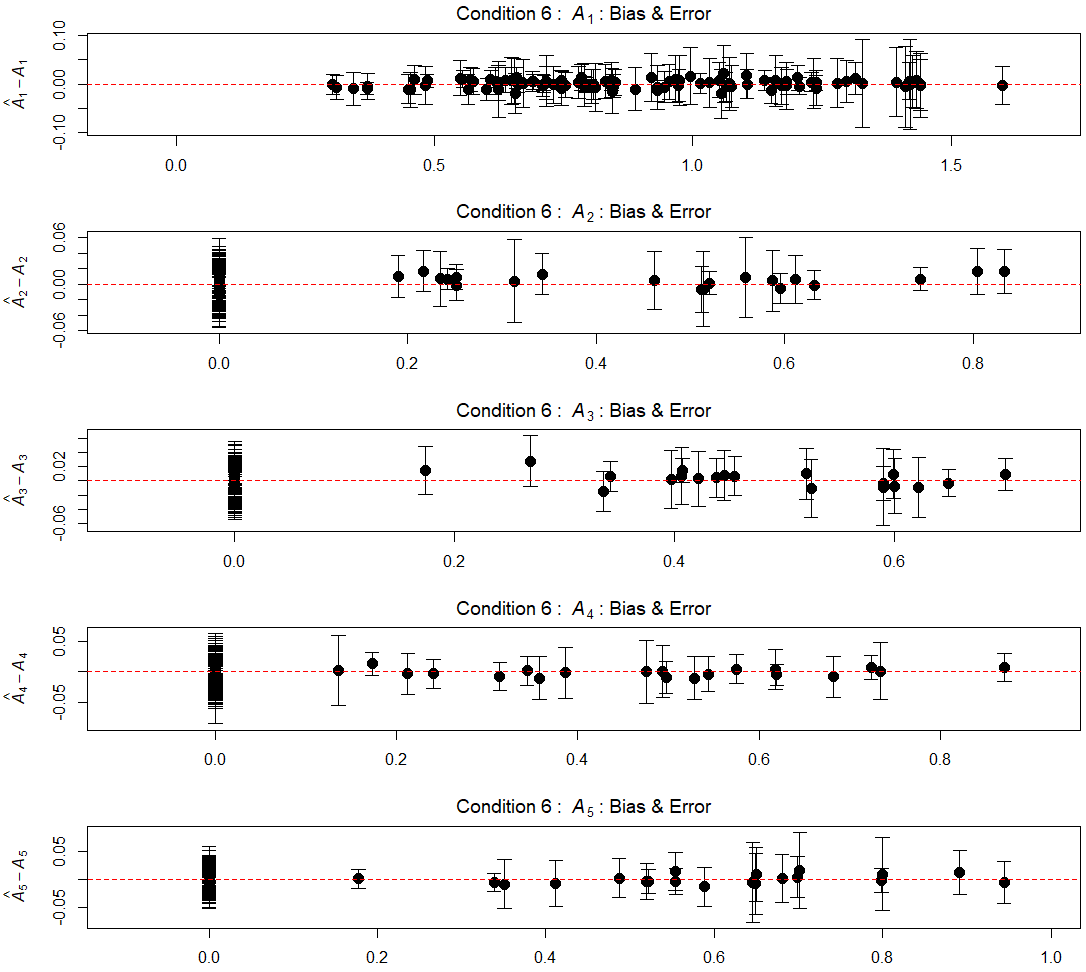}
\caption[Polytomous Multivariate Slope Parameter Estimation in Condition 6]{The results of 5 replications of parameter estimation using the SAEM algorithm coded in \textcite{GeisGit2019} show the five bifactor dimensions after a target rotation. Each of the 5 replications of structural parameter estimates minus the actual values are plotted along with error bars representing a 95\% confidence interval calculated from the RMSE. Note that axes are not on the same scale. Mean estimates for detection of bias are in Table~\ref{tab:S67}.}\label{fig:S6a}
\end{figure}

\begin{figure}
\centering
\includegraphics[width=5.5in]{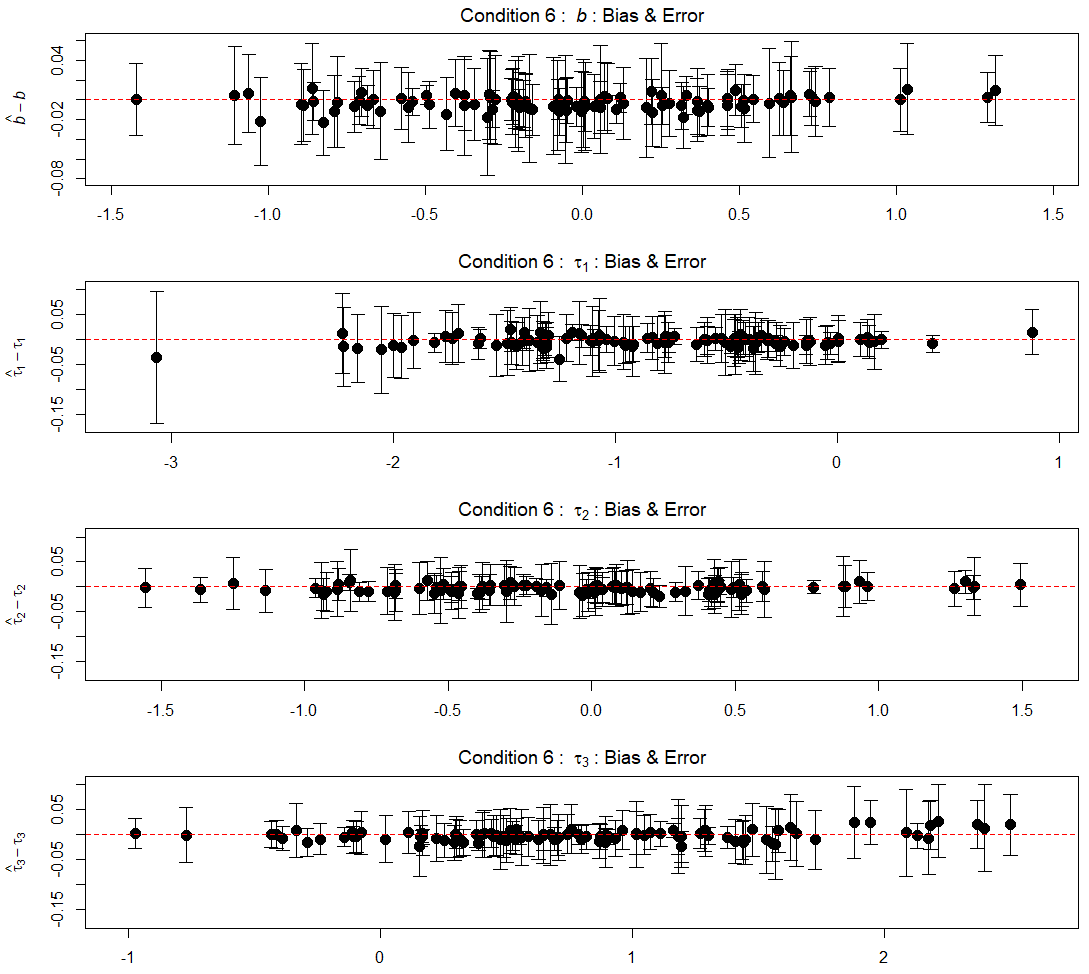}
\caption[Polytomous Intercept Parameter Estimation in Condition 6]{The results of 5 replications of parameter estimation using the SAEM algorithm coded in \textcite{GeisGit2019} show the intercepts and thresholds of the polytomous structure. Each of the 5 replications of structural parameter estimates minus the actual values are plotted along with error bars representing a 95\% confidence interval calculated from the RMSE. Note that axes are not on the same scale. Mean estimates for detection of bias results are in Table~\ref{tab:S67}.}\label{fig:S6b}
\end{figure}

The structural parameters' residuals are displayed along with their RMSEs over 5 replications for condition 6 in Figures~\ref{fig:S6a} and \ref{fig:S6b}. With the 100 items being sampled and the five dimensions of bifactor loadings, the slopes, intercepts and thresholds are quite stable. All estimates of bias at an item level indicates a negligible deviation from the generated parameters. In comparison to the plots in Figure~\ref{fig:S4a}, the RMSEs of condition 6 are about 35\% smaller in the first dimension and 45\% smaller in the bifactor dimensions. The uncertainty in the intercepts and thresholds have dropped by one-third.

\begin{table}
\centering
\begin{tabular}{lcrr}  
Condition & Paramater & $\overline{\mbox{bias}}$ & $\mbox{RMSE}$ \\ \hline
6 (Bifactor) & $A_1$ & .0008 & .0080 \\  
6 (Bifactor) & $A_2$ & .0010 & .0065 \\ 
6 (Bifactor) & $A_3$ & .0008 & .0072  \\ 
6 (Bifactor) & $A_4$ & -.0002 & .0065 \\ 
6 (Bifactor) & $A_5$ & -.0001 & .0065  \\ 
6 (Bifactor) & $b$ & -.0031 & .0068  \\ 
6 (Bifactor) & $\tau_1$ & -.0041 & .0096  \\ 
6 (Bifactor) & $\tau_2$ & -.0033 & .0070  \\  
6 (Bifactor) & $\tau_3$ & -.0019 & .0100  \\ \hline
7 (Subscale) & $A_1$ & -.0001 & .0054  \\  
7 (Subscale) & $A_2$ & -.0002 & .0051   \\
7 (Subscale) & $A_3$ & -.0006 & .0065  \\ 
7 (Subscale) & $A_4$ & .0015 & .0064   \\ 
7 (Subscale) & $A_5$ & -.0006 & .0066   \\  
7 (Subscale) & $b$ & .0041 & .0080   \\  
7 (Subscale) & $\tau_1$ & .0033 & .0094   \\  
7 (Subscale) & $\tau_2$ & .0044 & .0086   \\ 
7 (Subscale) & $\tau_3$ & .0046 & .0109   \\ \hline
\end{tabular}
\caption[Diagnostics of Bias and RMSE in Conditions 6 and 7]{Over 5 replications, condition 6 showed statistically significant bias in one slope and each of the difficulty parameters. Condition 7 showed statistically significant bias in two slope parameters and all diffculty parameters.} \label{tab:S67}
\end{table} 

\begin{figure}
\centering
\includegraphics[width=5.5in]{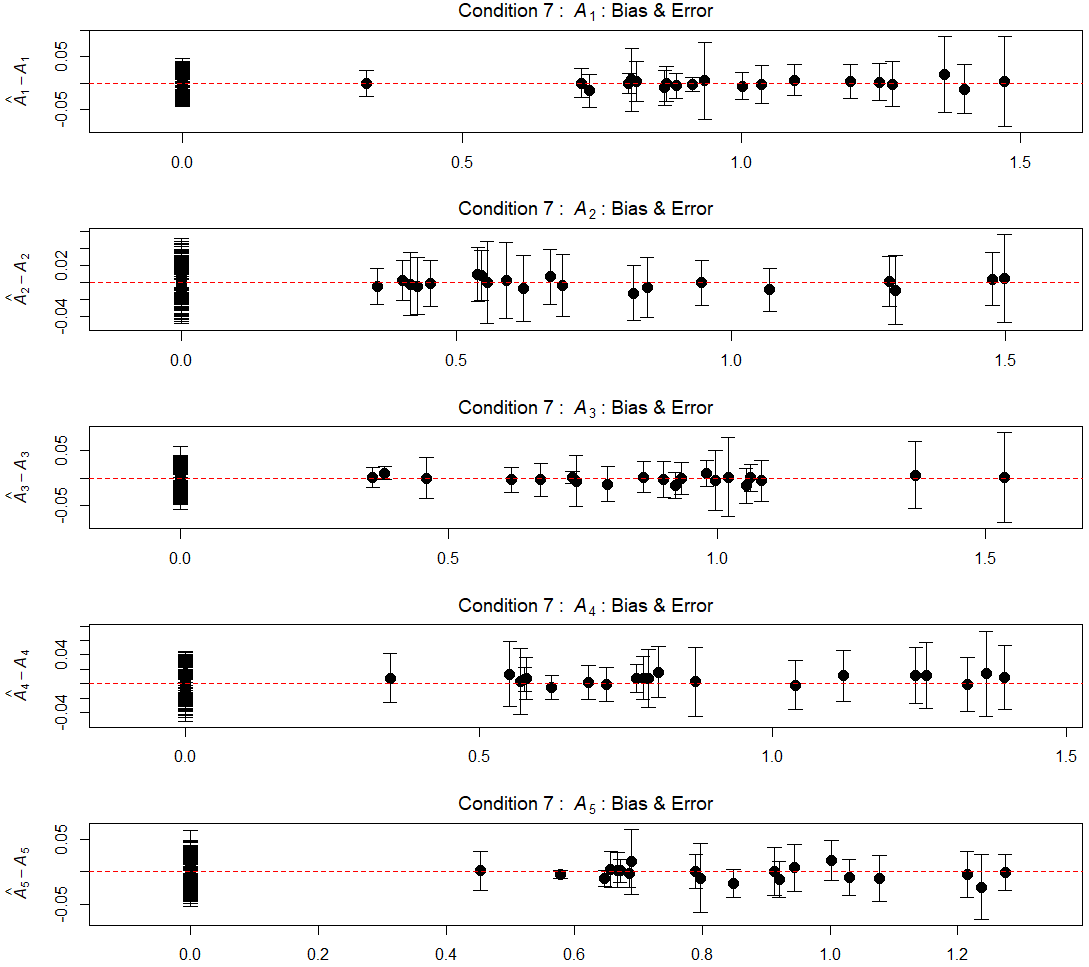}
\caption[Polytomous Multivariate Slope Parameter Estimation in Condition 7]{The results of 5 replications of parameter estimation using the SAEM algorithm coded in \textcite{GeisGit2019} show the five subscale dimensions after a target rotation. Each of the 5 replications of structural parameter estimates minus the actual values are plotted along with error bars representing a 95\% confidence interval calculated from the RMSE. Note that axes are not on the same scale. Mean estimates for detection of bias are in Table~\ref{tab:S67}.}\label{fig:S7a}
\end{figure}

\begin{figure}
\centering
\includegraphics[width=5.5in]{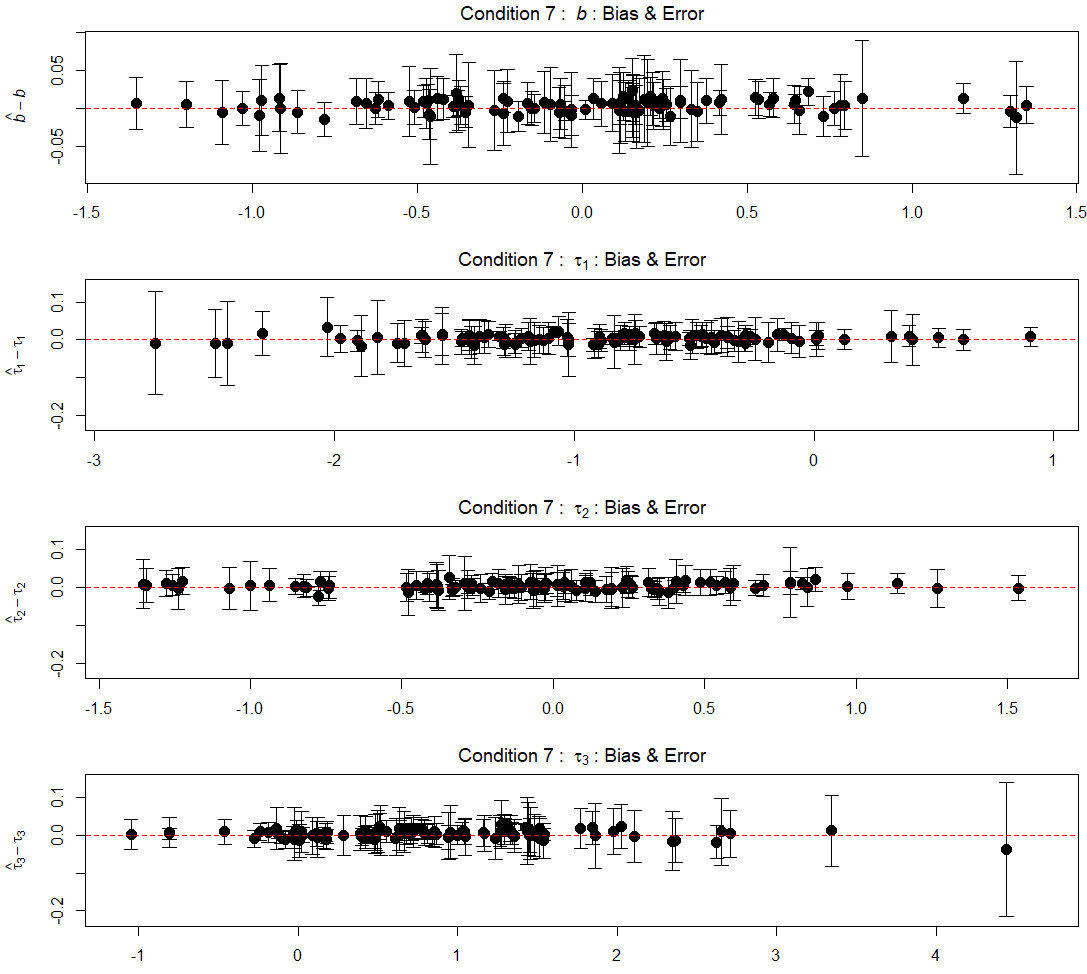}
\caption[Polytomous Intercept Parameter Estimation in Condition 7]{The results of 5 replications of parameter estimation using the SAEM algorithm coded in \textcite{GeisGit2019} show the intercepts and thresholds of the polytomous structure. Each of the 5 replications of structural parameter estimates minus the actual values are plotted along with error bars representing a 95\% confidence interval calculated from the RMSE. Note that axes are not on the same scale. Mean estimates for detection of bias results are in Table~\ref{tab:S67}.}\label{fig:S7b}
\end{figure}

For subscale simulations, Figures~\ref{fig:S7a} and \ref{fig:S7b} show the diagnostics of the residuals and RMSEs of slopes, intercepts, and thresholds. As compared to condition 5, the uncertainties in the slopes are decreased by an average of 38\% with the uncertainties of the difficulties a little more than 25\% smaller. Though it is not in the table of bias and errors, \ref{tab:S67}, the fourth dimension slope of condition 7 had a significant bias when regressed on the simulated parameter along with all of the difficulty parameters. Again, while statistically significant, these errors are a negligible proportion of the parameters being estimated, and the overall mean bias is not significant.

For condition 7, the benchmark plots demonstrating the RMSE and reconstruction accuracy of the 5 replications are shown in Figure~\ref{fig:S7a} and \ref{fig:S7b} with the statistics of the bias of these structural parameters in Table~\ref{tab:S67}. In extracting the final rotation required to target the generated loadings, the factor estimates can also be rotated to extract the examinee abilities. 

\subsection{Factor Estimates of Conditions 6 and 7}

As with the previous five conditions, 100 post-convergence gibbs cycles are continued skipping the maximization steps as the structural parameters have converged with slopes already rotated to the target configuration of positive non-zero slopes where discrimination parameters were generated. Examinee abilities are estimated from the mean of the 100 sampled multivariate latent factors.

As with condition 4, the errors in the first dimension of condition 6 significantly outperforms the reconstruction of the second thru fifth latent dimensions of abilities; all 100 items load on this first discrimination parameter from the larger space of absolute values of the Beta distribution. The bias that was also quite noticeable in Figure~\ref{fig:ThetaS45} is consistent in all of the four bifactor dimensions. Given that the bifactor dimensions load on twice the number of items in this condition, the slope of the bias decreased to approximately -.28 as compared to -.45 in condition 4; thus the underestimation and overestimation at low and high true abilities, respectively, is less severe but is still occuring. Condition 7's factor estimations are quite consistent with the behavior seen in condition 5 (Figure~\ref{fig:ThetaS45}), but the bias at the tails is slightly more than half the under- and overestimations as low and high abilities, respectively. The errors are decreased by a little more than 20\% throughout the range of each latent factor.




\subsection{Multidimensional Inferences from Alternate Configurations of Conditions 6 and 7}

\begin{figure}
\centering
\includegraphics[width=5.5in]{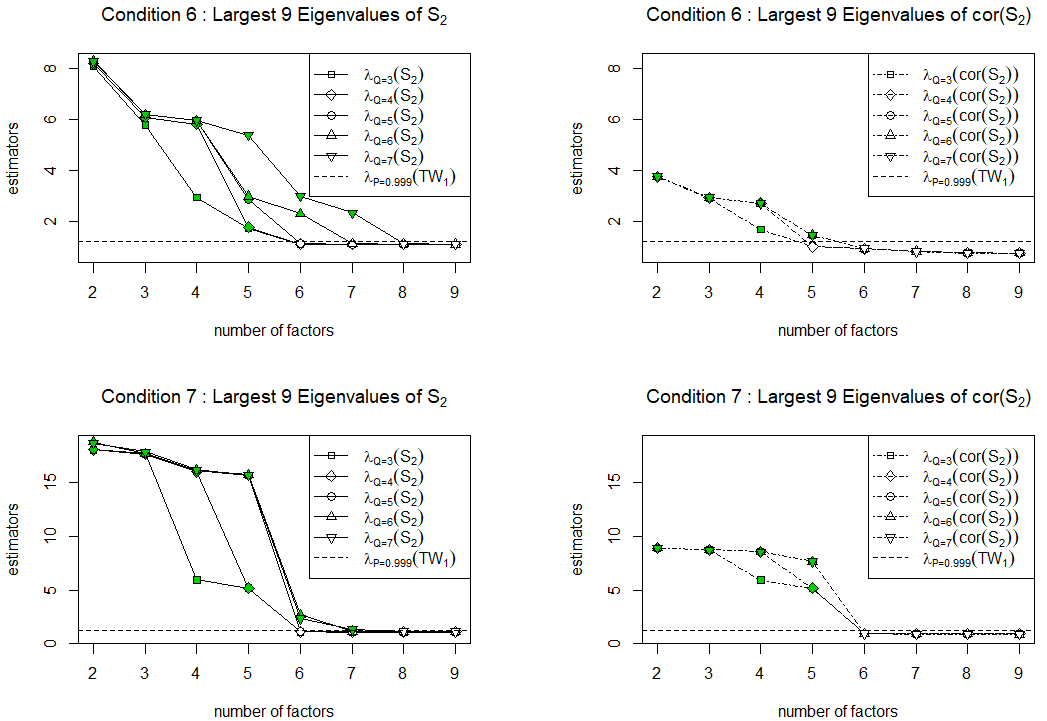}
\caption[Dimensional Sensitivity Analysis of Conditions 6 and 7]{The eigenvalues arising from the augmented data's covariance matrix at convergence is plotted for five configurations of dimensionality imposed on the SAEM algorithm's treatment of these simulation conditions. The true number of dimensions of the generated item loadings is five. In the upper left plot, the natural log of the eigenvalues is plotted to enable visualization of the bifactor case as the loadings of the upper dimensions are sampled from the restricted $B_4 (2.5, 3, .1, .9)$ rather than the first factor's distribution of $B_4 (2.5, 3, .2, 1.7)$. The eigenvalues of $\mathbf{S}_2$ are shown with a solid line, while those of the correlation matrix of $\mathbf{S}_2$ are shown with a dotted line. When the point is significant at a \textit{p}-value of .001 (the horizontal dotted line), the point is shaded. }\label{fig:EVTestsS6S7}
\end{figure}

As with the previous multidimensional conditions, an eigenanalysis of alternate configurations of spectral dimensions is explored. Three through seven dimensions are estimated and plotted in Figure~\ref{fig:EVTestsS6S7}. The Tracy-Widom test of the correlation matrix is proving to be robust against the choice of dimensionality when guessing is not an aspect of the simulated data. There is no aberrant behavior that differs from the results of the bifactor and subscale structures of conditions 4 and 5. This analysis will be explored further in the 10-dimensional conditions in the next section.

\subsection{Estimation in Conditions 8 and 9}

The final simulation conditions explored in this chapter are generated on 100,000 examinees and their simulated responses to 100 items in a bifactor and subscale configuration of 10 dimensions. The same approach as above will be used to explore the residuals and RMSEs of five replications of these conditions.
 
\begin{figure}
\centering
\includegraphics[width=5.5in]{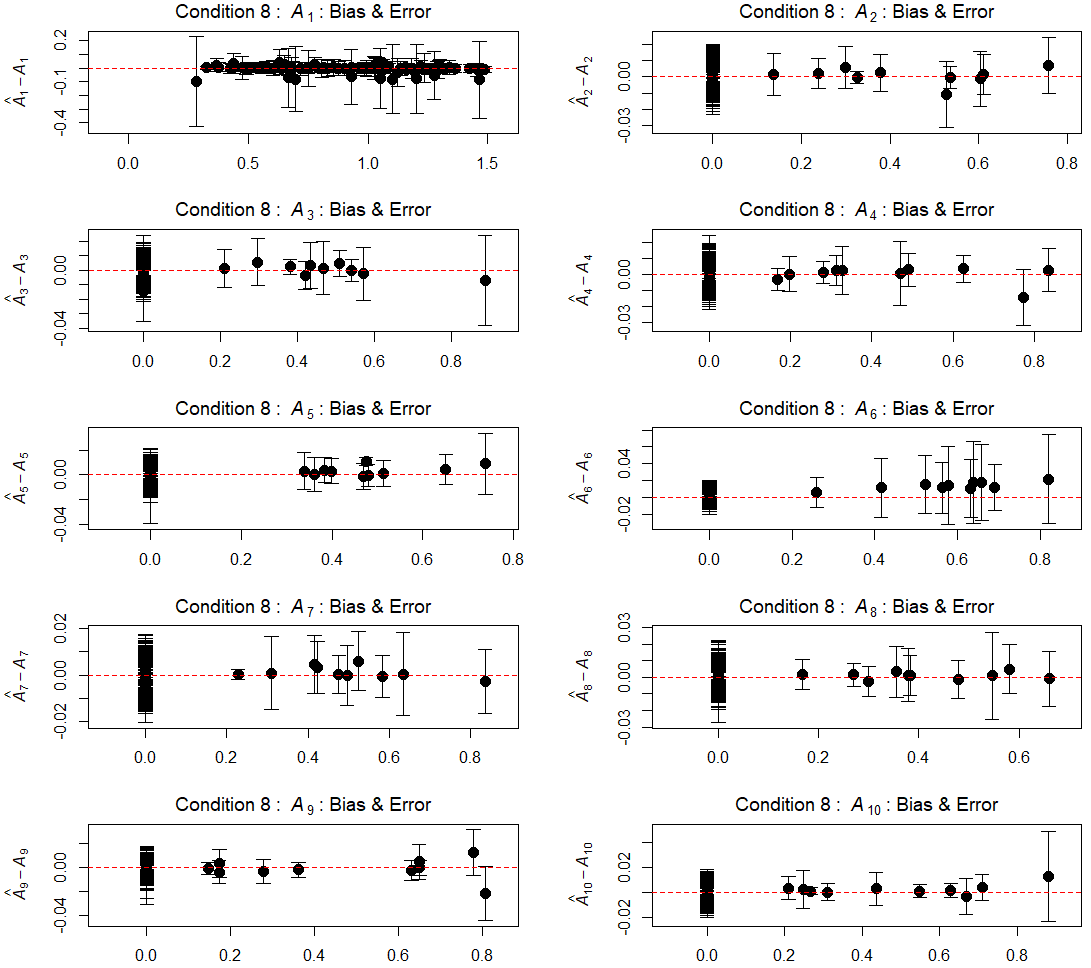}
\caption[Polytomous Multivariate Slope Parameter Estimation in Condition 8]{The results of five replications of parameter estimation using the SAEM algorithm coded in \textcite{GeisGit2019} show the ten bifactor dimensions after a target rotation. Each of the five replications of structural parameters' residuals are plotted along with error bars representing a 95\% confidence interval calculated from the RMSE. Note that axes are not on the same scale. Mean estimates for detection of bias are in Table~\ref{tab:S89}.}\label{fig:S8a}
\end{figure}

Condition 8's slope diagnostics in Figure~\ref{fig:S8a} are instrumental for delineating the importance of the target rotation. In the left uppermost residuals of the first dimension, there are approximately 10 estimates that are significantly underestimated and have large RMSEs. A closer inspection of the matrix of RMSEs reveals that they are highly correlated with the errors of the sixth factor, $A_6$, which are also substantially larger than the uncertainties in the other dimension and demonstrate the most significant deviation from zero bias in the table of diagnostics, \ref{tab:S89}. These results indicate the necessity of paying careful attention to the target rotation chosen at convergence of the SAEM algorithm, especially if a difficulty parameter of an item loading on this factor is at an extreme value.

\begin{figure}
\centering
\includegraphics[width=5.5in]{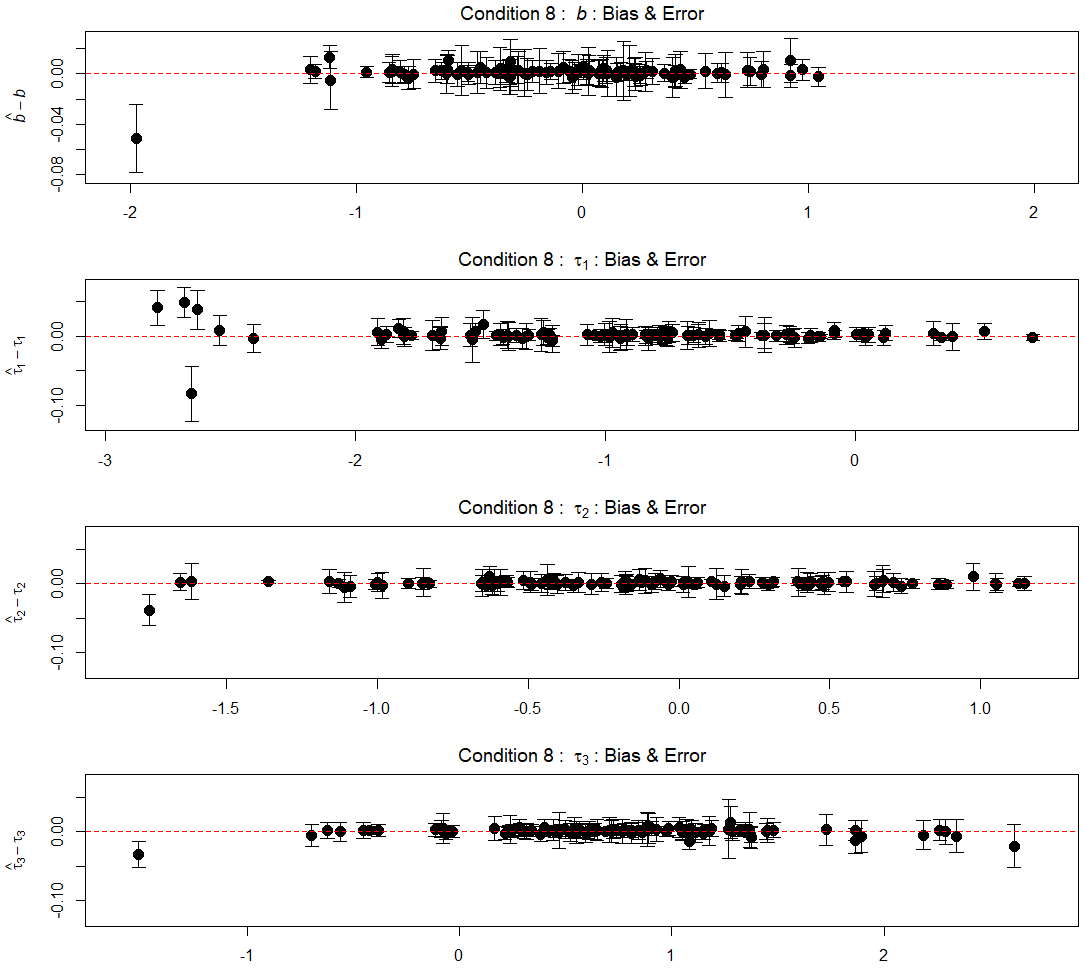}
\caption[Polytomous Intercept Parameter Estimation in Condition 8]{The results of five replications of parameter estimation using the SAEM algorithm coded in \textcite{GeisGit2019} show the intercepts and thresholds of the polytomous structure. Each of the five replications of structural parameter estimates' residuals are shown along with error bars representing a 95\% confidence interval calculated from the RMSE. Note that axes are not on the same scale. The results of mean estimates for detection of bias are in Table~\ref{tab:S89}.}\label{fig:S8b}
\end{figure}

The statistical significance of the bias diagnostics in Table~\ref{tab:S89} indicate that $A_6$ shows a large correlation of the bias to the simulated parameter, while the rest of the structural parameters show negligibly small uncorrelated deviations from zero. The residuals of the difficulty parameters are displayed in Figure~\ref{fig:S8b}; most notable is the extreme difficulties on the low end of $b$ and $\tau_1$ showing poor estimations of that average difficulty and the first threshold of four items.

\begin{table}
\centering
\begin{tabular}{lcrr} 
Condition & Paramater & $\overline{\mbox{bias}}$ & $\mbox{RMSE}$  \\ \hline
8 (Bifactor) & $A_1$ & -.0067 & .0248  \\  
8 (Bifactor) & $A_2$ & .0002 & .0030  \\  
8 (Bifactor) & $A_3$ & -.0001 & .0029 \\  
8 (Bifactor) & $A_4$ & .0000 & .0026  \\  
8 (Bifactor) & $A_5$ & .0003 & .0029  \\  
8 (Bifactor) & $A_6$ & .0015 & .0049 \\  
8 (Bifactor) & $A_7$ & .0002 & .0024  \\  
8 (Bifactor) & $A_8$ & .0001 & .0025  \\  
8 (Bifactor) & $A_9$ & -.0001 & .0035  \\  
8 (Bifactor) & $A_{10}$ & .0002 & .0027  \\
8 (Bifactor) & $b$ & .0010 & .0060  \\
8 (Bifactor) & $\tau_1$ & .0017 & .0117 \\ 
8 (Bifactor) & $\tau_2$ & .0009 & .0048  \\
8 (Bifactor) & $\tau_3$ & .0004 & .0055 \\ \hline  
9 (Subscale) & $A_1$ & -.0002 & .0021  \\ 
9 (Subscale) & $A_2$ & -.0002 & .0023 \\  
9 (Subscale) & $A_3$ & -.0002 & .0025  \\  
9 (Subscale) & $A_4$ & .0008 & .0098 \\ 
9 (Subscale) & $A_5$ & -.0000 & .0020  \\  
9 (Subscale) & $A_6$ & -.0004 & .0025  \\  
9 (Subscale) & $A_7$ & -.0001 & .0021 \\  
9 (Subscale) & $A_8$ & -.0001 & .0020  \\  
9 (Subscale) & $A_9$ & -.0001 & .0027  \\  
9 (Subscale) & $A_{10}$ & .0001 & .0028  \\  
9 (Subscale) & $b$ & .0018 & .0160  \\  
9 (Subscale) & $\tau_1$ & .0045 & .0291  \\  
9 (Subscale) & $\tau_2$ & .0010 & .0128  \\  
9 (Subscale) & $\tau_3$ & -.0002 & .0190  \\ \hline 
\end{tabular}
\caption[Diagnostics of Bias in Condition 8 and 9]{Over five replications, condition 8 showed small but statistically significant bias in two slopes.} \label{tab:S89}
\end{table} 

\begin{figure}
\centering
\includegraphics[width=5.5in]{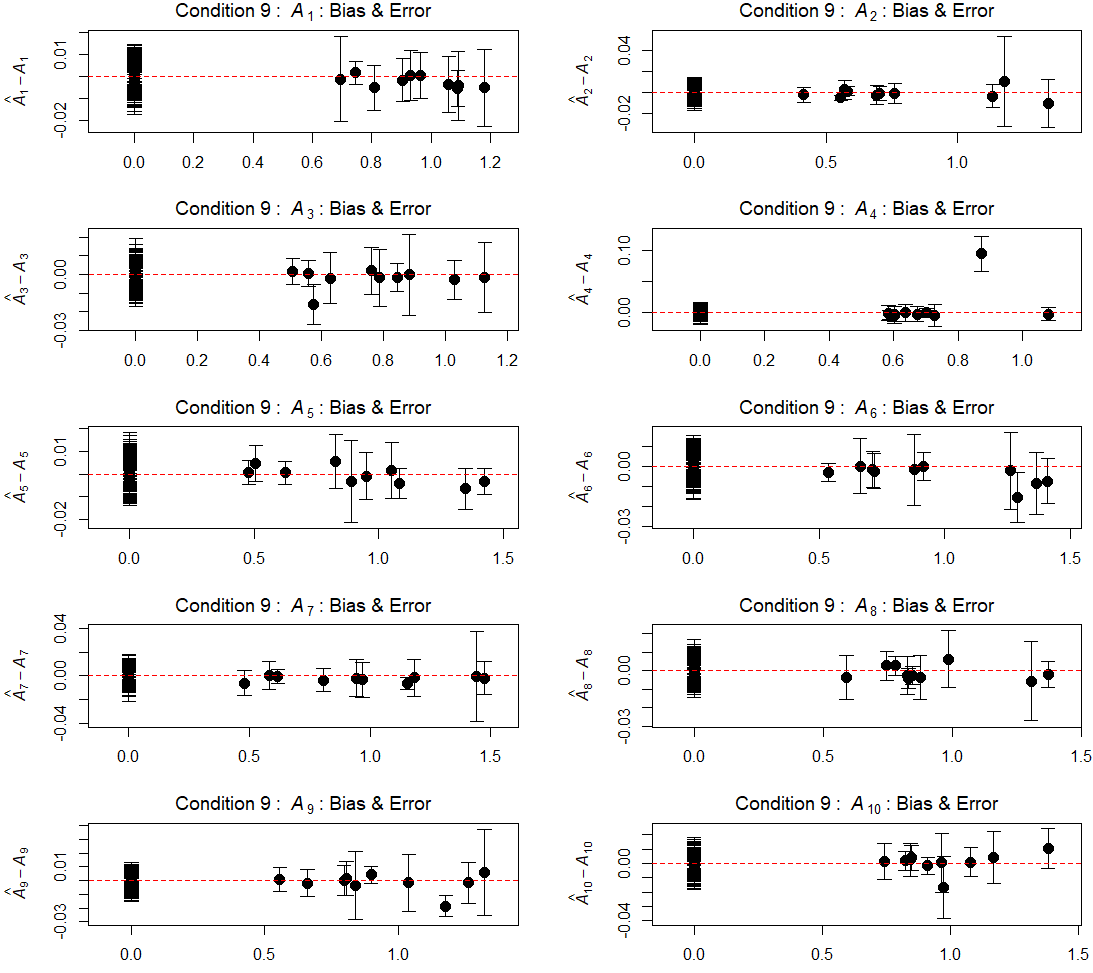}
\caption[Polytomous Multivariate Slope Parameter Estimation in Condition 9]{The results of five replications of parameter estimation using the SAEM algorithm coded in \textcite{GeisGit2019} show the five subscale dimensions after a target rotation. Each of the five replications of structural parameter estimates minus the actual values are plotted along with error bars representing a 95\% confidence interval calculated from the RMSE. Note that axes are not on the same scale. Mean estimates for detection of bias are in Table~\ref{tab:S89}.}\label{fig:S9a}
\end{figure}

\begin{figure}
\centering
\includegraphics[width=5.5in]{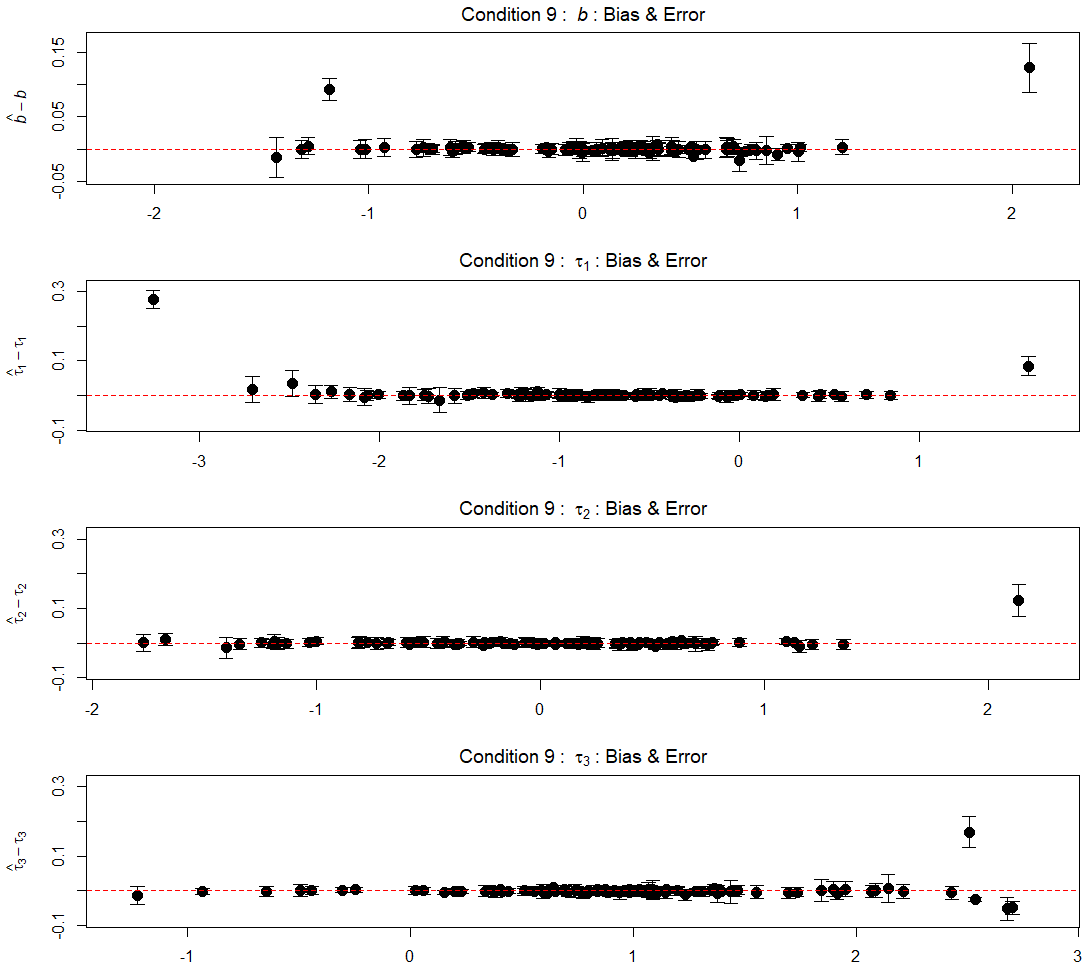}
\caption[Polytomous Intercept Parameter Estimation in Condition 9]{The results of five replications of parameter estimation using the SAEM algorithm coded in \textcite{GeisGit2019} show the intercepts and thresholds of the polytomous structure. Each of the five replications of structural parameter estimates minus the actual values are plotted along with error bars representing a 95\% confidence interval calculated from the RMSE. Note that axes are not on the same scale. The results for mean estimates for detection of bias are in Table~\ref{tab:S89}.}\label{fig:S9b}
\end{figure}

In the subscale simulations of condition 9, Figure~\ref{fig:S9a} shows the residuals and RMSEs of all 10 dimensions of slopes. One point stands out from the entire sample in latent factor $A_4$. It turns out that this slope is from item 34 and has an extreme average difficulty of 2.08 that is clearly visible in the diagnostic plots in Figure~\ref{fig:S9b}; the highest, poorly reconstructed outlier of the thresholds $\tau_1$ and $\tau_2$ are also artifacts of the same item resulting in the poor fit to the slope parameter in question. A tabulation of the response pattern for one replication of this condition shows that more than 88,000 simulated responses earned zero partial credit and thus made this item difficult to estimate.

Another item slope worth investigating is from item 26 and can be seen in the plot of $A_3$ as its RMSEs do not straddle zero, implying poor reconstruction of the generated slope. This item is from the same item as the poorly reconstructed, extremely low outlier visible in the plots of $\tau_1$. The third point to inspect is a biased estimate in $A_9$, and this point coincides with the most overestimated third threshold at approximately $\tau_3 = 2.7$. In the case of this subscale configuration, each poorly estimated slope comes from an item with difficulties at the tails of the examinees' latent abilities.

The factor estimates from these conditions are nearly equivalent to those in the cases of conditions 4 and 5 with the only exception being along the first dimension in condition 8. For the first factor estimates in condition 8, the results show a negligible difference from the distribution of the first factor's estimates in condtion 6; both condition 6 and condition 8 load on 100 items in a bifactor structure.

\subsection{Multidimensional Inferences from Alternate Configurations of Conditions 8 and 9} \label{sec:EValgo}

\begin{figure}
\centering
\includegraphics[width=5.5in]{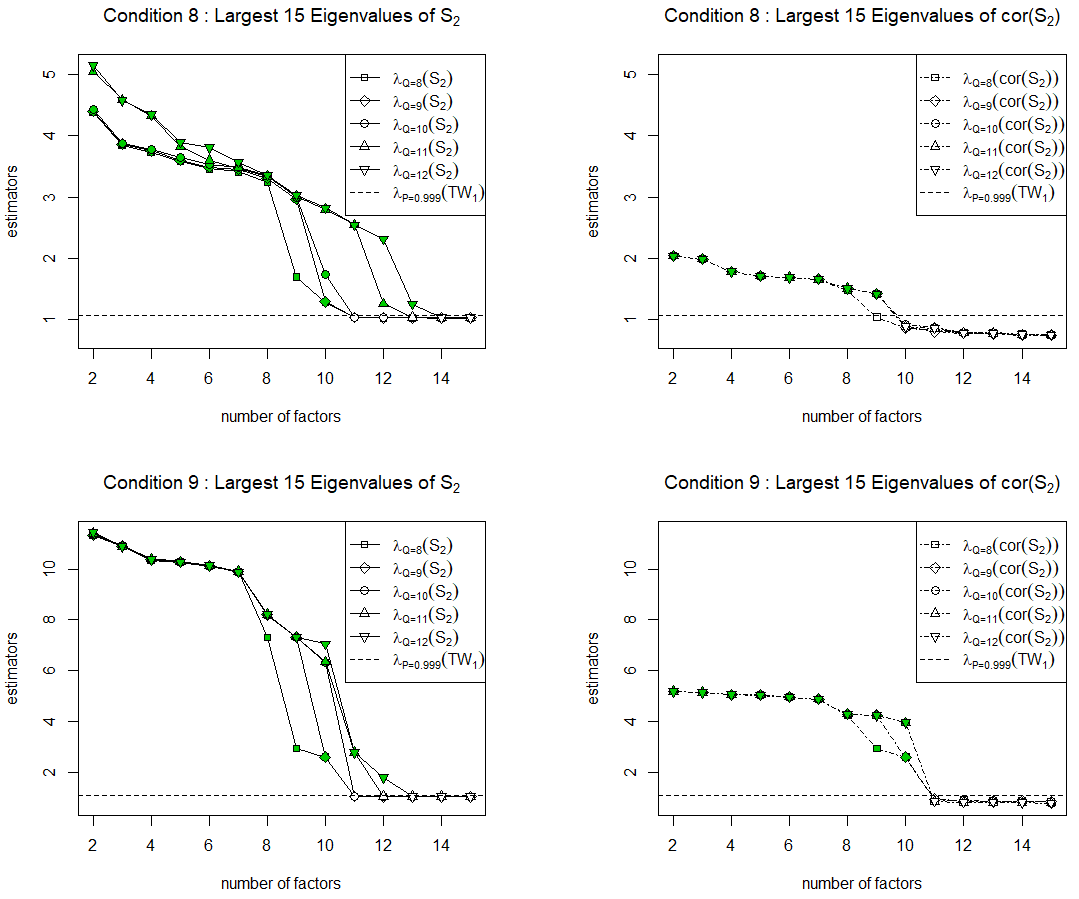}
\caption[Dimensional Sensitivity Analysis of Conditions 8 and 9]{The eigenvalues arising from the augmented data's covariance matrix at convergence is plotted for five configurations of dimensionality imposed on the SAEM algorithm's treatment of these simulation conditions. The true number of dimensions of the generated item loadings is ten. In the upper left plot, the natural log of the eigenvalues is plotted to enable visualization of the bifactor case as the loadings of the upper dimensions are sampled from the restricted $B_4 (2.5, 3, .1, .9)$ rather than the first factor's distribution of $B_4 (2.5, 3, .2, 1.7)$. The eigenvalues of $\mathbf{S}_2$ are shown with a solid line, while those of the correlation matrix of $\mathbf{S}_2$ are shown with a dotted line. When the point is significant at a \textit{p}-value of .001 (the horizontal dotted line), the point is shaded. }\label{fig:EVTestsS8S9}
\end{figure}

Given the analysis above, dimensionality of the simulations provide scope into the behavior of the augmented data. The \textit{edge universality} of random covariance \parencite{erdHos2011universality} and correlation matrices \parencite{pillai2012edge} motivates the avoidance of the parallel test, and demonstrates a pattern across all nine simulation conditions that provokes an algorithm for retaining factors that adheres to the law of parsimony, thus choosing the dimensionality of the response data; this algorithm is presented in Appendix~\ref{sec:MyAlgo}.

\section{Speed of this Codebase's Implementation of the SAEM Algorithm}

The advantage of the SAEM method comes from the efficiency of MCMC to simplify a hyperspace of parameters and the stochastic approximation of \textcite{delyon1999convergence} used to accelerate convergence. With the statistical community's gravity towards \texttt{R}, it is a convenient choice. Further, \texttt{R} is optimized for vectorization and allows the compilation of functions in \texttt{C++} if particular functions cannot be hastened. In particular, this code utilizes only one such function used for sampling from a multivariate normal \parencite{DickoMVNorm} as its performance nearly doubled the speed of \texttt{R}'s naive implementation of the Cholesky decomposition. 

Given the relatively straightforward conditional sampling in each cycle, the algorithm is relatively simple to code once the matrix math is properly calculated. The SAEM method is fast because the computational cost only depends on generating independent random values for missing information. Such algorithms can be easily coded for multiple processors and a host of technologies (e.g. GPU processing) are now coming into the mainstream. In computer science, an \textit{embarrassingly parallel} problem in which a computation can be easily divided into a number of completely independent components and each of these components can be executed on one of multiple cores. Thus, the most expensive computational aspect of the SAEM method is becoming substantially cheaper. Parallelization is more difficult with EM algorithms that require numerical integration.

Parallelization in \texttt{R} is simpler than the efforts required using \texttt{OpenMP}, \texttt{MPI}, or high performance clusters like Hadoop. Large scale parallelization offered by the DAG (directed acyclic graph) implementations of MapReduce, Spark, or other Hadoop services, are not expected to offer much assistance as the M-steps of SAEM are effectively a \textit{reduce} step; this implies that the DAG would simply be comprised of a sequence of mappers and reducers, and this computation would cancel nearly all of the advantages of the Hadoop ecosystem without novel data structures. 

In \texttt{R}, the embarrassingly parallel implementations are exploited differently depending on the operating system. Windows uses a package called \texttt{snow} and a \textit{socket} approach which launches a new version of \texttt{R} on each logical core. Linux and MacOSX uses a base package called \texttt{parallel} and involves \textit{forking} which copies the current instance of \texttt{R} to each logical core; care needs to be taken to insure that pseudo-random number generation on each parallelized instance does not source the same seed. The nuances of each approach are beyond the scope of this material, but it is noteworthy that the code in \textcite{GeisGit2019} has been implemented to operate on either type of operating system, facilitating the code's usability on any mainstream personal computing platform. It is worth a remark that the Linux implementation was expected to be faster but ran approximately 30\% slower with the same number of logical cores and the same speed CPU. 

The parallelization of the stochastic version of the Expectation steps as described in Section~\ref{sec:SAEMIRT} significantly speeds up the implementation of this code in multidimensional and polytomous calculations; the univariate dichotomous condition is difficult to improve as \texttt{rnorm} and \texttt{pnorm} are optimized functions. In step S1, the sampling of abilities is performed by parallelizing on the examinees. In next steps, the draws of $x$ in polytomous settings and $z$ in dichotomous settings (equations~\ref{eq24} and \ref{eq28}, respectively) are parallelized on each item. When the guessing parameter is being estimated, it can also be parallelized across the items since each draw is conditional on the item being correct. 

With both socket and forking methods of parallelization, data is required to be exported to each parallel instance operating on each logical unit. As the SAEM algorithm requires information from previous steps in order to run the M-steps, new data objects need to be exported at each iteration to facilitate the S-steps. Certain efficiencies are built into the code such that static data structures like the response pattern and functions are exported to each logical unit prior to the iterations of the Markov chain. 

Several improvements can be made to expand the code's functionality. First, a package implementation would allow it to be easily installed for any user sourcing \texttt{CRAN} (Comprehensive R Archive Network). Second, in Bayesian methods, it is useful to run multiple chains across multiple cores; this is not implemented. Third, the form of the items is currently expected to be homogeneous w.r.t. the probabilistic model of the response function for each item. In other words, dichotomous and polytomous items are not currently capable to be estimated in a single assessment. Most importantly, the functions have been built as the research has been pursued, rather than with a neophyte user in mind; thus a review of a user-friendly sequence of EFA-specific interactivity with a real response data set is the first priority before submitting a packaged version to CRAN. The github repository currently contains two tutorials, an \texttt{Rmarkdown} document that introduces a new user to a univariate dichotomous simulation, and a second that outlines the process for a multivariate polytomous simulation.

%




\chapter{Empirical Application of SAEM and Exploratory Tests of Multidimensionality} \label{ch:em}

Three empirical treatments using real response data will also enhance the learnings from the simulations. In the measurement of latent traits, a significant effort needs to devoted to the construction \textit{and} validation of multiple-choice items as they relate to the intent and purpose of an achievement test, concept inventory, or psychological assessment. Statistical methods under the umbrella of IRT are often applied to these ends (\cite{embretson2013item}; \cite{Haladyna2004}), and the methodology should advance the understanding of the student's competence within a domain of knowledge or the patient's psychological tendencies within a behavioral domain; the multiple-choice test is expected to be a sample of that domain. 

The three real response data sets in this chapter provide an opportunity to explore the utility of SAEM at small but significant sample size. The first and second assessments to be investigated are the Force Concept Inventory ($J=30$, $N=748$) and Chemistry Concept Inventory ($J=22$, $N=628$). The third assessment is a Pediatric Quality of Life questionnaire ($J=24$, $N=753$). \textcite{de2013theory} suggests a rule of thumb that factor analysis sample sizes should exceed 10 times the number of items to trust item calibration and the estimation of factor loadings, and a thorough exploration of the topic of sample size is covered in \textcite{sahin2017effects} for unidimensional models. In \textcite{de1994influence}, estimations are reasonably accurate at a 5:1 ratio of sample size to items, and \textcite{jiang2016sample} demonstrates a sample size of 500 is quite sufficient for multidimensional test conditions with 30 items, with correlations of estimated slopes and intercepts to their true parameters exceeding .94. Given the sample sizes of real response data in the following studies and the conclusions of the cited authors, there is ample reason to use the analyzed response data to make empirical judgements about the utility of SAEM. 

It is noteworthy that the size of these assessments and their response sample sizes are quite small relative to large-scale assessments. Despite the acceptability of the EFA approach to the response data in the analyses below, work is still to be done to demonstrate the utility of SAEM and EFA on assessments like PISA and NAEP. The security controls around this large-scale response data are also an impediment to access.

The FCI (Force Concept Inventory) and CCI (Chemistry Concept Inventory) response data are from the WIDER assessment project at Rutgers University \parencite{brahmia2014mathematization}; this project was used to assess learning and applications of mathematics across multiple domains in the first year of undergraduate STEM (science, technology, engineering, and math) coursework. Large public universities have been learning to adjust to a more conservative fiscal restraint from states and federal funding networks \parencite{saltzman2014economics}, forcing the transition from traditional classroom environments to cost-effective solutions such as an online format; despite the simple choice when presented with the budgeting, the student experience would fundamentally shift.

In order to properly gauge the changes to educational outcomes from such a transition, it is important to measure how students are performing \textit{ex ante}. The WIDER proposal was initiated for Fall 2013-Spring 2014 to diagnose the preparation of the incoming freshmen engineering class of 2017, as well as their understanding \textit{ex post} of the traditional coursework and instruction. Questions from the CLASS (Colorado Learning Attitudes about Science Survey) in Physics \parencite{adams2006new} and Chemistry \parencite{adams2008modifying}, as well as the FCI (Force Concept Inventory; \cite{hestenes1992force}) and CCI (Chemistry Concept Inventory; \cite{krause2004development}) were asked of the students before and after the traditionally taught courses, specifically Extended Analytical Physics 115, Analytical Physics 123, and General Chemistry for Engineers 159. Application data such as SAT scores and a socio-economic-status metrics were collated and anonymized for inclusion in subsequent analyses. The concept inventories administered to hundreds of students for pre and post-testing are well-researched, notably the FCI. In this application of the algorithm, the FCI and CCI response data will be analyzed and compared to outcomes in the literature. 

The third set of response data comes from a 24-item form for the pediatric QOL (quality of life) scale that assesses the social quality of life for children. The assessment was developed and factor analyzed by \textcite{dewalt2013promis} and was again estimated by \textcite{cai2010high}, making this an appropriate response data set to appraise the performance of SAEM. 

Before continuing further, it is important to recognize that all test designs founded upon subject matter expertise cannot account for the psychological perceptions of the students being tested. A PhD level physicist may fully appreciate the conceptual distinguishability between Netwon's Second and Third Laws, while a student may exercise intuitions that conform to that kinematic phenomena with no knowledge of Isaac Newton; the FCI requires no mathematical formulae and was specifically written to target conceptual understanding. With this in mind, the hypothetical dimensionality proposed by experts designing a test is highly unlikely to conform to the modes of perception of the sampled student population. The factor analytic approach to item response theory can only address the interaction of the students' perceptions of the items within the tests being analyzed here, rather than the intentions of test design.

\section{SAEM Applied to the Force Concept Inventory}

The FCI is an instrument designed to assess fundamental concepts to Newton’s Laws of Motion (\cite{hestenes1992force};\cite{savinainen2002force}). The FCI's development reinvigorated the interest in physics education and inspired the creation of multiple concept inventories to offer diagnostics that could provoke further reforms in STEM education \parencite{evans2003progress}. The full instrument has 30 multiple choice questions and looks at six areas of understanding: kinematics, Newton's First, Second, and Third Laws, the superposition principle, and types of force (e.g., gravitation and friction) with distractors designed to address common naïve conceptions. This 30-item version (revised form 081695R, \cite{halloun1995force}) of the FCI was administered to first-year physics students $N = 748$ as a pre-test and post-test, but only the pre-test is reported here. There is an expectation that the examined students run the gamut, as Rutgers is the state university of New Jersey, and New Jersey high schools typically require only three years of lab-based science; the latest enrollments in high schools reported by the NCES (National Center for Education Statistics) in 2012 are approximately 32\%\parencite{meltzer2015brief} with figures reported by the AIP (American Institute of Physics) closer to 40\% for the same period \parencite{white2014high}.

\subsection{Previous Research on FCI Response Data}

As with most assessments, the FCI has been a source of debate and controversy especially as performance on the assessment has been shown to be dependent on instructional practice \parencite{hake1998interactive} which seems to be the classic ``chicken and egg'' argument; many might argue this a classic example of the Hawthorne effect dubbed by \textcite{landsberger1958hawthorne}, or more likely in the halls of physics departments, cause and effect. Plenty of researchers have worked diligently to critique or advance the interpretability of the application of the FCI to assess student understanding (\cite{henderson2002common}; \cite{huffman1995does}; \cite{bao2001concentration}), while the focus of the authors has been to ``evaluate the effectiveness of instruction'' (\cite{hestenes1995interpreting}, pg. 1). 

Within the scope of the current research, the goal is to compare the quantitative properties made available from the application of SAEM and the Tracy-Widom test to those of quanititative studies that included some measure of dimensionality of the FCI. As the FCI has the serendipitous advantage of maintaining status as a psychometric frontrunner within the physics discipline, several authors are quite skilled in applied statistics and mathematics, especially matrix operations like rotations and eigenanalysis. Thus, the FCI becomes an apt test with which to expound upon the utility of SAEM.

In\textcite{hestenes1992force}, it is described that the FCI is a test on six dimensions: Kinematics, Newton's First Law, Newton's Second Law, Newton's Third Law, Superposition Principle, and Kinds of Force. One of the first attempts to disentangle statistical evidence of this psychometric structure, \textcite{huffman1995does} calculate the sample Pearson correlation coefficient, followed by a PCA (principal component analysis) where the common heuristic to retain factors that ``only account for factors that account for at least 5 to 10\% of the total variance'' (p. 142) is utilized. The PCA is performed using SPSS; the software extracted 10 factors, but only two accounted for enough variance to be considered significant despite the original hypothesis of six individual conceptual structures. In response to this treatment, \textcite{hestenes1995interpreting} critique the work of Huffman \& Heller, pushing for a more univariate interpretation that enforces their suggested use case of the evaluation of effective instruction, quoting: the ``total FCI score is the most reliable single index of student understanding, because it measures cogerence across all dimensions of the Newtonian force concept'' (p. 505). 

Recently, \textcite{scott2012exploratory} runs an exploratory factor analysis on the tetrachoric correlation matrix of the response data. Tetrachoric correlations are calculated based on the assumption that the binary data of a dichotomous response pattern are but bernoulli random variables sampled from underlying continuous normally distributed variables \parencite{schmitt2011current}. In this analysis, the authors choose five factors resulting from the ``parallel test'' motivated by \textcite{horn1965rationale}; a didactic on this approach is developed in \textcite{hayton2004factor}. Essentially, a random $J \times N$ matrix is generated many times and the distribution of rank-ordered eigenvalues is generated. This technique conforms well to the approach in this dissertation, but it is good to repeat that specific differences apply. The work here: (1) applies a structural model using the results from convergence of an IRT parameterization, (2) applies an eigenanalysis of the latent, augmented propensity matrix at convergence, and (3) compares the eigenvalue distribution to the asymptotic Tracy-Widom ($\beta=1$) distribution. Interestingly, this application of SAEM and its exploit of the augmented data of \textcite{MengSchilling1996} might be considered as direct access to the underlying continuous latent propensities of the examinees implied by a tetrachoric correlation. 

\subsection{Descriptives of the FCI Data}

\begin{figure}
\centering
\includegraphics[width=5in]{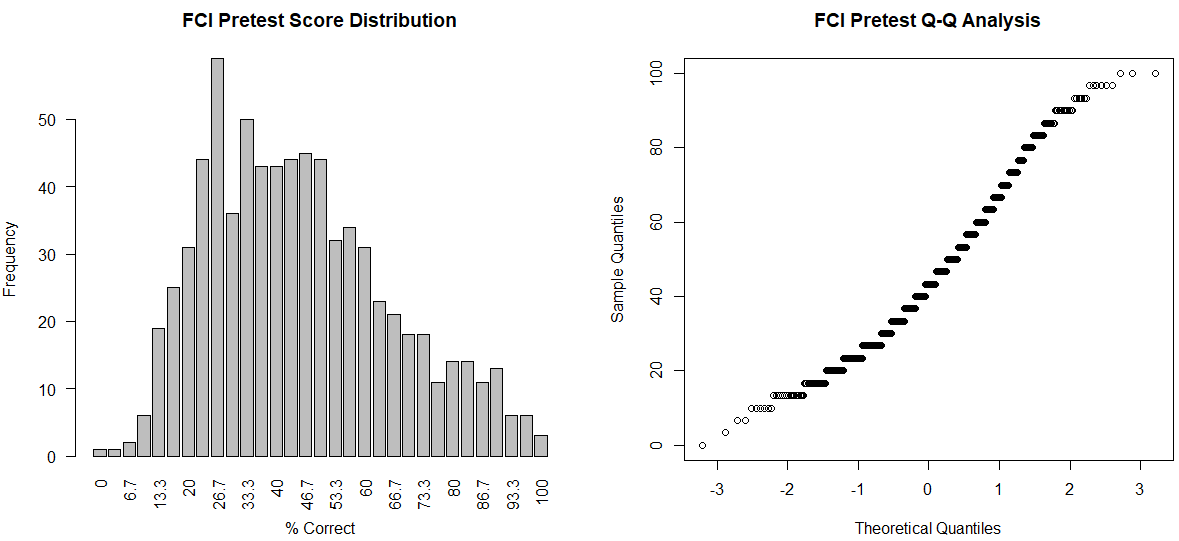}
\caption[WIDER FCI Pretest Scores]{Using the percent correct as a score, the distribution of FCI pretest scores are shown here.}\label{fig:FCI_ScoreDist}
\end{figure}

\begin{figure}
\centering
\includegraphics[width=5.5in]{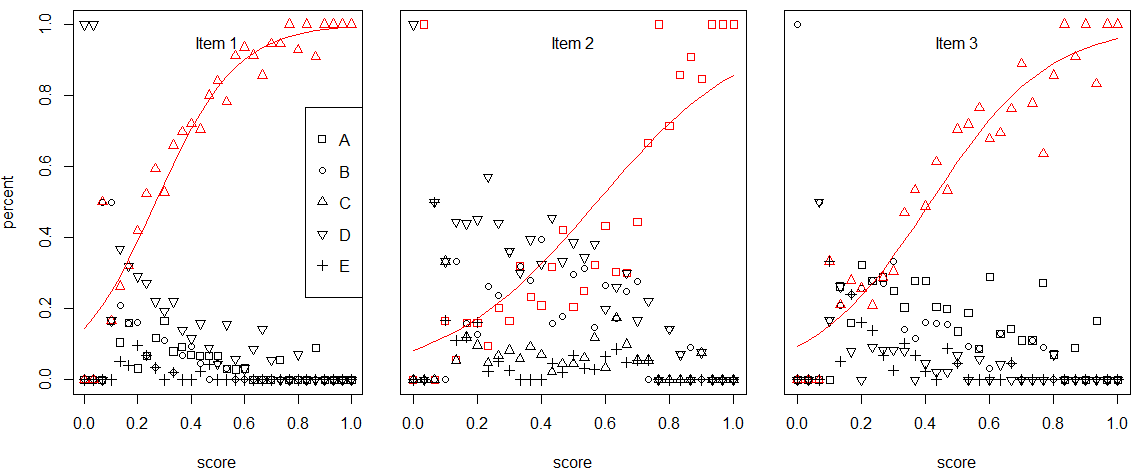}
\caption[Response Patterns of Items 1-3 of FCI Data]{Within each of the 31 possible scores on the FCI, the respective proportions of categorical responses is shown for items 1, 2, and 3. A logistic regression is fit to the correct response demonstrating the monotonic increase in probability of a correct response as score increases.}\label{fig:FCI_RP}
\end{figure}

The distribution of scores is displayed in Figure~\ref{fig:FCI_ScoreDist}. The mean score is 45.2\%, has a median of 43.3\%, a standard deviation of 20.8\%, and shows positive skewness. In general, this assessment meets the students at or near the sample population's central ability. In Figure~\ref{fig:FCI_RP} are three plots of specific items showing the proportion of answers being selected by students within each score bucket; the correct selections conform well to a two-parameter logistic regression on score (no assumption of a parameter for guessing). 

There are not many distinguishable patterns beyond the monotonic increase in proportion of correct answers chosen, but these figures allow visualization of the most commonly chosen distractors rather easily. This will prove useful in item analysis. The ICCs (item characteristic curves) also provide direct observations of difficulty and discrimination. In the appendix, all 30 items are shown in order to facilitate comparisons with other research into the FCI. A quick scan of the figures demonstrates that item 1 shows up as a relatively simple problem and item 13 shows very strong discrimination. 

Part of the motivation of IRT is that ability estimates are better approximations of ability than traditional univariate test scores. After applying a one-dimensional SAEM IRT analysis to the data, the regressions as a function of the discrete scores can be compared to a logistic regression on discrete bins of the latent ability that can be approximated using the converged structural item parameters. The residuals of regressions on 31 scores as compared 31 discrete bins of latent ability scores can be performed to show that the IRT approach outperforms the stratification created by simple scores. With a 2PNO latent ability approximation, 17 mean residuals are closer to zero and 22 of the standard deviations are smaller than those from regressions on simple scores. When using the 3PNO for approximating latent ability, 20 mean residuals are closer to zero and 18 of the standard deviations are smaller.

As several physics education research papers have included IRT approximations to the FCI's item response data, it is important to include the univariate SAEM fits and the transformation of the difficulties to the 2PL parameterization (scaling the difficulty parameter by the constant 1.7). Those results, for both estimations with and without a guessing parameter, have also been provided in Appendix~\ref{sec:TestDiagnostics} in Table~\ref{tab:IRT1Dpars}. Before any comparison with the literature, the data demonstrates a substantive difference between 2PNO and 3PNO parameters, especially as the discriminations show a stark contrast. Model fit statistics for the 2PNO and 3PNO are summarized in Table~\ref{tab:FCI1Dfits} along with multidimensional models that will be further assessed in the following subsection. Favoring the 3PNO, a $p$-value less than $10^{-13}$ is calculated from the likelihood ratio test, and is consistent with the AIC suggesting that the 2PNO is less than $10^{-15}$ as probable as the 3PNO to minimize information loss. The BIC favors the 2PNO. 

\begin{table}
\centering
\begin{tabular}{lrrrr}  
Model & LogL & AIC & BIC & $dof$ \\ \hline
2PNO & -11390 & 22900 & 23177  &  60  \\ 
3PNO & -11325 & 22830 & 23246  &  90 \\ 
2D & -10611 & 21403 & 21819   &  90  \\ 
2D w/ guessing & -10544 & 21328 & 21882  &  120 \\ 
3D & -9992 & 20224 & 20778  & 120  \\ 
3D w/ guessing & -10165 & 20630 & 21323  &  150  \\ \hline  
\end{tabular}
\caption[FCI SAEM Model Fit Statistics]{2PNO, 3PNO, and Multivariate estimations are performed. The log likelihood is calculated along with the AIC and BIC. } \label{tab:FCI1Dfits}
\end{table} 

Two particular studies list one-dimensional parameters that can be compared to the application of SAEM here. \textcite{wang2010analyzing} collected pre-test data on 2,802 students from 2003-2007 and applied a 3PL; the average score in that data is 52.7\% with a standard deviation of 20.2\% as compared to the 45.2\% and 20.8\%, respectively, from the sample population estimated here. \textcite{chen2011comparisons} used data that overlaps with the data from \textcite{wang2010analyzing} and employed three available software platforms to fit a 3PL on response data from 3,139 examinees with a mean score of 49.3\% and a standard deviation of 18.1\%. The software tested included MULTILOG, PARSCALE, and the package \texttt{LTM} in \texttt{R}. 

The converged parameters of the SAEM fits are presented in appendix~\ref{sec:TestDiagnostics}. While exact parameter comparisons are not useful without a much deeper analysis of the sample populations, it is worth noting the consistency of specific results of the two published FCI estimation studies and the 3PNO results here. All three analyses were consistent on the three items with the highest discrimination, items 5, 13, and 18. This is very interesting as the absolute size of the discrimination parameter directly correlates with the statistical information of the item; thus these three items are the most ``informative'' items of the test across each student sample. Of the five items with the lowest difficulty, two were consistent across all three analyses, items 12 and 1; among the five of highest difficulty, only 1 item is common to all analyses, namely item 26. Items 9 and 6 were present in the five highest estimations of guessing for all fits. 

\subsection{Dimensionality of the FCI Data}

\begin{figure}
\centering
\includegraphics[width=5.5in]{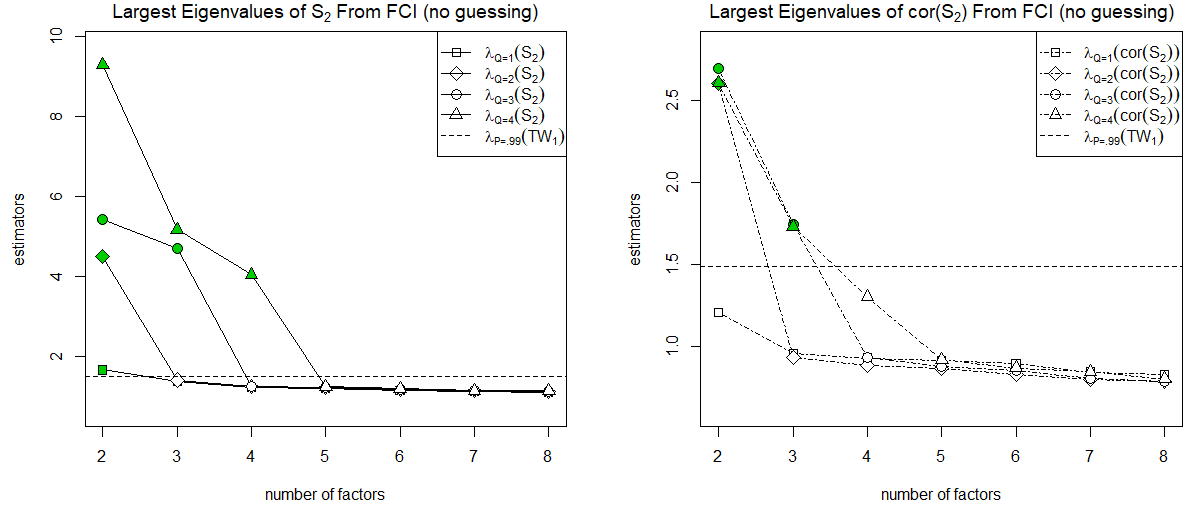}
\caption[Dimensional Sensitivity Analysis of FCI Data]{The eigenvalues calculated from the augmented data's covariance and correlation matrix at convergence are plotted for five configurations of dimensionality imposed on the SAEM algorithm's treatment of the FCI data. Eigenvalues from one, two, three, four, and ten dimensions are displayed along with a line targeting a cumulative probability of 99\% of the Tracy-Widom distribution.}\label{fig:FCI_EV}
\end{figure}

In Table~\ref{tab:IRT1Dpars}, the fit statistics support a guessing parameter until a three-dimensional structure is estimated. Using the comparative eigenvalue methodology explored in the simulations, the plots in Figure~\ref{fig:FCI_EV} are constructed, with and without a guessing parameter. Configurations from one to four dimensions were plotted, but the data was estimated for retaining up to 12 dimensions. When running model fit statistics, it is important to mention that the likelihood ratio test and information criterion metrics favor more and more dimensions. This favorability toward less degrees of freedom has been noted in the EFA literature (\cite{velicer1990component};\cite{floyd1995factor}) and makes goodness of fit statistics a bit troublesome.

The key enhancement provided by the SAEM approach to item factor analysis is the access to the augmented data covariance, a sufficient statistic of this statistical model. It was described in the simulations of chapter~\ref{ch:ss} that the proposed algorithm (see Appendix~\ref{sec:MyAlgo}) for selecting factors correctly assessed the dimensionality of the simulated loadings for all conditions except the 10-dimensional bifactor structure, of which nine factors were calculated to be significant at a $p$-value less than .001. It is more than reasonable to explore this real data applying the same lens with which to interpret the dimensionality of the data. 

Figure~\ref{fig:FCI_EV} shows that the one-dimensional configuration of either those fits with or without guessing results in $\Lambda_S = 2$. Then when adjusting $\hat{\Lambda}$ to use two dimensions, with or without guessing, the result gives $\Lambda_S = \Lambda_C = 2$. This gives confidence in stopping at two dimensions, despite the results in the fit at three dimensions where both $\Lambda_S = \Lambda_C = 3$. This may be evidence that there is a weak third factor, but in the two dimensional configuration, only 2.8\% of variance of $\mathbf{S}_2$ is explained with a third factor, and with guessing the third factor only explains 1.1\%. Following the algorithm from the appendix explained in section~\ref{sec:EValgo}, the result is two dimensions.

Once a number of dimensions has been decided upon, it is important to rotate to facilitate interpretation. It is appropriate to look for a rotation that is \textit{simple}, i.e. ``each variable loads highly on as few factors as possible; preferably, each variable will have only one significant or primary loading'' (\cite{floyd1995factor}, pg. 292). Using a varimax rotation can be suitable in this situation, but another orthogonal approach can be applied here as it is simple to visualize in two dimensions. For the analyses below, the 2D fit \textit{with} guessing is selected for two reasons: (1) the discriminations show much stronger variation in the 3PNO columns from Table~\ref{tab:IRT1Dpars}, and (2) the fit statistics favor the inclusion of guessing.

\begin{figure}
\centering
\includegraphics[width=5.5in]{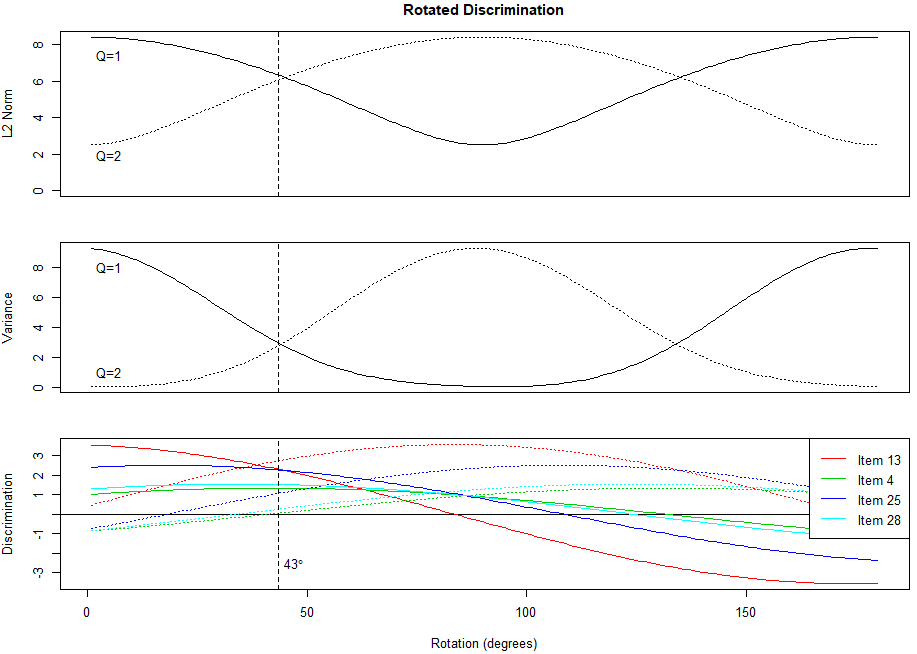}
\caption[FCI Loadings Rotated]{The loadings extracted from the two-dimensional fit are rotated through 180 degrees. The top plot shows the euclidean distance of the item vectors of the first and second dimensions. The middle plot shows the variance of the loadings in each vector of factor loadings at each degree of rotation. The lower plot shows the first and second dimensions of the four items of the largest discrimination in each dimension as the loadings are rotated. In each plot, the solid line shows the first dimension, and the dotted line is the second dimension. Finally, a vertical dashed line is drawn where the variance of both vectors of loadings are maximized and the discriminations of the four items show the largest separation. The normal scaled loadings corresponding to this rotation at 43 degrees are in Table~\ref{tab:FCI_2D_Loadings}.}\label{fig:FCI_Rotations}
\end{figure}

It is simple to use a continuous orthogonal rotation on two dimensions as this only requires a two dimensional rotation matrix, 
\begin{equation}
R = \left[ \begin{array}{cc} \cos \theta & -\sin \theta \\ \sin \theta & \cos \theta \end{array} \right].
\end{equation} \label{eq:Rotate}
\noindent Three visualizations of the results of such a rotation.The euclidean distance or L2-norm ($\norm{A_i}_{L_{2}}$) is useful to optimize the information available from each dimension. Secondly, the variance of the loadings in each vector help us to visualize the concept of a varimax rotation, especially with two dimensions. Last, the change in the discrimination of the three items with the largest slopes can be useful for visualizing the effects of a rotation. All three of these plots, as function of angle of rotation, are shown in Figure~\ref{fig:FCI_Rotations}.

In the top plot of this figure, the L2-norm shows that information from both dimensions are spread about evenly where the two lines intersect. The variance among the loadings of each dimension should also not favor either ability, and the intersection of the two lines in the middle plot provide that visualization of the rotation angle. A vertical dashed line is drawn and two of the three items in the lowest plot show that the chosen angle of rotation does well to maximize one factor's influence over the other.

\begin{table}
\centering
\begin{tabular}{crr}  
Item & First Loading & Second Loading \\ \hline
1  &  .453 & .409  \\   
2  &  .902 & .772  \\   
3  &  .314 & .468  \\   
4  &  .800 & .034  \\   
5  &  .880 & .877  \\   
6  &  .324 & .533  \\   
7  &  .413 & .556  \\   
8  &  .279 & .627  \\   
9  &  .264 & .567  \\   
10  &  .364 & .544  \\   
11  &  .757 & .873  \\   
12  &  .312 & .572  \\   
13  &  .919 & .938  \\   
14  &  .362 & .611  \\   
15  &  .720 & -.092  \\   
16  &  .577 & .235  \\   
17  &  .856 & .489  \\   
18  &  .906 & .874  \\   
19  &  .369 & .655  \\   
20  &  .352 & .536  \\   
21  &  .156 & .711  \\   
22  &  .549 & .810  \\   
23  &  .366 & .673  \\   
24  &  .252 & .468  \\   
25  &  .916 & .724  \\   
26  &  .865 & .782  \\   
27  &  .180 & .577  \\   
28  &  .836 & .231  \\   
29  &  .256 & .358  \\   
30  &  .849 & .907  \\ \hline
\end{tabular}
\caption[FCI 2D Rotation]{With the choice of two dimensions and a rotation that creates positive loadings and maximizes variation across the two dimensions, this table provides the normal scaled discriminations on each of the 30 items.} \label{tab:FCI_2D_Loadings}
\end{table} 

After choosing this angle of approximately 43 degrees, there are a few items that now require inspection. Items 4, 15, and 28 strongly load on one dimension while items 21, 22, 23, and 27 should give insight into the second dimension. Items 5, 13, and 18 provide an interesting look into the two abilities as those items very strongly load on both dimensions.


\subsection{Qualitative Discussion and Comparisons to Previous Research} \label{sec:FCILit}
In assessing any qualitative interpretation of these items, it is important to reference the classification of misconceptions as delineated by the authors of the FCI. \textcite{hestenes2007} provides a thorough table of the motivations that may motivate students to choose each of the distractors specific to each item. Physics education researchers have used this taxonomy and probed large samples of FCI data and qualitative interviews (\cite{poutot2015exploration}; \cite{wells2019exploring}) to further unpackage the diagnostic properties of this assessment. Autoethnographically, my background includes teaching science to sixth, seventh, and ninth grade physical science, eleventh and twelfth grade physics, and undergraduate physics recitations as a graduate student of physics at ASU (Arizona State University). I have also had a considerable amount of exposure to David Hestenes' modeling instruction as facilitated by the physics education research group at ASU and had been trained and participated in PUM (physics union math, \cite{etkina2010pedagogical}) research as a high school teacher in Edison, NJ. With this background, it is important that I include a disclaimer that there is potential for a signficant personal bias in the interpretive context of the items of the FCI.

Beginning with clear indicators of the loading on the first dimension, item 4 posits a question about the equal and opposite forces of an automobile collision, item 15 discusses the equal and opposite forces of contact between two automobiles where one is pushing the other, and item 28 repeats the concept by asking about the equal and opposite forces of two students pushing off of each other. There is a clear pattern to each of these questions, and the distractors of each item pair strongly in the table of taxonomy of misconceptions associated with the distractors of each question \parencite{hestenes2007}.

From the perspective of a former physics instructor, within the first few days of the introduction of \textit{forces}, the conception of equal and opposite forces is also delivered. It is a necessary foundational perspective required for understanding: (1) the reason we do not fall through the floor, (2) the reason we feel as if we are not moving when cruising forward in a car or plane, (3) the basics of a \textit{normal force}, as well as (4) the consequences of Newton's third law. The misconceptions implied the distractors are classic examples of intuitive falsehoods, specifically ``only active agents exert forces'', ``greater mass implies greater force'', the ``most active agent produces greatest force'', and ``obstacles exert no force''. These types of misconceptions are necessary to dispell in order to proceed to more complex problem solving tasks. It is most interesting that the items isolated here show difficulties greater than zero, implying that a majority of this collegiate level sample of examinees has yet to firmly adhere to this axiom of physical law.

For assessing the second dimension extracted from the FCI, items 21, 22, and 23 refer to the motion of a rocket that is at first floating horizontally to the right, followed then by a constant thrust (\textit{constant force}) in an upward direction; finally this thrust is then turned off. In item 27, a constant horizontal force is suddenly stopped, just like the last event in the questions about the rocket. Again, the pattern is quite consistent among these four questions: constant unbalanced forces are applied before or after there is no net force. This particular set of items carries a sequence of downstream consequences that include but are not limited to (1) constant acceleration, (2) changing velocity, (3) perpendicular vectors being treated independently, and (4) parabolic motion. All four of these questions have misconceptions bundled under the taxonomy \textcite{hestenes2007} classifies as Impetus or Active Forces. The paths in items 21 and 23 include images that resemble parabolas; there are also trajectories with discontinuities of slope that are quite attractive to intuitive falsehoods that derive from ego-centered reference frames\footnote{An example of an ``ego-centered reference frame'' is the sensation of being pulled to the right when turning left in a vehicle at a constant speed. Our intuition is that we are being pulled to the right when in fact, we are being pulled to the left by the seat of the car while the physical law of inertia would have our massive bodies move in a straight line.}.

\textcite{wells2019exploring} shows a novel graph analysis isolating \textit{communities} of examinees choosing combinations of distractors of different items. Curiously, items 4, 15, and 28 have a cluster of distractors that significantly correlate ($r>.25$), and items 21, 22, 23, and 27 have a series of distractors that imply ``related incorrect reasoning'' (pg. 11) with significant correlations ($r >.2$). These results add validity to the statistical outcomes of the factor rotation chosen here. While \textcite{scott2012exploratory} retained 5 factors, it is notable that items 4, 15, and 18 were isolated to their fourth factor, items 21, 22, 23, and 27 (along with items 19 and 20) all loaded together on their fifth factor, and the three items (5, 13, and 18) that strongly load on both factors in this analysis load strongly on their second factor. 

It is worth noting that the items showing the first dimension's strongest signals are specific to Newton's third law and those from the second dimension are specific to Newton's second law. Personally, I have come to the conclusion that Newton's first law is a philosophical statement of the \textit{null hypothesis} as applied to all objects in all reference frames, reinterpreted as ``when nothing is happening to an object that may or may not be moving, there is nothing that will change about its motion;'' this statement is required before analyzing any aspect of an object's motion. In fact, it can be said that the second law contains the first law because zero acceleration and zero force are implicit to a balanced equation where nothing is changing about an object's motion. 

Examining items that load very strongly on both factors, items 5, 13, and 18 include the force of gravity on objects in motion. Items 5 and 18 also invoke circular motion which comes much later in the force and motion curriculum. The median of the test scores is 43\%, but if a student answers any of these three questions correctly, the conditional median of FCI test scores rises to 60\%. 

\section{SAEM Applied to the Chemistry Concept Inventory}

In the wake of the FCI, researchers backed by the NSF (National Science Foundation) reported their progress in developing concept inventories in STEM-associated domains \parencite{evans2003progress}. One of these deliverables addresses conceptual foundations in chemistry and is known as the CCI (chemistry concept inventory; \cite{krause2004development}). In the WIDER project described above, the CCI was administered as a pre- and post-test to first-year engineering students in their freshman chemistry course. Only the pre-test is analyzed here. The analysis being done here does not allow for much comparison to ongoing research as a search of the archives results in very little applications of IRT to the CCI. This may simply be an artifact of the long history of Physics Education Research, but the exercise still serves to illustrate a quantitative application of the SAEM to the dimensional analysis of a concept inventory. 

As it may be pertinent to the analysis here, \textcite{krause2004development} designed the CCI where ``the goal was to make a reliable instrument that is easy to use, administered in a short period of time and can accurately assess student understanding of general chemistry topics'' (pg. 1). The test covers chemistry topics that fall into six categories with each category including one to four sub-topics. The six categories include Thermochemistry, Bonding, Intermolecular Forces, Equilibrium, Acids and Bases, and Electrochemistry. As this test is far more broad than the FCI from the previous section, it may be the case that this data will not conform well to a qualitative treatment of the statistical outcomes of the factor analytic approach. Coming from the background as a teacher of introductory chemistry in lab-based high school physical science, I find it quite daunting to compartmentalize chemistry concepts into a clean conceptual scaffold; the topics range from physical properties of density to abstract metrics like gibbs free energy and entropy, not to mention implausibly large numbers invoked throughout stoichiometry, e.g. Avogadro's constant and its application to the ideal gas laws.

\subsection{Descriptives of the CCI Data}

\begin{figure}
\centering
\includegraphics[width=5in]{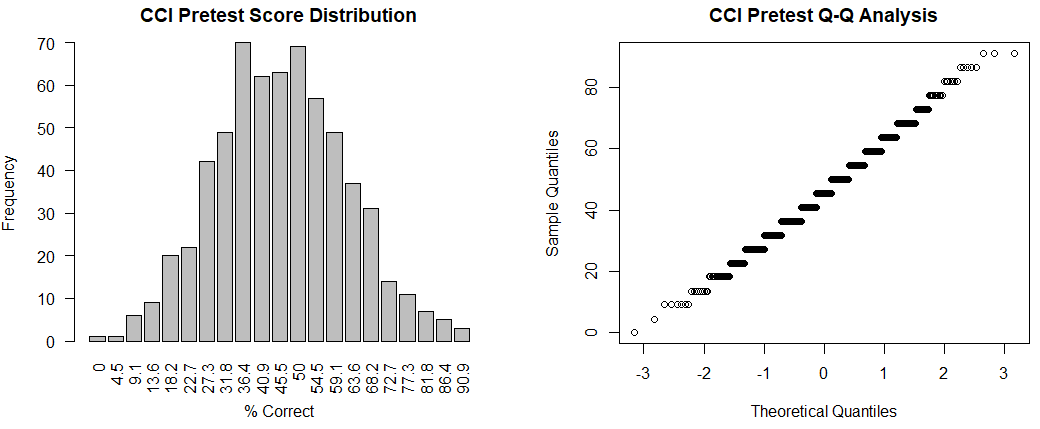}
\caption[CCI Pretest Scores]{Using the percent correct as a score, the distribution of CCI pretest scores are shown here along with a Q-Q analysis.}\label{fig:CCI_ScoreDist}
\end{figure}

The scores on the 22 items of the CCI are almost normal as seen in Figure~\ref{fig:CCI_ScoreDist}. The mean of the 628 scores is 45.8\% and the standard deviation is 16.3\%. No individual scored 100\%. In the appendix (section~\ref{sec:HDCCI}), the ICCs are plotted, each showing a clear monotonic increase in the proportion of correct answers chosen as scores increase. Several items have remarkably similar curves (items 7, 8, and 13) in terms of discrimination, difficulty, and small residuals. Items 5, 9, 14, and 22 show a very high difficulty. 

In the interest of qualitative review of the content, the low difficulty items provide some insight to the test. Item 7 is a True/False question addressing whether ``matter is destroyed'' when a match burns. Item 8 asks for the student to choose a reason for their answer to item 7. Several pairs of items are presented in this fashion, i.e. one item specific to a chemistry concept is followed by another item that is used to probe \textit{why} the student chose their answer. Items 12 and 13 conceptually conform well to items 7 and 8; a closed tube with a specific mass of solid matter is heated and evaporated and the examinee is asked if the mass changes, then to provide the reason for their answer. A researcher may hypothesize that these four questions will load on the same factor.

\begin{table}
\centering
\begin{tabular}{lrrrr}  
Model & LogL & AIC & BIC & $dof$ \\ \hline
2PNO &  -6675 & 13438 & 13633  &  44  \\ 
3PNO & -6684 &  13501  &  13794  &  66 \\ 
2D & -5946 & 12025 & 12318   &  66  \\ 
2D w/ guessing & -5743 &  11661 &  12052  &  88 \\ 
3D &  -5335 &  10846 &  11237   & 88  \\ 
3D w/ guessing & -5242 &  10704 &  11193  &  110  \\   \hline
\end{tabular}
\caption[CCI SAEM Model Fit Statistics]{2PNO, 3PNO, and Multivariate estimations are performed. The log likelihood is calculated along with the AIC and BIC. } \label{tab:CCI1Dfits}
\end{table} 

The fit statistics for the CCI data were compiled for 3 dimensions with and without guessing and are listed in Table~\ref{tab:CCI1Dfits}. Using the likelihood ratio test can give paradoxical results. The 2PNO fit has a higher likelihood than the 3PNO and definitively implies that guessing is not required, and the univariate logistic regressions in most of the plots from Figures~\ref{fig:CCI_RP1} and \ref{fig:CCI_RP2} seem to confirm little need for a guessing parameter. But as dimensions are added, the inclusion of a guessing parameter results in a significant likelihood ratio test suggesting the less parsimonious models. As with the FCI, as more dimensions are added, each addition of another 22 degrees of freedom results in a `better' model fit statistic. These results seem to confirm, again, that these traditional fit statistics are highly suspect with real data.

\subsection{Dimensionality of the CCI Data}

\begin{figure}
\centering
\includegraphics[width=5.5in]{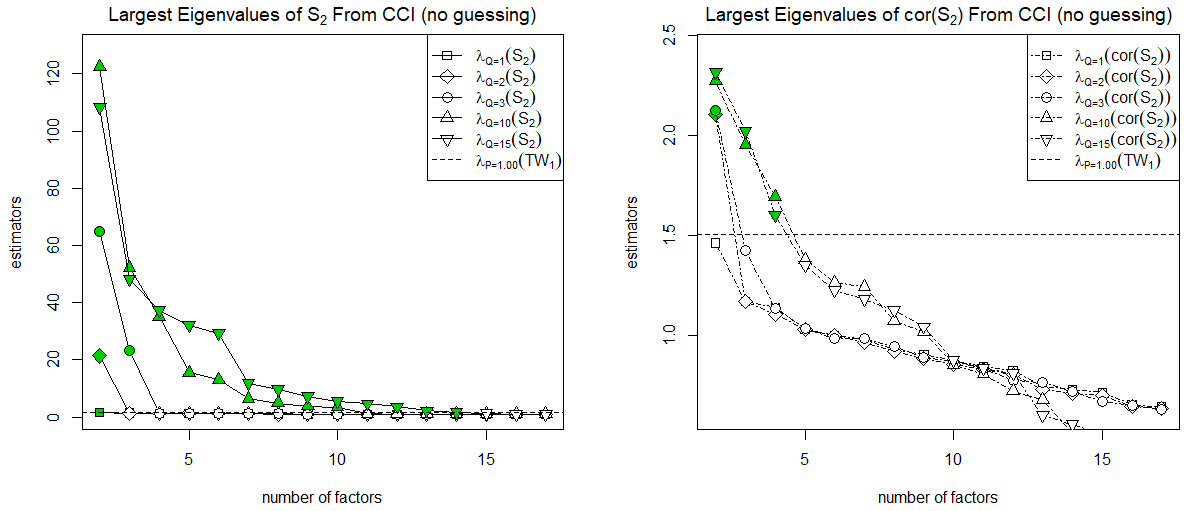}
\caption[Dimensional Sensitivity Analysis of CCI Data]{The eigenvalues calculated from the augmented data's covariance and correlation matrix at convergence are plotted for four configurations of dimensionality imposed on the SAEM algorithm's treatment of the CCI data. Eigenvalues from one, two, three, and four dimensions are displayed along with a line targeting a cumulative probability of 99\% of the Tracy-Widom distribution.}\label{fig:CCI_EV}
\end{figure}

The major contribution from the utility of the IRT methodology proposed in this dissertation is the extraction of the augmented data matrix. The covariance of $\mathbf{S}_2$ from the CCI data is subjected to the algorithm to be used for retaining the number of dimensions to be factor analyzed. In Figure~\ref{fig:CCI_EV}, the results from one through four dimensions \textit{without} guessing are shown. While the algorithm doesn't require us to use more than two dimensions, it is still useful to assess the general behavior of the eigenvalues of higher dimensional configurations due to the behavior of the model fit statistics, and the nature of the assessment being analyzed. Guessing was neglected due to the confluence of indications in the fit statistics of the univariate 2PNO favored over the 3PNO in Table~\ref{tab:CCI1Dfits}, as well as the strong asymptotic convergence to zero in the plots of the item characteristic curves. 

The factor retention algorithm from section~\ref{sec:MyAlgo} applied to this data results in two dimensions. Before diving further into the two dimensions implied by a subsequent rotation, it is noteworthy that the dimensions retained would differ if the guessing parameter was included; while the FCI showed a consistent result for the dimensionality regardless of the inclusion of a guessing parameter, the CCI would only result in one dimension to be retained if guessing is included. A sensitivity analysis will not be pursued here, but the multidimensionality of this assessment is suspect due to this empirical anomaly. 

When rotating these factors, as it was approached with the FCI, there is a substantial difference as the five items with highest discrimination do not show strong differential loading on the two factors near the intersection of the lines that correspond to balanced L2-norms and variance. Instead, at approximately 10 degrees there is very strong separation between the absolute values of the two dimensions of loadings for the five items shown.


At a rotation of 10 degrees, the factor loadings are extracted and presented in Table~\ref{tab:CCI_2D_Loadings}. When glancing at this table, it is clearly visible that two pairs of items \{10,11\} and \{12,13\} are the strongest items to load on each of the two factors, respectively. From the point of view of a former teacher, it is humorous to review the substance of these pairs of questions as they were deeply controversial in the classroom. Items 12 and 13 as discussed above, is a question about a solid that is evaporated within a closed container and the student is asked whether the weight has changed, followed by a question about the student's reasoning for their answer. In item 10, an ice cube is placed in a glass of water and the examinees are asked if the water level will change, with item 11 questioning the reasoning for the answer to item 10. 

\begin{table}
\centering
\begin{tabular}{crr}  
Item & First Loading & Second Loading \\ \hline
1 & -.117 & .342 \\ 
2 & -.079 & .413 \\ 
3 & .000 & .507 \\ 
4 & .085 & .562 \\ 
5 & .037 & .283 \\ 
6 & -.027 & .523 \\ 
7 & -.029 & .564 \\ 
9 & -.057 & .101 \\ 
8 & -.078 & .578 \\ 
10 & .984 & .571 \\ 
11 & .989 & .504 \\ 
12 & .208 & .906 \\ 
13 & .420 & .961 \\ 
14 & .002 & .100 \\ 
15 & -.064 & .322 \\ 
16 & .035 & .196 \\ 
17 & .010 & .225 \\ 
18 & -.117 & .618 \\ 
19 & -.112 & .682 \\ 
20 & -.136 & .264 \\ 
21 & -.045 & .355 \\ 
22 & .059 & .098 \\  \hline
\end{tabular}
\caption[CCI 2D Rotation]{With the choice of two dimensions and a rotation that creates mostly positive loadings and maximizes variation across the two dimensions, this table provides the normal scaled discriminations on each of the 22 items.} \label{tab:CCI_2D_Loadings}
\end{table} 

First, these pairs of questions are highly crystallized misconceptions; it is doubtful that these questions can be fully expanded into a continuum of ability. It is not a surprise that the discriminations within each of these dimensions, especially as they target a clearly defined fallacy, demonstrate an unambiguous signal. Due to the content of this assessment and the results above, the data was estimated with larger and larger dimensionality to scrape for signals of crystallized information by using the SAEM algorithm followed by an InfoMax rotation. This analysis was done in Appendix~\ref{sec:HDCCI}. It starts from mapping the conceptual domains of items, apply the SAEM method followed by factor rotations, examine the output, and iterate. This kind of interactivity and speed for working with empirical data is what makes the method in this dissertation pragmatic for a researcher. 

In the appendix in section~\ref{sec:HDCCI}, a 13-dimensional result gave remarkably distinct signals where each dimension had one or a pair of items that showed loadings well-differentiated from the rest. As a statistical researcher and trained expert in physical science, this assessment feels quite granular in its knowledge map; in taking the test myself, ten cursory conceptual domains were identified (described in section~\ref{sec:HDCCI}) but may split further justifying 13 or 14 dimensions. In using the language of CDMs (cognitive diagnosis models; \cite{de2014cognitively}), this test may work well for isolating crystallized attributes and running an efficient diagnostic that could enable an instructor to target misconceptions that are highly specific to the domains of the items in this test. The application of a CDM could do well to isolate latent classes of students that can benefit from similar remediation. The eigenvalue algorithm (see Appendix~\ref{sec:MyAlgo}) that resulted in two retained dimensions may not be a qualitatively effective approach for an assessment such as the CCI, but the ability to scale the dimensionality up and down with the availability of the resulting covariance matrix is proven quite useful in this analysis. 

\section{SAEM Applied to the Pediatric Quality of Life Scale}

In the third empirical exercise, the SAEM method is applied to response data for a tryout form of 24 polytomous items (five categories) assessing the social quality of life for the pQOL (pediatric quality of life) scale. This pQOL is intended as a health assessment of the social quality of life for children and includes items such as “I wished I had more friends” and “I felt comfortable with other kids my age”. The pQOL items have a Likert response scale that includes the alternative selections: “Never,” “Almost never,” “Sometimes,” “Often,” and “Almost Always.” Item response polarity was reversed, if necessary, to obtain positive valence. These data on 753 children will be instrumental as they were previously analyzed by \textcite{dewalt2013promis} and \textcite{cai2011irtpro}. Cai used the logistic form of item response functions and obtained solutions for one and five dimensions comparing two methods (EM with adaptive quadrature, MHRM); specifically, Cai compared thresholds in the univariate fit, factor structure in the five dimensional configuration, and computation time. The SAEM algorithm will be applied to this pQOL data with the ogive response function, then transforming the thresholds and factors to facilitate a strong comparison between commercially available methods and the code implemented here. 

\subsection{Descriptives of the pQOL Data}

Use of a quality-of-life scale implies a latent dimensionality that is qualitatively different from the purpose of the assessments studied in the previous sections; the inferential treatment of a \textit{score} is \textit{not} to be interpreted to quantify an examinee's state specific to a longitudinal progression toward mastery of a domain of knowledge. Typical deterministic assumptions of a \textit{difficulty parameter} do not map to qualitative social domains. Review of the nuances of the domain of social competence are beyond the scope of this dissertation, but it bears metion that one of the methodological challenges for assessments of social-emotional development is construct validity, as cohorts measured across time may respond to items relating to interpersonal relationships or social health differently \parencite{denham2009assessing}; cultural and parental context, as well as cognitive development, play significant roles in measures of social competence \parencite{rose1997nature}.

\begin{figure}
\centering
\includegraphics[width=5in]{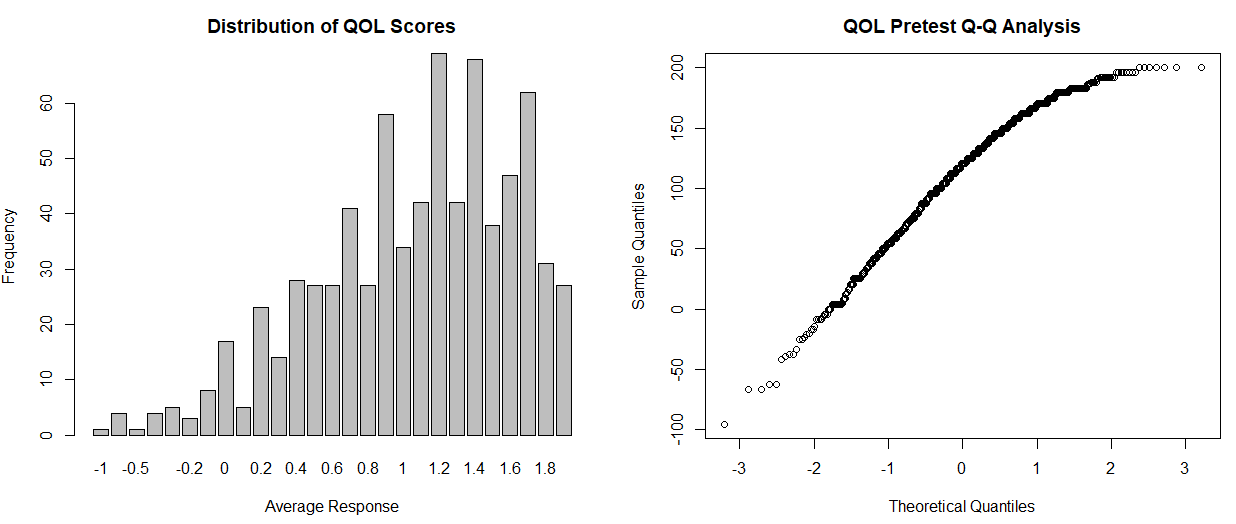}
\caption[pQOL Scores]{Using the average response as a score where the responses are ordinally ranked \{-2,-1,0,1,2\} to positively correlate with \textit{social health}, the distribution of pQOL scores are shown here.}\label{fig:QOLscores}
\end{figure}

A brief view of the outcomes in these data is performed to investigate the distribution of scores, as seen in Figure~\ref{fig:QOLscores}. In this histogram, the 5-point Likert scale for each item is quantitatively treated as a set of ordinal scores \{-2,-1,0,1,2\}; thus the mean is 1.12, the standard deviation is .55, and the distribution has strong negative skewness. The response curves are less useful in the context of the pQOL as there is no \textit{correct} response. For visual inspection, the plots of response curves are included in Figure~\ref{fig:QOLICC} in the appendix (section~\ref{sec:QOLComp}).

\subsection{Dimensionality of the pQOL Data}

\textcite{dewalt2013promis} began designing items of this assessment to be classified as either social function or sociability. The authors performed EFA (exploratory factor analysis) on data from four tryout forms, of which two contained sufficient samples of items for EFA. For one of these two forms, four dimensions were retained, while five dimensions were retained for the other with three common factors to both forms. The form comprising the data analyzed here was reported with four factors that were ``almost a bifactor model'' (\cite{dewalt2013promis}, pg. 7). 

\begin{figure}
\centering
\includegraphics[width=5.5in]{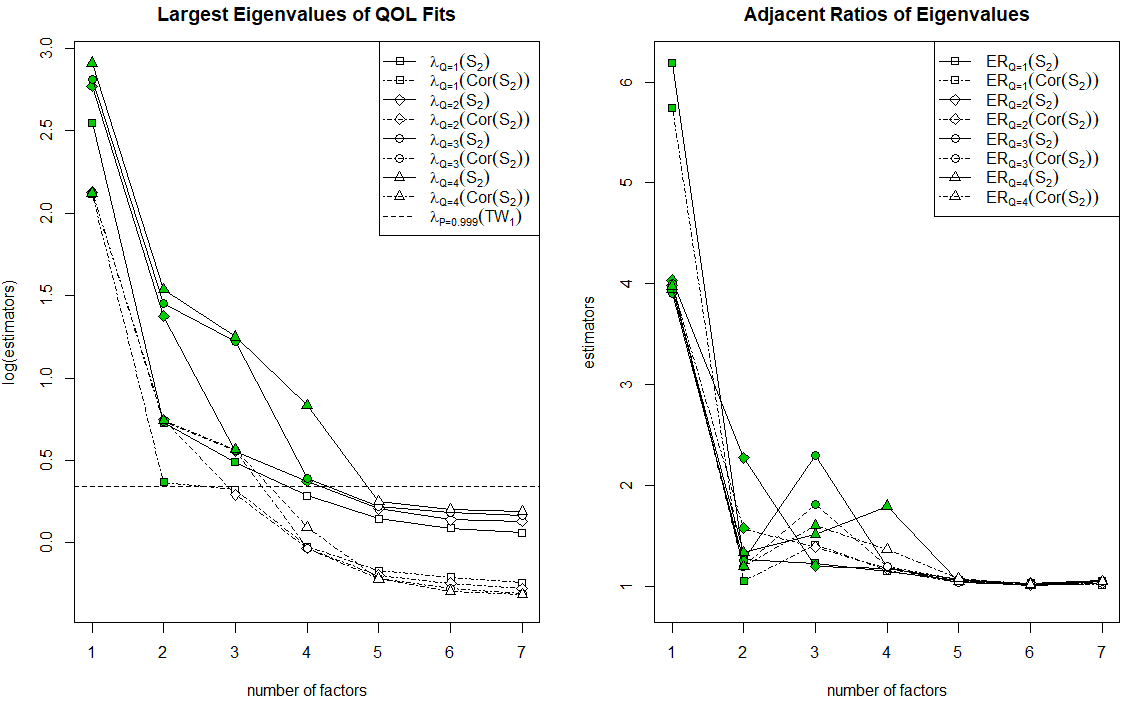}
\caption[Dimensional Sensitivity Analysis of pQOL Data]{The eigenvalues calculated from the augmented data's covariance and correlation matrix at convergence are plotted for four configurations of dimensionality imposed on the SAEM algorithm's treatment of the pQOL data. In the left plot, eigenvalues from one, two, three, and four dimensions are displayed along with a line targeting a cumulative probability of 99.9\% of the Tracy-Widom distribution. The right plot is a visualization of the adjacent eigenvalue ratios. Shaded points are significant with respect to the Tracy-Widom test.}\label{fig:QOL_EV}
\end{figure}

Using SAEM followed by the dimension retention algorithm presented in section~\ref{sec:EValgo}, the following steps apply:
\begin{enumerate} \label{eq:QOLDims}
 \item the data is configured for one dimension giving $\Lambda_C = 2,\Lambda_S = 3$,
 \item two dimensions are configured resulting in $\Lambda_C = 2,\Lambda_S = 4$,
 \item four dimensions are configured resulting in $\Lambda_C = 3,\Lambda_S = 4$, and finally
 \item three dimensions are configured resulting in $\Lambda_C = 3,\Lambda_S = 4$.
\end{enumerate}
\noindent Using this method gives a minimum of three dimensions to be retained. The visualization of this eigenvalue analysis is shown in Figure~\ref{fig:QOL_EV}. With the three factors extracted here, an oblimin transformation is performed and the final loadings are listed in Table~\ref{tab:QOL3loadings}. Oblimin was chosen as it is likely these dimensions may correlate. In fact, the first and second dimensions did correlate at .52, the first and third at .30, and the second and third at .27. 

The 3D solution was found to be highly interpretable. In Table~\ref{tab:QOL3loadings}, loadings are given for the unstandardized loadings rotated to simple structure by the oblimin transformation. All items loading on Factor 1 involve negative affect, with regard to either other kids, friends, or adults. This factor is primarily marked by two items: ``Other kids were mean to me,'' and ``Other kids made fun of me'' (polarity was reversed for negative statements). In contrast, the highest loadings on Factor 2 involve positive affect. Distinct factors for items having positive and negative valence is a common finding, especially with Likert-type items \parencite{tomas1999rosenberg}.  As noted by \textcite{alexandrov2010characteristics} the inclusion of mixed-valence wording can change the dimensionality of the construct. Because this is often considered to be a method effect, Alexandrov recommends the exclusive use of positively worded Likert items with a high level of intensity.

\begin{landscape}
\begin{table}
\centering
\begin{tabular}{lrrrrrrl}  
Item & $Q_1$ & SE($Q_1$) & $Q_2$ & SE($Q_2$) & $Q_3$ & SE($Q_3$) & Description\\ \hline
1 & .43 & (.06) & .33 & (.06) & -.04 & (.05) & I wished I had more friends. \\ 
2 & -.36 & (.05) & .26 & (.13) & .38 & (.11) & I want to spend more time with my family. \\ 
3 & -.01 & (.07) & .44 & (.05) & .07 & (.06) & I was good at talking with adults. \\ 
4 & .14 & (.09) & .69 & (.07) & .06 & (.11) & I was good at making friends. \\ 
5 & .46 & (.05) & .23 & (.05) & .15 & (.05) & I have trouble getting along with other kids my age. \\ 
6 & .36 & (.21) & -.06 & (.17) & .70 & (.05) & I had trouble getting along with my family. \\ 
7 & .84 & (.06) & .11 & (.18) & -.04 & (.18) & Other kids were mean to me. \\ 
8 & .46 & (.11) & -.07 & (.07) & .32 & (.08) & I got into a yelling fight with other kids. \\ 
9 & .12 & (.11) & .68 & (.04) & -.08 & (.09) & I felt accepted by other kids my age. \\ 
10 & -.37 & (.10) & .65 & (.16) & .51 & (.11) & I felt loved by my parents or guardians. \\ 
11 & .54 & (.07) & .13 & (.12) & .00 & (.13) & I was afraid of other kids my age. \\ 
12 & .82 & (.10) & -.05 & (.12) & .11 & (.11) & Other kids made fun of me. \\ 
13 & .64 & (.06) & .15 & (.08) & .18 & (.08) & I felt different from other kids my age. \\ 
14 & .38 & (.06) & .11 & (.06) & .04 & (.06) & I got along better with adults than other kids my age. \\ 
15 & .16 & (.12) & .70 & (.04) & -.07 & (.09) & I felt comfortable with other kids my age. \\ 
16 & .00 & (.05) & .43 & (.07) & .29 & (.06) & My teachers understood me. \\ 
17 & .17 & (.19) & -.05 & (.30) & .70 & (.10) & I had problems getting along with my parents or guardians. \\ 
18 & .13 & (.09) & .70 & (.06) & .23 & (.10) & I felt good about how I got along with my classmates. \\ 
19 & .42 & (.07) & .16 & (.05) & .29 & (.04) & I felt bad about how I got along with my friends. \\ 
20 & .53 & (.06) & .21 & (.08) & .10 & (.08) & I felt nervous when I was with other kids my age. \\ 
21 & .44 & (.04) & .31 & (.04) & .17 & (.05) & I did not want to be with other kids. \\ 
22 & .06 & (.12) & .72 & (.04) & -.05 & (.11) & I could talk with my friends. \\ 
23 & .15 & (.15) & .71 & (.05) & -.19 & (.10) & Other kids wanted to be with me. \\ 
24 & .07 & (.20) & .70 & (.07) & -.22 & (.14) & I did things with other kids my age. \\  \hline
\end{tabular}
\caption[pQOL 3D Oblimin Rotation]{With the choice of three dimensions and a rotation that creates mostly positive loadings and allows for correlations between dimensions, this table provides the discriminations on each of the 24 items of the pQOL response data.} \label{tab:QOL3loadings}
\end{table} 
\end{landscape}

The strongest loadings in Factor 3 were primarily items regarding family, with the two that loaded most strongly being specific to ``getting along with'' family. Two items that stick out in Factor 1 refer to kids being mean and making fun. When scanning the questions that load heaviest on Factor 2, it is clear that these refer to the disposition of peer relationships. Factors 1 and 2 were correlated, but nearly uncorrelated with Factor 3. This may suggest that social health is more heavily influenced by our peer-to-peer relationships more than our relationships with family. 

When focusing on items that bridge factors, item 16 is of some interest: ``My teachers understood me.'' Its loadings are relatively small and split across Factors 2 and 3. This may suggest a dual role of teachers as caregivers and friends. Item 10 ``I felt loved by my parents or guardians'' also suggests a similar dual role of parents. Items 2 and 10 seem to indicate that there may be an inverse relationship between the affinity of a child to adults and their peers; the items have significant negative loadings on the first Factor, but show strong loadings on the third Factor. 

For the standard errors in this run, let ${\bf{A}}^{R}$ represent the value of the rotated factor loading matrix in the Table~\ref{tab:QOL3loadings}. Standard errors for pQOL factor loadings were obtained by employing the central limit theorem of the multivariate MCMC. A chain of length 300 for each coefficient in $\bf{A}$ was generated post-convergence without invoking the Robbins-Monro procedure, i.e. $\gamma_t$ is fixed at $1$. On each cycle, each sampled value of $\bf{A}$ was rotated to the target ${\bf{A}}^{R}$ and saved. Standard errors (SE) were then obtained, as in chapter~\ref{ch:ss}, from the chain of rotated loadings with the \texttt{R} package \texttt{mcmcse} using the Tukey method, which provides slightly more conservative SEs than other methods(\cite{dai2017multivariate};\cite{flegal2012applicability}). The largest loadings exceed 4 SEs. Standard error computation as described above can be easily incorporated into any gibbs or Metropolis-Hastings algorithm for factor analysis. Note that flexMIRT\textsuperscript{\textregistered} does not yet provide standard errors for rotated coefficients.

This analysis illustrates the beneficial role of an exploratory approach. While this qualitative treatment can proceed further from the retention of three loadings, it is important to produce the outputs useful for comparison to the literature. For this purpose, five dimensions are extracted in the following section for straightforward comparison to the MHRM algorithm in \textcite{cai2010high}.

\subsection{Review of Quatitative Research Methods Citing Estimation of pQOL Data}

Using a univariate fit and transforming appropriately to match the graded response parameterization to the normal factor scale presented in \textcite{cai2011irtpro}, comparisons to the results of MHRM are shown in Table~\ref{tab:QOLThresholds}. Despite the use of a logistic in the MHRM algorithm, the parameters show good agreement. The slopes and thresholds from the ogive using SAEM are systematically smaller, especially where the response tend towards the tails of the score distribution, and this is expected as the logistic probability distribution has more probability mass in the tails of the latent variable being integrated. While the numbers may be different, the directional nature of the estimates is preserved. The ranking of highest discrimination is preserved in the first four items with the same ordinal position among all but seven items with the greatest shift being three (item 19). In studying the ordinal positioning of the thresholds, the results are similar; no item's threshold is rank-ordered differently by more than three positions. The spearman rank correlation is above .98 for the comparisons of all five parameters.

\begin{landscape}
\begin{table}
\centering
\begin{tabular}{ccccccccccc} 
Item &  \multicolumn{2}{r}{Threshold 1} &  \multicolumn{2}{r}{Threshold 2} &  \multicolumn{2}{r}{Threshold 3} &  \multicolumn{2}{r}{Threshold 4} &  \multicolumn{2}{r}{Slope}\\
\hline
{} & SAEM & MHRM & SAEM & MHRM & SAEM & MHRM & SAEM & MHRM & SAEM  & MHRM  \\ \hline
1 & 3.01 & 3.32 & 2.43 & 2.60 & .83 & .82 & -.08 & -.09 & 1.18 & 1.31 \\  
2 & 2.88 & 3.07 & 2.18 & 2.19 & .60 & .56 & -.61 & -.59 & .12 & .11 \\  
3 & 3.24 & 3.42 & 2.60 & 2.70 & .78 & .77 & -.40 & -.40 & .71 & .73 \\  
4 & 4.14 & 4.53 & 3.57 & 3.72 & 1.82 & 1.88 & .35 & .35 & 1.45 & 1.62 \\  
5 & 2.99 & 3.28 & 2.36 & 2.58 & .93 & 1.03 & -.40 & -.39 & 1.19 & 1.32 \\  
6 & 3.19 & 3.42 & 2.40 & 2.46 & .80 & .79 & -.42 & -.41 & .89 & .93 \\  
7 & 3.74 & 4.14 & 2.90 & 3.14 & 1.02 & 1.10 & -.53 & -.60 & 1.73 & 1.96 \\ 
8 & 3.30 & 3.45 & 2.63 & 2.66 & 1.18 & 1.13 & .12 & .11 & .82 & .82 \\  
9 & 2.98 & 3.58 & 2.47 & 2.94 & 1.26 & 1.54 & .09 & .17 & 1.34 & 1.67 \\  
10 & 3.68 & 4.04 & 3.55 & 3.77 & 2.73 & 2.74 & 1.66 & 1.62 & .85 & .84 \\  
11 & 3.95 & 4.65 & 3.60 & 3.91 & 2.50 & 2.65 & 1.32 & 1.33 & 1.08 & 1.22 \\  
12 & 3.81 & 4.22 & 3.09 & 3.30 & 1.63 & 1.70 & .21 & .21 & 1.53 & 1.71 \\  
13 & 3.47 & 3.88 & 2.55 & 2.71 & 1.03 & 1.07 & -.01 & -.05 & 1.48 & 1.70 \\  
14 & 2.46 & 2.52 & 1.63 & 1.60 & -.28 & -.28 & -1.26 & -1.24 & .80 & .87 \\  
15 & 3.27 & 3.93 & 2.81 & 3.26 & 1.24 & 1.50 & .08 & .19 & 1.4 & 1.69 \\  
16 & 3.34 & 3.61 & 2.83 & 2.91 & 1.17 & 1.20 & .09 & .12 & .85 & .91 \\  
17 & 3.38 & 3.65 & 2.65 & 2.69 & 1.25 & 1.23 & -.14 & -.14 & .68 & .71 \\  
18 & 3.77 & 4.46 & 3.39 & 3.85 & 1.60 & 1.77 & .22 & .26 & 1.56 & 1.88 \\  
19 & 3.46 & 3.98 & 2.84 & 3.12 & 1.73 & 1.81 & .43 & .47 & 1.10 & 1.33 \\  
20 & 4.09 & 4.72 & 3.30 & 3.61 & 1.69 & 1.74 & .32 & .33 & 1.30 & 1.48 \\  
21 & 3.93 & 4.42 & 3.20 & 3.43 & 1.76 & 1.81 & .54 & .55 & 1.30 & 1.48 \\  
22 & 3.91 & 4.64 & 3.56 & 4.03 & 2.46 & 2.78 & 1.27 & 1.38 & 1.33 & 1.62 \\  
23 & 3.69 & 4.20 & 3.33 & 3.61 & 1.17 & 1.22 & -.51 & -.55 & 1.33 & 1.51 \\  
24 & 3.65 & 4.17 & 2.95 & 3.20 & 1.55 & 1.60 & .03 & .04 & 1.18 & 1.32 \\   \hline 
\end{tabular}
\caption[pQOL Thresholds Compared to MHRM]{Using a one dimensional estimation, this table provides the structural parameters on a normal factor scale for each of the 24 items of the pQOL response data along with the estimates derived from the MHRM analysis in \textcite{cai2011irtpro}.} \label{tab:QOLThresholds}
\end{table} 
\end{landscape}

\begin{figure}
\centering
\includegraphics[width=5in]{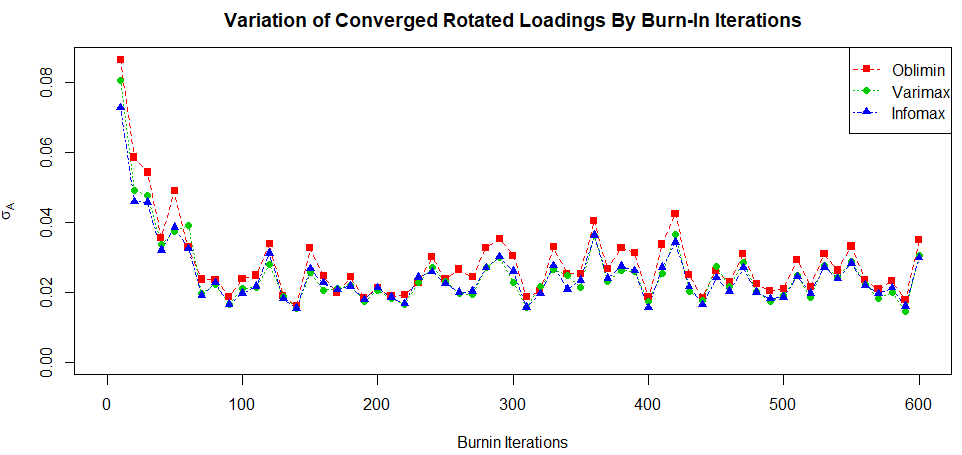}
\caption[Diagnostics of Univariate Slope Variability in pQOL Burn-in Iterations]{The SAEM algorithm was run on the QOL data with progressive burn-in iterations prior to engagement of the Robbins-Monro. The RMSE of the approximated slopes is plotted against the number of iterations used for burn-in.}\label{fig:QOL_Burn}
\end{figure}

In this univariate condittion, \textcite{cai2011irtpro} informs the reader that the MHRM implementation required 10 seconds of run time and took 128 cycles on a 2 GHz Intel Duo Core CPU with 2 GB of RAM. While Cai does not discuss a burn-in period for this specific estimation, a convergence criterion threshold is defined such that inter-iteration maximum parameter difference of .0001 over a window of three iterations must be met. The same convergence criterion was utilized here with the SAEM algorithm; reminding the reader, this implies that the extracted eigenvalues of the covariance matrix do not change more than .0001 for three full iterations. The overall time including 80 iterations of burn-in took 31 seconds and 433 cycles. The burn cycle may seem quite meager, but a study of multiple runs in Figure~\ref{fig:QOL_Burn} on one dimension of slopes is shown to give very little change in discrimination variability after the first 70 iterations.

This analysis was performed without any parallelization on a 3.6 GHz Intel Quad Core CPU with 32 GB of RAM. The RAM is somewhat inconsequential as the data set is small, but the increase in clock speed is nearly a factor of two. There are two likely explanations for the increased time: (1) the gibbs method is more free to probe the edges of the hyperspace than MHRM, and (2) the SAEM convergence criterion is not a result of a second-order derivative approximation, i.e. the computation of the Hessian. The convergence in SAEM is the result of a linear decomposition of the sufficient statistic. A third explanation is that this is not a compiled and optimized code package.

\begin{landscape}
\begin{table}
\centering
\begin{tabular}{ccrrcrrcrrcrrcrr}  
Item &  \multicolumn{3}{r}{Factor 1} &  \multicolumn{3}{r}{Factor 2} &  \multicolumn{3}{r}{Factor 3} &  \multicolumn{3}{r}{Factor 4} &  \multicolumn{3}{r}{Factor 5}  \\ \hline
{}  & TP & SAEM & MHRM & TP & SAEM & MHRM & TP & SAEM & MHRM & TP & SAEM & MHRM & TP & SAEM & MHRM \\ \hline
1 & X & .62 & .62 & 0 & .11 & .14 & 0 & -.12 & -.12 & 0 & .08 & .04 & 0 & -.06 & .02  \\  
2 & X & -.04 & .00 & 0 & .25 & .19 & X & .40 & .40 & 0 & .16 & .18 & 0 & -.08 & -.13  \\  
3 & X & .30 & .28 & X & .40 & .38 & 0 & .11 & .09 & X & .32 & .32 & 0 & .15 & .06  \\  
4 & X & .63 & .57 & X & .53 & .47 & 0 & .04 & .01 & 0 & .20 & .16 & 0 & -.03 & -.06  \\  
5 & X & .62 & .61 & 0 & .04 & .08 & 0 & .06 & .02 & 0 & .05 & -.01 & 0 & -.02 & .06  \\  
6 & X & .59 & .52 & 0 & -.12 & -.08 & X & .63 & .55 & 0 & .08 & .02 & 0 & -.01 & .05  \\  
7 & X & .86 & .76 & 0 & .05 & .05 & 0 & -.17 & -.09 & 0 & .12 & .06 & X & .73 & .55  \\  
8 & X & .46 & .44 & 0 & -.04 & -.04 & 0 & .27 & .24 & 0 & -.02 & -.01 & 0 & .26 & .27  \\  
9 & X & .60 & .55 & X & .55 & .52 & 0 & -.06 & -.08 & 0 & .14 & .07 & 0 & -.04 & .01  \\  
10 & X & .45 & .29 & X & .60 & .44 & X & .63 & .43 & X & .58 & .34 & 0 & -.13 & -.03  \\  
11 & X & .68 & .73 & 0 & -.14 & -.16 & 0 & -.11 & -.17 & 0 & .30 & .30 & 0 & -.04 & -.09  \\  
12 & X & .82 & .72 & 0 & -.02 & -.01 & 0 & .02 & .02 & 0 & -.10 & .00 & X & .75 & .51  \\  
13 & X & .72 & .75 & 0 & -.01 & -.01 & 0 & .08 & .06 & 0 & -.13 & -.08 & 0 & .18 & .13  \\  
14 & X & .52 & .50 & 0 & .00 & .01 & 0 & -.06 & -.03 & X & -.50 & -.51 & 0 & -.09 & -.05  \\  
15 & X & .61 & .56 & X & .57 & .54 & 0 & -.06 & -.06 & 0 & .06 & .04 & 0 & .03 & .01  \\  
16 & X & .36 & .34 & X & .37 & .34 & 0 & .30 & .28 & 0 & .00 & -.01 & 0 & .10 & .10  \\  
17 & X & .48 & .43 & 0 & -.03 & -.08 & X & .71 & .68 & 0 & -.07 & -.05 & 0 & .00 & .00  \\  
18 & X & .63 & .56 & X & .63 & .56 & 0 & .24 & .16 & 0 & -.04 & -.03 & 0 & .12 & .09  \\  
19 & X & .60 & .65 & 0 & .02 & .01 & 0 & .21 & .19 & 0 & -.25 & -.19 & 0 & -.07 & -.04  \\  
20 & X & .69 & .75 & 0 & -.02 & -.03 & 0 & -.01 & -.03 & 0 & -.05 & -.01 & 0 & -.11 & -.15  \\  
21 & X & .65 & .69 & 0 & .13 & .09 & 0 & .08 & .04 & 0 & -.18 & -.15 & 0 & -.08 & -.08  \\  
22 & X & .60 & .53 & X & .60 & .56 & 0 & -.08 & -.05 & 0 & -.08 & -.04 & 0 & -.06 & -.05  \\  
23 & X & .58 & .50 & X & .59 & .56 & 0 & -.17 & -.12 & 0 & -.05 & -.04 & 0 & .05 & .01  \\  
24 & X & .55 & .46 & X & .59 & .57 & 0 & -.19 & -.13 & 0 & -.15 & -.11 & 0 & -.05 & -.07  \\  \hline 
\end{tabular}
\caption[pQOL 5D Loadings Compared to MHRM]{With the choice of five dimensions and a target rotation that creates mostly positive loadings, this table provides the discriminations on each of the 24 items of the pQOL response data. The SAEM loadings transformed to the normal factor scaling is compared to the reported results \parencite{cai2010high} of the MHRM algorithm applied to the same data.} \label{tab:QOL5DLoadings}
\end{table} 
\end{landscape}

Moving to multidimensional estimations, \textcite{dewalt2013promis} presented a 4-factor solution, and \textcite{cai2010high} presented a 5-factor solution target rotated to a partially specified structure. The SAEM algorithm was also run on this pQOL data with a five-factor configuration followed by a linear transformation to the normal factor scale. Then the loadings are subjected to the same partially specified target rotation to advance a straightforward comparison to MHRM. Those results are in Table~\ref{tab:QOL5DLoadings}. For each of the 24 items, four thresholds were estimated along with factor coefficients, i.e. $24*(K-1) = 96$ intercept parameters were estimated along with $24*Q$ factor loadings where $K=5$ and $Q=5$. While the estimation methods from previous research provided factor loadings for a $Q = 4$ factor solution, \textcite{cai2010high} motivates a choice of five dimensions.

The selection of five dimensions is useful for outlining the advantages of these bayesian methods over EM estimation as EM suffers the curse of dimensionality, i.e. the exponential increase in grid points required for estimation. To this point, Cai reports the EM algorithm requiring over 50 times more CPU time than MHRM, which required only 95 seconds. The SAEM algorithm is configured for a confirmatory analysis to replicate the MHRM configuration. Again, there is no discussion of a burn-in period for this specific estimation, but the same maximum parameter difference of .0001 over a window of three iterations applies. In this exercise for SAEM, the computation time on the same 3.6 GHz machine took 140 seconds and 1717 cycles including the 80 burn-in iterations. It is notable that moving from 1D to 5D increased the overall computation time by only a factor of four.

%




\chapter{Discussion} \label{ch:dis}

This work is written with a focus toward a quantitatively apt but unfamiliar reader. Several applied mathematical methods are being utilized that exploit advanced techniques from linear algebra, probability, factor analysis, monte carlo, matrix rotations, time series, and regression. Where appropriate, these techniques are introduced with an explanation rather than an assumption of prior knowledge. 

\section{Research Summary}

Starting from a summary of item response theory, response functions are constructed using the ogive function. A linear expansion is applied to the latent ability to generalize to multiple dimensions interpretable as domains of information specific to a psychometric instrument. This model is constructed in the bayesian framework building on the gibbs method introduced by \textcite{Albert1992}, and exploits the augmented data matrix introduced by \textcite{MengSchilling1996}. Convergence of the MCMC chain was hastened using the algorithm of \textcite{robbins1951stochastic} on the sufficient statistics proven to converge to a global optimum by \textcite{delyon1999convergence}. 

The techniques above are applied to nine diverse conditions of simulated data to demonstrate the accuracy necessary for a reliable modeling methodology. The dimensionality was varied from small to large sample population sizes, from univariate to 10-dimensional configurations. A novel annealing technique is implemented to mitigate risks of engaging the RM within a local maximum of the likelihood's hyperspace. Standard errors are produced using several methods, from the analytic form of the likelihood function as well as empirically from the MCMC chain's variance.

A thorough exploration of dimensionality is also demonstrated; the psychometric application to multidimensional data requires inferential treatment of the values of the structural parameters output at convergence. The properties of the augmented data's covariance matrix allow the application of random matrix theory for retention of latent factors; properties of random matrices summarized by \textcite{johnstone2001distribution} built on the derivations of \textcite{tracy1996orthogonal}, facilitate algorithmic decisioning founded on fundamental distributional statistics instead of \textit{rules of thumb}. An algorithm for factor retention specific to this method is also proposed leveraging the insights gathered in the simulation studies. Unlike the most widely accepted parallel test, this approach uses the analytic form of the distribution of the largest eigenvalue and thus does not require simulations of similar covariance matrices in order to test a null hypothesis against imputed randomness.

This research was motivated by two factors: (1) the computational inefficiencies of estimation and analysis of multidimensional item response theory, and (2) the challenges of factor analysis when applied to psychometric instruments. One advantage of a bayesian framework is the avoidance of the exponential growth of a frequentist grid of point estimates when shifting from univariate to multivariate estimations. The disadvantage of the classical approach to MCMC is the extended duration of iterations required to reach a criterion for convergence; the uncertainty of the mean estimate is inversely proportional to the square root of the number of iterations. With the application of the Robbins-Monro algorithm and the proof of \textcite{delyon1999convergence}, the convergence of MCMC cycles is dramatically accelerated. With this combination of tools, computation is far more efficient. As the problems in chapter 3 and 4 were being investigated, the speed of this algorithm was helpful in the triage of the simulated and real data.

The process of running a factor analysis was also quite efficient as the extraction of the covariance is an artifact of this routine. The loadings are output as a matrix which can be manipulated using an assortment of strategies available in open source packages that leverage the \texttt{R} platform. Computational statistical methods leveraging functional programming enable a researcher to quickly implement quantitative inspection. Commercially available software packages typically feel `clunky' when shifting from one type of process, say MCMC, and then using the outcomes for another computation, for example factor analysis. Applying complex arithmetic, especially several methods or computations in tandem, is not an efficient activity when there is limited ability to manipulate code. Development of the functionality that was made available in one codebase made for a potent ecosystem in pursuit of these investigations. The code \parencite{GeisGit2019} is made publicly available along with all analyzed data (anonymized where necessary).

\section{Limitations and Further Research}

In the past few years, stochastic approximation and annealing methods are becoming more widely used in Bayesian modeling. The diagnostics of the MCMC chains are important to characterize in relation to the test information and the observed data. Parameter estimates are strongly influenced by large magnitudes in the intercepts, small magnitudes of factor loadings, and the prior of a guessing parameter. Gibbs sampling does not reject unlikely draws of parameters, and thus cross correlations and autocorrelations of structural parameters can destabilize the updates of the analytical forms of the Jacobian or Hessian. 

Several topics can be pursued for further investigation. The influence of correlation in the latent factors requires investigation. Inclusion of support for more item probability functions such as a nominal response function would substantially change the ordinal approach that has been developed. In reference to random matrix theory, the bulk distribution of eigenvalues of real data is a very interesting problem; multiple authors are currently researching RMT approaches to variability arising from empirical applications of spectral analysis\footnote{A good example is the work of \textcite{yeo2016random} who demonstrate a method of fitting the Marchenko-Pastur law to empirical data and using the parameters to empirically drive the Tracy-Widom test.}. This inferential approach for retaining factors and addressing dimensionality of assessments should also be applied rigorously to real large-scale assessment data. Confirmatory analyses that motivate inferences based on a univariate score do not adhere to a widely accepted diagnostic that justifies the collapse onto a univariate latent scale.

Missing response imputation is another field that could greatly benefit from a method that relies on the covariance matrix as a sufficient statistic. The SAEM procedure assumes marginal sufficiency, and incomplete response vectors do not pose a serious challenge for the method to model missing responses. Because the sufficient statistics are updated on each SAEM iteration, these statistics are available for missing data imputation under the assumption of MAR (missing at random; \cite{little2019statistical}). This leads to an integrated imputation-based procedure as opposed to a model-based procedure. The latter approach is typically employed in the full information approach of \textcite{BockAitkin1981}.

In regards to the codebase, there is ample room for improvement in tightening up the usability and generalizing the accepted format of response data. As this was implemented for the exploration of this research without a team sharing best practices in software development, expedience was prioritized. The implementation of multiple chains in a parallelized environment would also be useful for convergence diagnostics. Numerical approximations to the Hessian would be a substantial addition, requiring a flexible but robust method. Multidimensionality adjustments within the gibbs cycles can be approached programmatically for a set-it-and-forget-it user experience. As with all \texttt{R} packages, a vignette is also called for.

\newpage

\printbibliography
\newpage

\appendix

\chapter{Algorithm for Extraction of Assessment Dimensionality} \label{sec:MyAlgo}

The results in Chapter~\ref{ch:ss} allowed for an algorithmic approach to the number $\Lambda$ of factors to extract. The results for each of the simulation conditions (Table~\ref{tab:sims}) are valuable to list in order to provoke a logical selection strategy. For simplicity, the \textit{true} number of dimensions simulated is designated $\Lambda$, the number being estimated will be denoted as $\hat{\Lambda}$, the number of significant eigenvalues from the converged covariance matrix $\mathbf{S}_2$ is $\Lambda_S$, and the number of significant eigenvalues of $\mathbf{S}_2$ transformed to a correlation matrix is $\Lambda_C$. Note, this first enumeration is a survey of the results of the first five conditions in chapter~\ref{ch:ss} and is not meant to be imposed as rules or proofs.

\begin{enumerate} \label{eq:EVRules}
   \item When $\hat{\Lambda} = \Lambda$ , $\Lambda_S = \Lambda_C = \Lambda$.
   \item When $\hat{\Lambda} > \Lambda$,  $\Lambda_C = \Lambda$ for all but condition 1 and 2, and $\Lambda_S > \Lambda$
   \begin{enumerate}
     \item In condition 1, only the second eigenvalue remained significant, even when imposing the retention of 5 dimensions.
     \item Condition 2 showed a very noisy profile and implies that care is required as the Tracy-Widom is used on real data, where guessing clearly plays a role.
   \end{enumerate}   
   \item When $\hat{\Lambda} < \Lambda$, 
      \begin{enumerate}
      \item Condition 4 showed $\Lambda_S = \Lambda$, and $\Lambda_C = \Lambda - 1$
      \item Condition 5 showed $\Lambda_S = \Lambda_C = \Lambda$
     \end{enumerate}
  \end{enumerate}

From the listed results, the focus of the analysis using Tracy-Widom tests centralized on the use of $\Lambda_C$ over $\Lambda_S$. While $\Lambda_S$ has been accurate when $\hat{\Lambda} \leq \Lambda$, scaling the covariance to a correlation seems to clean the covariance matrix substantially, preventing confusion when $\hat{\Lambda} > \Lambda$. Highlighting these results, specifically as it relates to conditions 4 and 5, the data in Figure~\ref{fig:EVTestsS6S7} is confirmation that the results in \ref{eq:EVRules} are also sufficient to describe the results of conditions 6 and 7; when $\hat{\Lambda} \leq \Lambda$, $\Lambda_S = \Lambda$, and when $\hat{\Lambda} \geq \Lambda$,  $\Lambda_C = \Lambda$ except for the univariate dichotomous conditions. For the 5 choices of dimensionality in the configurations of the estimations of conditions 6 and 7, $\Lambda_C$ is within 1 factor of the true number of simulated factors, $\Lambda$. 

The application of the Tracy-Widom tests from the first seven conditions and dimensional configurations allows for comparison to the 10-dimensional conditions with only one new feature made evident in condition 8. In Figure~\ref{fig:EVTestsS8S9}, the $\Lambda_C$ provides a new result where it differs from $\Lambda$ by two eigenvalues when $\hat{\Lambda} = 8$. It is notable that in this condition, $\Lambda_S = \Lambda$. When increasing the dimensions retained to 9 and 10, $\Lambda_C = 9$ and $\Lambda_S = 10 = \Lambda$. Still, the rule is retained that when $\hat{\Lambda} \leq \Lambda$,  $\Lambda_S = \Lambda$. The 10-dimensional bifactor is the only result that does not have a result where $\Lambda_C = \Lambda_S = \Lambda$, but it is still an important one as it demonstrates the sensitivity of the weak bifactor structure. Condition 9 is consistent with the eigenvalue tests seen in the previous subscale configurations.

The consistency of the behavior of this analysis allows for the proposal of an algorithm for retaining factors. It starts by setting $\hat{\Lambda}$ to one, and the difference $\Lambda_S - \Lambda_C$ is set to infinity. The goal is to increase the factors as statistical information is extracted, and decreasing its upper limit in the interest of parsimony. The SAEM-IFA (stochastic approximation expectation maximization, item factor analysis) should be run using this univariate configuration. If $\Lambda_C$ implies a larger number of dimensions, the configured dimensionality should be increased to $\Lambda_C$ and re-fit. If $\Lambda_S$ is larger than the configured dimensionality, the SAEM-IFA algorithm should be run again with dimensionality set to $\Lambda_S$. At this point, it is important to look at the difference of $\Lambda_S$ and $\Lambda_C$, as each fit seems to favor the smallest $\Lambda_S$ where the difference between $\Lambda_S$ and $\Lambda_C$ is at a minimum. When this difference is non-zero, the results imply a weak factor exists. 

\begin{algorithm} \label{al:EValgo}
\caption{Set Dimensionality}\label{euclid}
\begin{algorithmic}
\Procedure{ResponseDimension}{}
\State $\hat{\Lambda} \gets 1$
\State $d_{i} \gets \infty$
\BState \emph{FitData}:
\State Run \textbf{SAEM-IFA}
\State $l_S \gets \Lambda_S$
\State $l_C \gets \Lambda_C$
\State $d_{f} \gets \Lambda_S - \Lambda_C$
\BState \emph{Test}:
\If {$l_C > \hat{\Lambda}$} 
    \State $d_{i} \gets d_{f}$
    \State $\hat{\Lambda} \gets l_C$
    \State \textbf{goto} \emph{FitData}
\ElsIf {$l_S > \hat{\Lambda}$}
    \State $d_{i} \gets d_{f}$
    \State $\hat{\Lambda} \gets l_S$
    \State \textbf{goto} \emph{FitData}
\Else
    \If {$d_f = 0$} \Return $l_C$
    \ElsIf {$d_f = d_i$ } \Return $l_C, d_f$
    \Else 
            \State $d_{i} \gets d_{f}$
    	 \State $\hat{\Lambda} \gets l_S - 1$
    	 \State \textbf{goto} \emph{FitData}
    \EndIf
\EndIf
\EndProcedure
\end{algorithmic}
\end{algorithm}

Applying this algorithm to each simulation condition in chapter~\ref{ch:ss} would result in the correct number of dimensions for all simulation conditions except condition 8, but it would return nine dimensions along with the difference $\Lambda_S - \Lambda_C$ which would imply one weak factor. This algorithm is also employed on the real data in chapter~\ref{ch:em} to allow a statistical motivation for the retention of factors from real item response data.

\chapter{Diagnostics of the FCI, CCI, and pQOL Data} \label{sec:TestDiagnostics}

In this section, the diagnostics of the data, response curves, and univariate fits are presented to facilitate comparison to published literature. It is common to see response curves in psychometric papers written by science educators, especially as the FCI is a foundational instrument. 

\section{FCI Data: Response Curves and Single Factor Estimations}

\begin{figure}
\centering
\includegraphics[width=5.5in]{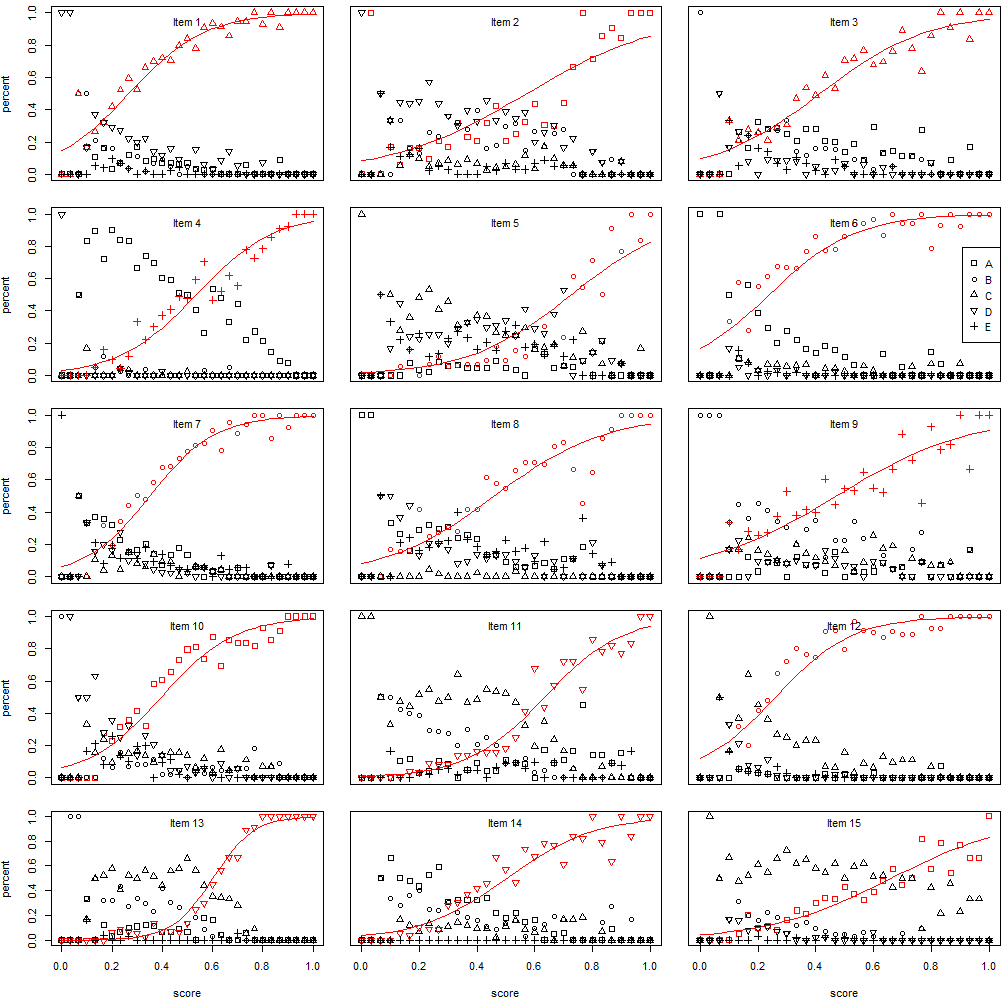}
\caption[Response Patterns of Questions 1-15 of FCI Data]{Within each of the 31 possible scores on the FCI, the respective proportions of categorical responses is shown for items 1-15. A logistic regression is fit to the correct response demonstrating the monotonic increase in probability of a correct response as score increases.}\label{fig:FCI_RP1}
\end{figure}

\begin{figure}
\centering
\includegraphics[width=5.5in]{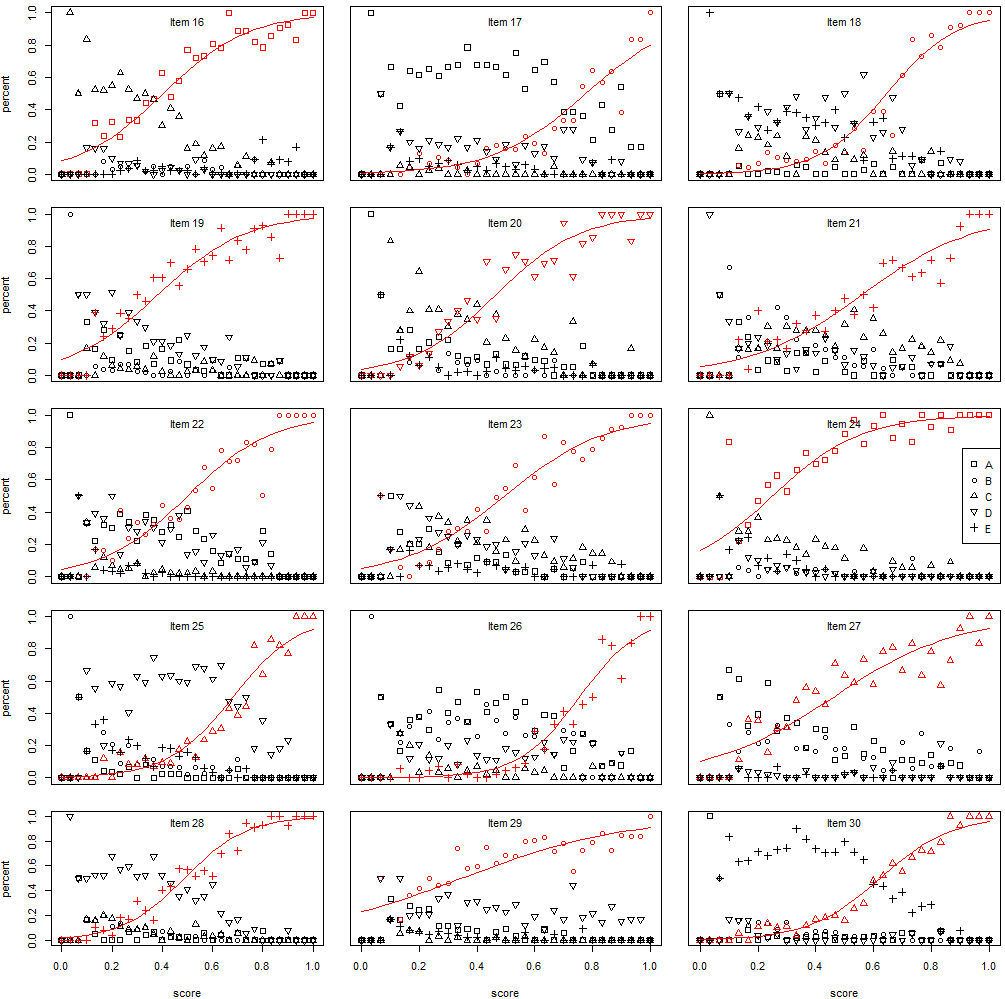}
\caption[Response Patterns of Questions 16-30 of FCI Data]{As with Figure~\ref{fig:FCI_RP1}, the respective proportions of categorical reponses for items 16-30 are displayed. Correct reponses are fit to a logistic on score.}\label{fig:FCI_RP2}
\end{figure}

Response curves of the FCI are shown in Figures~\ref{fig:FCI_RP1} and \ref{fig:FCI_RP2}. Each set of plots shows the proportion of answers being selected by students within each score bucket; the correct selections conform well to a two-parameter logistic regression on score (no assumption of a parameter for guessing). 

Univariate fits using a 2PNO and 3PNO are done to allow comparison to real FCI data estimations published in physics education research journals (\cite{wang2010analyzing}; \cite{chen2011comparisons}). The intercepts and slopes are scaled to approximate a 2PL and 3PL intercept as the cited articles use the logistic rather than an ogive response function.

\begin{table}
\centering
\begin{tabular}{crrrrrrr} 
Item & $a_{2PNO=2PL}$ & $b_{2PNO}$ & $b_{2PL}$ & $a_{3PNO=3PL}$ & $b_{3PNO}$ & $b_{3PL}$ & $g$ \\ \hline
$1$ & $.658$ & $-.703$ & $-.413$ & $.869$ & $-.229$ & $-.135$ & $.345$ \\  
$2$ & $.537$ & $.492$ & $.289$ & $2.791$ & $3.482$ & $2.048$ & $.239$ \\  
$3$ & $.575$ & $-.151$ & $-.089$ & $.836$ & $.330$ & $.194$ & $.235$ \\  
$4$ & $.712$ & $.273$ & $.160$ & $.689$ & $.288$ & $.169$ & $.000$ \\  
$5$ & $.668$ & $1.046$ & $.616$ & $2.995$ & $3.913$ & $2.302$ & $.089$ \\  
$6$ & $.543$ & $-.86$ & $-.506$ & $.613$ & $-.663$ & $-.390$ & $.238$ \\  
$7$ & $.698$ & $-.445$ & $-.262$ & $.962$ & $-.068$ & $-.040$ & $.218$ \\  
$8$ & $.507$ & $-.016$ & $-.009$ & $.685$ & $.306$ & $.180$ & $.137$ \\  
$9$ & $.382$ & $.009$ & $.005$ & $.861$ & $.97$ & $.570$ & $.332$ \\  
$10$ & $.702$ & $-.294$ & $-.173$ & $.721$ & $-.256$ & $-.151$ & $0$ \\  
$11$ & $.882$ & $.765$ & $.450$ & $1.431$ & $1.381$ & $.813$ & $.082$ \\  
$12$ & $.633$ & $-.834$ & $-.491$ & $.642$ & $-.785$ & $-.462$ & $.000$ \\  
$13$ & $1.439$ & $1.015$ & $.597$ & $4.569$ & $3.852$ & $2.266$ & $.060$ \\  
$14$ & $.787$ & $.141$ & $.083$ & $.834$ & $.181$ & $.107$ & $.000$ \\  
$15$ & $.497$ & $.521$ & $.307$ & $.738$ & $.964$ & $.567$ & $.131$ \\  
$16$ & $.59$ & $-.207$ & $-.122$ & $.681$ & $-.032$ & $-.019$ & $.113$ \\  
$17$ & $.581$ & $1.074$ & $.632$ & $1.229$ & $1.961$ & $1.154$ & $.078$ \\  
$18$ & $.867$ & $.88$ & $.518$ & $2.862$ & $2.974$ & $1.749$ & $.098$ \\  
$19$ & $.503$ & $-.279$ & $-.164$ & $.737$ & $.153$ & $.090$ & $.259$ \\  
$20$ & $.718$ & $-.001$ & $-.001$ & $.712$ & $.019$ & $.011$ & $.000$ \\  
$21$ & $.460$ & $.285$ & $.167$ & $.571$ & $.465$ & $.274$ & $.110$ \\  
$22$ & $.569$ & $.132$ & $.078$ & $1.056$ & $.756$ & $.444$ & $.201$ \\  
$23$ & $.739$ & $.224$ & $.132$ & $.801$ & $.307$ & $.181$ & $.000$ \\  
$24$ & $.575$ & $-.732$ & $-.431$ & $.664$ & $-.502$ & $-.295$ & $.182$ \\  
$25$ & $.736$ & $.913$ & $.537$ & $2.201$ & $2.64$ & $1.553$ & $.101$ \\  
$26$ & $.869$ & $1.379$ & $.811$ & $2.221$ & $2.893$ & $1.702$ & $.038$ \\  
$27$ & $.429$ & $-.068$ & $-.040$ & $.525$ & $.12$ & $.071$ & $.108$ \\  
$28$ & $.869$ & $.162$ & $.095$ & $1.212$ & $.608$ & $.358$ & $.116$ \\  
$29$ & $.306$ & $-.317$ & $-.186$ & $.375$ & $-.098$ & $-.058$ & $.125$ \\  
$30$ & $.896$ & $.773$ & $.454$ & $1.731$ & $1.661$ & $.977$ & $.080$ \\      \hline
\end{tabular}
\caption[FCI Structural Parameter Estimates From SAEM Univariate Fits]{Univariate 2PNO and 3PNO estimations are performed. The slope and difficulty coefficients are listed for each item. The estimation of the guessing parameter is also listed for the 3PNO. Slopes in the 2PL and 3PL are already scaled properly for comparison but the difficulty parameter must be divided by a scaling constant of 1.7 before attempting a general comparison to IRT fits from physics education research literature.} \label{tab:IRT1Dpars}
\end{table} 

Note, the parameters listed in Table~\ref{tab:IRT1Dpars} are not normal scaled as the published references described in section~\ref{sec:FCILit} do not scale their parameterizations.

\section{CCI Data: Response Curves and High Dimensional Exploratory Factor Analysis} \label{sec:HDCCI}

The response curves for the CCI data are presented in the following Figures, \ref{fig:CCI_RP1} and \ref{fig:CCI_RP2}. As it was quite difficult to locate a psychometric analysis of the CCI, it is believed that this is the first attempt to publish real data from this instrument. The response patterns conform well to a two parameter logistic regression on score, though a few items are noted to be rather difficult for the sample population. 

\begin{figure}
\centering
\includegraphics[width=5.5in]{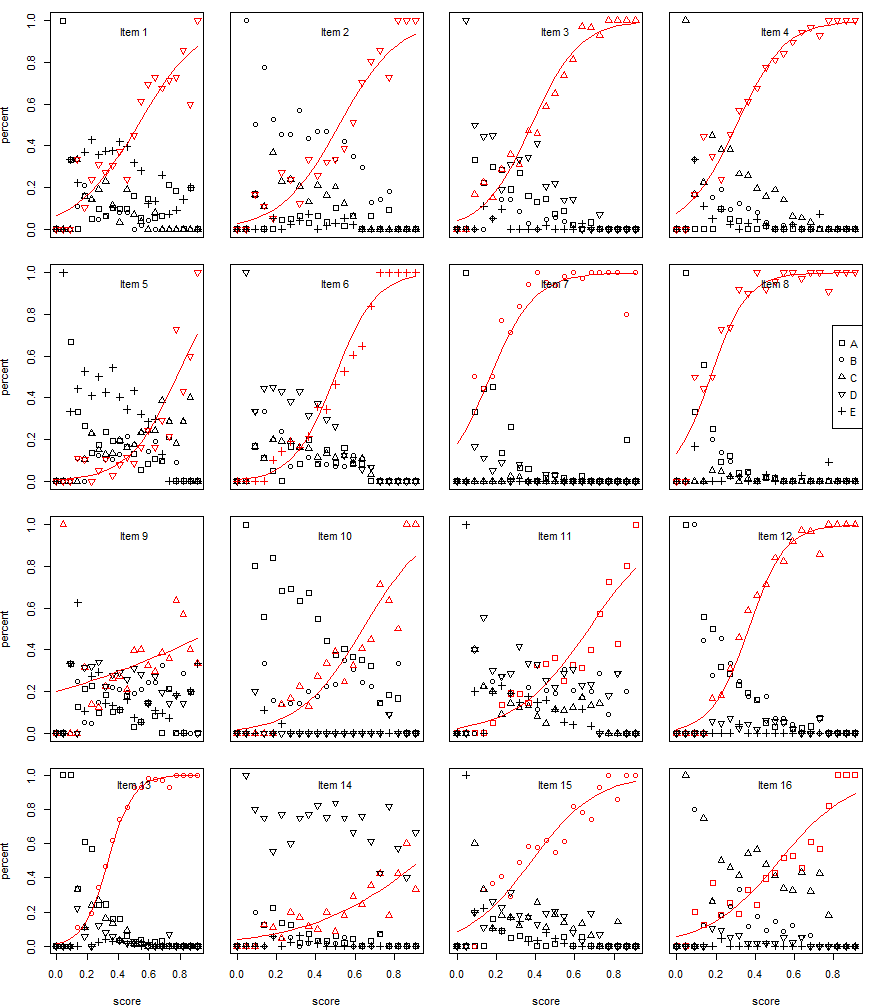}
\caption[Response Patterns of Questions 1-16 of CCI Data]{Within each of the 23 possible scores on the CCI, the respective proportions of categorical responses is shown for items 1-16. A logistic regression is fit to the correct response demonstrating the monotonic increase in probability of a correct response as score increases.}\label{fig:CCI_RP1}
\end{figure}

\begin{figure}
\centering
\includegraphics[width=5.5in]{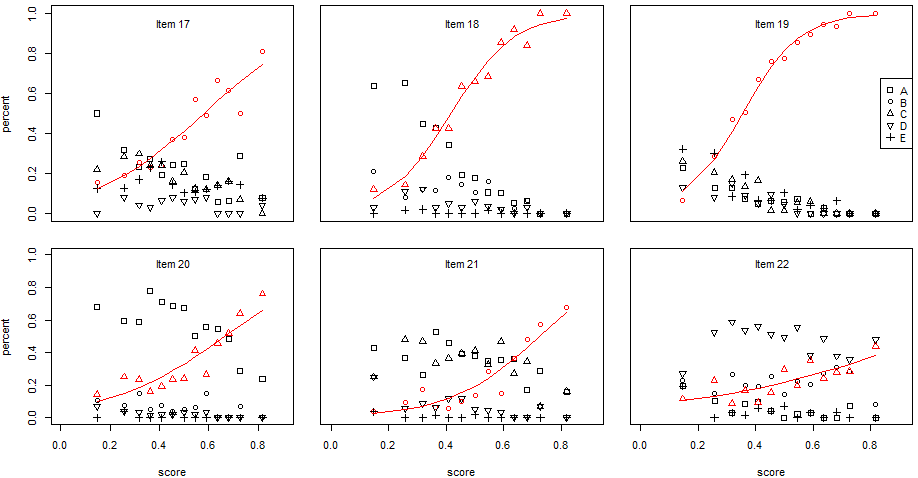}
\caption[Response Patterns of Questions 17-22 of CCI Data]{As with Figure~\ref{fig:CCI_RP1}, the respective proportions of categorical reponses for items 17-22 are displayed. Correct reponses are fit to a logistic regression on score.}\label{fig:CCI_RP2}
\end{figure}

To begin the qualitative investigation of the CCI, I took the test to map the conceptual domains of each question. As a trained physicist and former high school science teacher, it was informative to reflect on my academic experience to segment the conceptual map of the instrument. In taking the test, I identified \textit{a priori} 10 separate domains being assessed in the 22 items; listing them for qualitative inspection, they are: (1) conservation of matter and atoms, (2) phases of matter, (3) chemical reactions and their equations, (4) heat from reactions, (5) bouyancy and liquid displacement, (6) size of atoms in units of length, (7) proportional reasoning, (8) heat capacity, (9) solubility, and (10) bulk matter properties.

As this analysis progresses into higher dimensions, it is important to note that the SAEM method as applied to this data took less than a half hour to (a) crunch estimations of one through fifteen dimensions, (b) inspect their eigenvalue distributions, (c) rotate each output of loadings using an InfoMax rotation, and (d) examine the output. This kind of interactivity and speed for working with this empirical data is what makes this method pragmatic for a researcher. The fit statistics continue to favor higher and higher dimensions until the fifteenth dimension failed the likelihood ratio test. It is also notable that the fifteenth eigenvalue of the covariance matrix was the first configuration where one of the number of retained eigenvalues did not pass the Tracy-Widom test. Up to the fourteenth dimension, all retained eigenvalues of the covariance were found to be significant, while only four eigenvalues of the correlation matrix are significant for all configurations of retained dimensions greater than or equal to four. 

Since ten domains were selected for investigation, Table~\ref{tab:CCI_Qstudy} shows the results of a 13-dimensional confirmatory analysis with certain elements bolded to visually isolate the signal for each item. There is also an effort to map the converged estimates of loadings as they may approximate the domains I have considered.

\begin{landscape}
\begin{table}
\begin{tabular}{lcrrrrrrrrrrrrr} 
Item & Domain & $Q_1$ & $Q_2$ & $Q_3$ & $Q_4$ & $Q_5$ & $Q_6$ & $Q_7$ & $Q_8$ & $Q_9$ & $Q_{10}$ & $Q_{11}$ & $Q_{12}$ & $Q_{13}$   \\ \hline
1  &  Conservation 			& -.046 & .206 & .123 & .140 & .140 & .144 & .200 & .057 & -.040 & .126 & -.057 & .046 & .147 \\  
2  &  Phases				& -.074 & .325 & .343 & .598 & -.037 & .161 & .303 & .107 & -.124 & $\mathbf{.881}$ & -.176 & .022 & -.094  \\  
3  &  Phases 				& .328 & .791 & .368 & $\mathbf{.986}$ & .383 & .086 & .738 & .365 & .174 & .272 & .216 & .093 & .181  \\  
4  &  Conservation 			& .177 & .512 & .200 & .444 & .214 & .122 & .259 & -.183 & -.273 & .074 & -.062 & .089 & -.385  \\  
5  & Conservation \& Reactions 	& -.003 & .040 & .535 & .460 & -.156 & .594 & .494 & .169 & -.115 & -.193 & $\mathbf{.868}$ & .014 & -.143 \\  
6  &  Phases 				& .040 & .325 & .178 & .355 & .387 & .209 & .254 & -.005 & .019 & .609 & .264 & -.088 & .054  \\  
7  &  Conservation 			& .437 & .704 & -.248 & .351 & $\mathbf{.968}$ & .233 & .347 & -.312 & -.319 & -.088 & .119 & .168 & -.213  \\  
8  &  Conservation 			& .215 & .609 & .109 & .242 & $\mathbf{.963}$ & .081 & .598 & -.195 & .073 & .143 & -.276 & .001 & .145 \\  
9  &  Heat \& Reactions 		& -.106 & .091 & .079 & .057 & .021 & .023 & -.040 & -.052 & .022 & .104 & .119 & -.069 & .034  \\  
10  &  Bouyancy 			& $\mathbf{.985}$ & .350 & .003 & .327 & .097 & .182 & -.076 & .514 & .072 & .097 & .290 & -.132 & .251 \\  
11  &  Bouyancy 			& $\mathbf{.993}$ & .379 & -.059 & .072 & .260 & -.006 & .133 & -.035 & -.030 & -.065 & -.078 & .051 & -.231 \\  
12  &  Conservation \& Phases 	& .094 & $\mathbf{.977}$ & .416 & .533 & .354 & -.175 & .501 & -.009 & .138 & .020 & .317 & .213 & .165  \\  
13  &  Conservation \& Phases 	& .687 & $\mathbf{.985}$ & .310 & .595 & .588 & .459 & .764 & .490 & -.015 & .141 & -.073 & .022 & -.049 \\  
14  &  Size 				& .003 & -.038 & .037 & .076 & -.171 & .078 & .126 & .072 & -.076 & .057 & .308 & .482 & .123  \\  
15  &  Proportions 			& -.040 & .360 & .078 & .191 & .296 & .031 & .073 & .044 & .243 & -.016 & -.037 & $\mathbf{.803}$ & .009 \\  
16  &  Heat Capacity  		& -.006 & .142 & .283 & .434 & -.224 & $\mathbf{.814}$ & .303 & .044 & $\mathbf{.927}$ & -.104 & -.062 & .119 & -.304 \\  
17  &  Heat Capacity  		& .143 & .279 & .061 & .074 & .406 & $\mathbf{.972}$ & .369 & .026 & .559 & .170 & .382 & .018 & .205 \\  
18  &  Conservation \& Reactions  & .016 & .664 & .069 & .605 & .293 & .295 & $\mathbf{.971}$ & -.315 & .061 & .086 & .289 & .537 & -.081 \\  
19  &  Conservation \& Reactions  & .001 & .812 & -.048 & .585 & .542 & .167 & $\mathbf{.971}$ & -.030 & .116 & .066 & .059 & -.084 & .252  \\  
20  &  Solubility  			& -.198 & .055 & $\mathbf{.935}$ & .399 & -.044 & .229 & .183 & -.127 & -.444 & -.095 & -.113 & .013 & $\mathbf{.903}$ \\  
21  &  Solubility  			& -.018 & .583 & $\mathbf{.983}$ & .287 & -.012 & .036 & .008 & .655 & .311 & .282 & .392 & .024 & .547  \\  
22  &  Bulk Matter  			& .216 & .323 & .538 & .395 & -.424 & .032 & -.188 & $\mathbf{.975}$ & .033 & .030 & .080 & -.001 & -.035 \\ \hline  
\end{tabular}
\caption[CCI Multidimensional Item Analysis]{A qualitative inspection of the most apparent signal of information arising from each item.} \label{tab:CCI_Qstudy}
\end{table}
\end{landscape}

There are some items that have no clear mapping to one or more dimensions, specifically items 1, 3, 9, and 22. Items 9 and 22 were only answered correctly by 29\% and 17\% of students respectively; item 9 is an advanced concept about the release of heat when water is formed from combustion, and item 22 isolates whether a student understands that singular atoms do not exhibit the bulk properties of matter. Item 1 asks whether mass, molecules, and/or atoms are conserved through a reaction, and item 3 asks for the student to recall the origin of condensation outside of a cold glass of milk. Item 22 clearly sits alone conceptually from the rest of the test, but item 9 is a subtopic of chemical reactions, albeit the energy released from molecules forming which is not a topic of any other question. Item 1 has some weak loadings on dimension 1 and 3 which are also evident some of the other items (4, 7, 8, 12, 13, 18, and 19) that have been identified as relating to \textit{conservation of matter and atoms}.

Several pairs of items show very significant dimensional segmentation from other items. The item pairs \{7,8\}, \{10,11\}, \{12,13\}, \{16,17\}, \{18,19\}, and \{20,21\} are questions that probe a particular problem with a follow-up question that is phrased: ``What is the reason for your answer to the previous question?'' It is very interesting that each of these pairs show very strong isolation from the other dimensions. Item 18 is a notable outlier where the ninth dimension also shows a significant signal. This question is unique as it probes the change in mass of a nail that completely rusts. The mass will increase due to the combination with oxygen from air. Scanning the column of dimension 9, there are a few questions which also load on this dimension; most clearly item 5 shows a diagram of sulfur atoms and diatomic oxygen before chemically combining into sulfur trioxide with some leftover sulfur atoms. The ninth dimension is interesting as it seems to isolate item 14. Item 14 asks the student approximately how many carbon atoms would have to be strung side by side to make a dot this size $\bullet$; this is very different from every other question, but implies that there must be some domain around the concept of atoms, their size, their numbers, and their interactions with other atoms.

Dimension 7 and 10 also deserve some inspection. The content of item 4 involves 20 pounds of water and 1 pound of salt dissolved into it; the student is asked the weight of the final mixture. This is partially \textit{solubility}, but most certainly \textit{conservation of mass}. Item 13 shows some loading on this dimension and is a follow-up question to the equivalence of weight before and after iodine is evaporated in a closed tube, but the signal in item 13 from other dimensions implies that there is not a definitive statement that can be made other than some correlation in the knowledge required to answer both item 4 and 13 correctly. For the tenth dimension, items 2 and 6 are well represented with their strongest loadings on this domain. Item 2 asks about the content of the bubbles in boiling water. Item 6 asks the student which of five diagrams most accurately shows molecules of water after evaporation. It is worth noting that items 3 and 10 also involve the phases of water, and item 12 includes a phase change of solid iodine.

\section{pQOL Data: Response Curves and Comparisons to MHRM} \label{sec:QOLComp}

\begin{figure}
\centering
\includegraphics[width=5.5in]{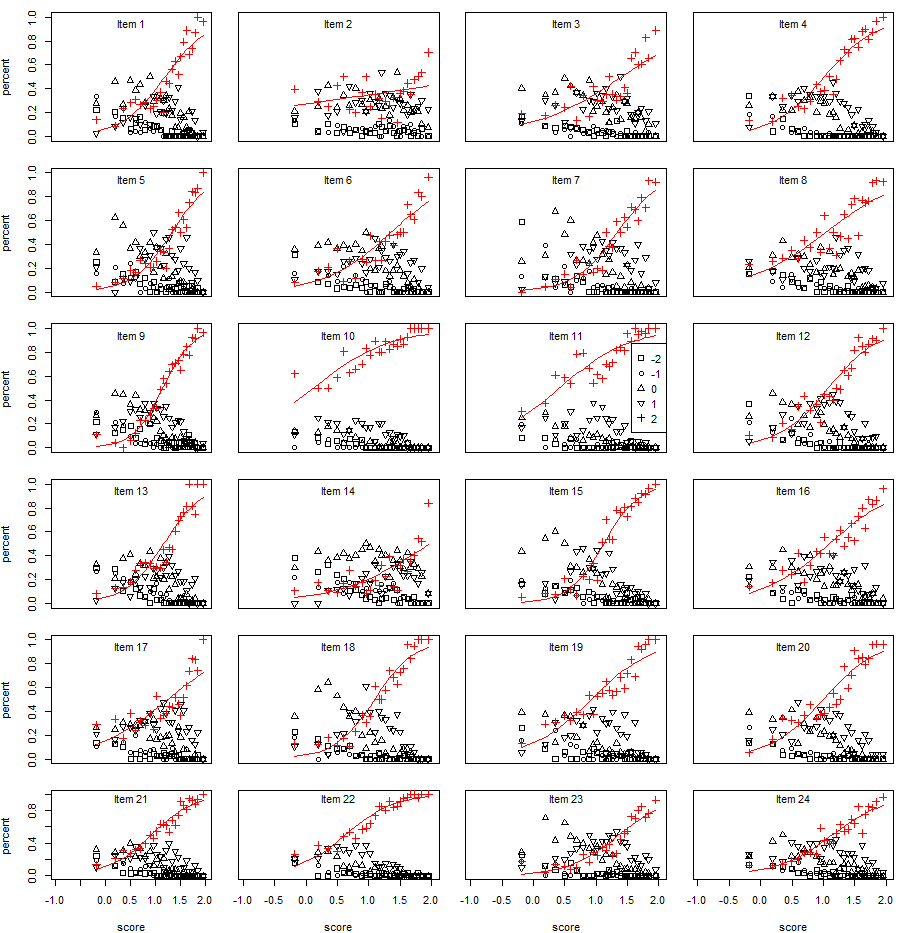}
\caption[Response Patterns of pQOL Data]{Within 25 bins of scores on the pQOL, the respective proportions of categorical responses is shown. A logistic regression is fit to the response that correlates the highest with social health, demonstrating the monotonic increase in probability of a this response as the pQOL score increases.}\label{fig:QOLICC}
\end{figure}

Using an ordinal breakdown and modeling the highest scored distractor as ``correct'', the response curves are shown in Figure~\ref{fig:QOLICC}. An independent treatment of each item allows a simple logistic regression to be performed. When viewing the items in Table~\ref{tab:QOL3loadings}, the distractors associating with abundance and scarcity can already be seen to be flipped to directionally orient the polarity of each item, in other words, where the Likert response most associated with social health is scored the highest.

\end{document}